\documentclass[11pt,letterpaper]{article}
\usepackage[dvipsnames]{xcolor}
\usepackage{authblk}
\usepackage{arxiv}
\usepackage{url}
\usepackage{amsmath,amssymb}
\usepackage{mathtools}
\usepackage{graphicx}
\usepackage{caption,subcaption}
\usepackage{authblk}
\usepackage{amsthm}
\usepackage{nameref}
\usepackage{float}
\usepackage{rotating}
\usepackage{nicematrix}
\usepackage[colorlinks,allcolors=red]{hyperref}
\usepackage{makecell}
\usepackage{cleveref}  

\usepackage{tikz}
\usetikzlibrary{shapes.geometric, shapes.misc, arrows, positioning, patterns}

\newsavebox\TBox
\def\textoverline#1{\savebox\TBox{#1}%
  \makebox[0pt][l]{#1}\rule[1.1\ht\TBox]{\wd\TBox}{0.4pt}}

\usepackage[section]{placeins}
\let\Oldsection\section
\renewcommand{\section}{\FloatBarrier\Oldsection}
\let\Oldsubsection\subsection
\renewcommand{\subsection}{\FloatBarrier\Oldsubsection}
\let\Oldsubsubsection\subsubsection
\renewcommand{\subsubsection}{\FloatBarrier\Oldsubsubsection}

\usepackage[normalem]{ulem}

\usepackage{booktabs}

\usepackage{enumitem}

\newcommand{\typeOne}{Kinetic (Type 1)}
\newcommand{\typeTwo}{Stoichiometric (Type 2)}

\newcommand{\secref}[1]{\cref{#1}}
\newcommand{\tabref}[1]{\cref{#1}}
\newcommand{\figref}[1]{\cref{#1}}

\setenumerate[0]{label=\roman*)}

\newcommand{\macro}[1]{#1}

\newcommand{\changedText}[2]{#2}

\theoremstyle{definition}

\theoremstyle{remark}

\newcommand{\mat}[1]{\boldsymbol{\mathrm{#1}}}
\renewcommand{\vec}[1]{\boldsymbol{#1}}

\setenumerate[1]{label=\arabic*.}
\setenumerate[2]{label=(\alph*)}

\newcommand{\argmin}{\mathrm{argmin}}

\begin{document}
\title{\textbf{Data-driven and Physics Informed Modelling \\ of Chinese Hamster Ovary Cell Bioreactors}}
\author[1,4]{Tianqi Cui}
\author[1,4]{Tom Bertalan}
\author[1]{Nelson Ndahiro}
\author[1]{Pratik Khare}
\author[1]{\authorcr Michael Betenbaugh}
\author[2]{Costas Maranas}
\author[1,3]{Ioannis G. Kevrekidis}

\affil[1]{Department of Chemical and Biomolecular Engineering, Johns Hopkins University, Baltimore, MD 21218, USA}
\affil[2]{Department of Chemical Engineering, The Pennsylvania State University, University Park, PA 16802, USA}
\affil[3]{Department of Applied Mathematics and Statistics, Johns Hopkins University, Baltimore, MD 21218, USA}
\affil[4]{These authors contributed equally.}
\date{}
\maketitle

\newcommand{\ddt}[1]{\frac{\mathrm{d}#1}{\mathrm{d}t}}

\newcommand{\numState}{\macro{K}}
\newcommand{\numParam}{\macro{P}}
\newcommand{\numFlux}{\macro{N}}
\newcommand{\numInt}{\macro{I}}
\newcommand{\numExt}{\macro{E}}
\newcommand{\numExtir}{\macro{E_{ir}}}
\newcommand{\numExtr}{\macro{E_r}}
\newcommand{\numMeta}{\macro{M}}
\newcommand{\rolloutLength}{\macro{L}}
\newcommand{\numRollouts}{\macro{N_\mathrm{run}}}

\newcommand{\state}{\macro{\vec{C}}}
\newcommand{\statePred}{\Tilde{\state}}
\newcommand{\param}{\macro{\vec{\alpha}}}
\newcommand{\netParam}{\macro{\vec{\theta}}}
\newcommand{\flux}{\macro{\vec{v}}}
\newcommand{\vI}{\macro{\vec{v}_I}}
\newcommand{\vE}{\macro{\vec{v}_E}}
\newcommand{\vEir}{\macro{\vec{v}_{E, ir}}}
\newcommand{\vEr}{\macro{\vec{v}_{E, r}}}
\newcommand{\LM}{\macro{\vec{\gamma}}}
\newcommand{\res}{\macro{\vec{r}}}

\newcommand{\Smat}{\macro{\mat{S}}}
\newcommand{\SmatI}{\macro{\mat{S}_I}}
\newcommand{\SmatE}{\macro{\mat{S}_E}}
\newcommand{\BI}{\macro{\mat{B}_I}}
\newcommand{\BE}{\macro{\mat{B}_E}}
\newcommand{\Bir}{\macro{\mat{B}_{ir}}}

\newcommand{\vIhat}{\macro{\hat{\vec{v}}_I}}

\newcommand{\fODE}{\macro{\vec{f_\mathrm{eqn:ode}}}}
\newcommand{\fkin}{\macro{\vec{f_\mathrm{kin}}}}
\newcommand{\foptim}{\macro{\vec{f}_{\mathrm{optim}}^{(\ell)}}}
\newcommand{\fint}{\macro{\vec{f_\mathrm{int}}}}

\begin{abstract}
    Fed-batch culture is an established operation mode for the production of biologics using mammalian cell cultures. Quantitative modeling integrates both kinetics for some key reaction steps and optimization-driven metabolic flux allocation, using flux balance analysis; this is known to lead to certain mathematical inconsistencies. 
    Here, we propose a physically-informed data-driven hybrid model (a “gray box”) to learn models of the dynamical evolution of Chinese Hamster Ovary (CHO) cell bioreactors from process data. 
    The approach incorporates physical laws (e.g. mass balances) as well as kinetic expressions for metabolic fluxes. 
    Machine learning (ML) is then used to  (a) directly learn evolution equations 
    (black-box modelling); (b)  recover unknown physical parameters (“white-box” parameter fitting) 
    or---importantly---(c) learn partially unknown kinetic expressions (gray-box modelling).
    We encode the convex optimization step of the overdetermined metabolic biophysical system as a differentiable, feed-forward layer into our architectures, connecting partial physical knowledge with data-driven machine learning.

\end{abstract}

\section{Introduction}
Chinese hamster ovary (CHO) cells are broadly used in biological and medical research, acting as the most common mammalian cell line used for the production of therapeutic proteins \cite{butler2005animal}. 
The advantage of using CHO cells is that the correct (i.e., mammalian-specific) glycosylation patterns are achieved for the protein therapeutics (e.g., therapeutic antibodies).  
Compared with conventional batch culture, fed-batch fermentation is more commonly used in this type of cell line, since it allows for easier control of the concentrations of certain nutrients that can affect the yield or productivity of the desired protein therapeutic molecule by ensuring the availability of precursor amino acids \cite{ma2009single}. 
However, lack of a complete, clear, quantitative model of the metabolism becomes an obstacle to achieving accurate and precise system simulation and control.

In the past several decades, mathematical models that incorporate physical knowledge have been extensively applied in the analysis of cell metabolism \cite{maranas2016optimization, Stephanopoulos1998Metabolic}. 
Metabolic Flux Analysis (MFA) leveraging stable carbon (i.e., 13C) labelled substrates  techniques is the only technique that can provide information on internal fluxes \cite{boghigian2010metabolic, goudar2010metabolic, quek2010metabolic}.  
Flux Balance Analysis (FBA), on the other hand provides a global inventory of carbon and energy resources throughout metabolism. 
By applying optimization principles, maximum theoretical yields for biomass formation or other products (e.g., metabolites or proteins) can be derived \cite{chassagnole2002dynamic, mahadevan2002dynamic}. 
Sometimes, for certain metabolic steps, detailed kinetic expressions are available that given the metabolite concentrations and enzyme levels can accurately estimate the flux through the metabolic reaction \cite{NOLAN2011108}. 
This requires the identification of the values of a number of enzymatic parameters. 
However, these expressions are usually available only for a subset of reactions,
necessitating a hybrid modeling approach, where optimization is used to identify metabolic fluxes for the remainder of reactions that lack kinetic expressions. 
This gives rise to a system of ordinary differential equations (ODEs) determined by the stoichiometry of the reactions.
In addition, given the fact that the metabolic reactions usually have relatively fast time constants (e.g. in the order of milliseconds to seconds) compared with other cellular processes like growth and death of cells, the pseudo-steady-state assumption (PSSA) suggests that the accumulation rate of any and every intracellular metabolite can be usefully approximated as zero. For some reactions, (ir)reversibility can be posited based on thermodynamics considerations; for others, reaction rates can be estimated from chemical kinetics considerations \cite{maranas2021recent}.

Data-driven approaches are today increasingly employed for identification of complex system dynamics, including traditional regression methods as well as neural networks and their variants 
\cite{adomaitis1990application,hudson1990nonlinear, rico1993discrete, Kutz2013, brunton_kutz_2022}. 
It is known, since the early 1990s, that neural networks embedded within numerical integrators can fruitfully approximate differential equations, and even learn corrections to approximate physical models, supplementing/enhancing them \cite{hudson1990nonlinear, adomaitis1990application, rico1992discrete, ricoMartinezAndersonKevrekidisGreyBox1994, rico1993discrete}. They can also be used to directly infer the evolution of the system variables when underlying physics are unclear \cite{martin2023physics, psarellis2022data, kemeth2022black, kemeth2022learning, lee2022learning}. Physics-Informed Neural Networks (PINNs) \cite{raissi2019physics}, Systems-Biology-Informed Neural Networks (SBINNs) \cite{SBINN, daneker2022systems}, and similar architectures \cite{DeepXDE}, can and have been used to solve supervised learning tasks while respecting known laws of physics, system biology, et al \cite{karniadakis2021physics}.
Nevertheless, as we will discuss below, the ambiguous structure of metabolic models creates nontrivial technical difficulties in exploiting partially known physical information from experimental fed-batch culture metabolic data; and can drastically affect the training process for gray box neural networks trying to infer such models from experiments. 
Our goal in this paper is to elucidate the nature of these modeling ambiguities,  demonstrating the ways in which they necessitate modifications of the architectures -and of the training- of traditional neural networks used for the identification task; and to implement networks capable of usefully identifying metabolic kinetics/parameters exploiting a synergy between physical modeling and scientific computation in neural network training.

\section{Methods}
\subsection{Structure of the Biophysical Model}
\label{sec:whiteboxStructure}

In a nutshell, the hybrid Chinese hamster ovary (CHO) bioreaction model we will use below (incorporating certain modifications (see \secref{sec:fkin} and \secref{sec:Nolan},) to the model presented in \cite{NOLAN2011108}, which constitutes our starting point) describes a continuous-time dynamical system (the terms are defined in \cref{tab:notation}:
\begin{equation}
    \ddt{\state} = \fODE(\state; \flux(\state; \param)); \label{eqn:ode}
\end{equation}

These evolution equations appear at first sight as simple ordinary differential equations 
(see \secref{sec:fODE} for expressions of \cref{eqn:ode});
yet, since evaluating the right-hand-side involves---as we will see---solving an optimization problem, 
we need another temporary label for the nature of the equations.
Connecting with existing literature
\cite{barton2002,barton2014,mahadevan2002dynamic}
we will here refer to these as {\bf D}ynamic {\bf F}lux {\bf B}alance {\bf A}nalysis (DFBA) equations.

Here, $\state \in \mathbb{R}^{\numState} \ (\numState = 14)$ 
are variables tracked by experiments
(which, though they might include concentrations of metabolites, cell densities, or other variables,
we will simply refer to as ``concentrations'' for simplicity, see \tabref{tab:concentrations});
$\flux \in \mathbb{R}^{\numFlux} \ (\numFlux = \numExt + \numInt = 35)$
are all fluxes (reaction rates, see \secref{sec:reactions} for all reaction expressions) including $\numInt$ intracellular fluxes (which can be precomputed from the $\state$)
    $\vI \in \mathbb{R}^{\numInt} \ (\numInt = 14)$
and $\numExt$ extracellular fluxes
    $\vE \in \mathbb{R}^{\numExt} \ (\numExt = 21)$.
Some of extracellular fluxes are assumed to be irreversible ($\vEir \in \mathbb{R}^{\numExtir} \ (\numExtir = 14), \vEir \geq 0$), while others are assumed reversible ($\vEr \in \mathbb{R}^{\numExtr} \ (\numExtr = 7)$); $v$ is a function of $\state$ and $\param$, where $\param \in \mathbb{R}^{\numParam} \ (\numParam = 45)$ are the kinetic parameters.

\begin{table}
\centering
\begin{tabular}{c|c|c}
    \toprule
    Notation & Variable & Dimension \\
    \midrule
    $\state$ & Variables tracked by experiments & $\numState = 14$ \\
    $\param$ & Kinetic parameters & $\numParam = 45$ \\
    $\vI$ & Intracellular fluxes & $\numInt = 14$ \\
    $\vEr$ & Reversible (extracellular) fluxes & $\numExtr = 7$ \\
    $\vEir$ & Irreversible (extracellular) fluxes & $\numExtir = 14$ \\
    $\vE$ & Extracellular fluxes & $\numExt = \numExtir + \numExtr = 21$ \\
    $\flux$ & All fluxes & $\numFlux = \numExt + \numInt = 35$ \\
    $\Smat$ & Stoichiometric matrix & $\numMeta \times \numFlux = 24 \times 35$ \\
    \bottomrule
\end{tabular}
\caption{Notation and dimensions for all variables used (the fact that here, $\numState = 14, \numInt = 14$ and $\numExtir = 14$ is a coincidence).}
\label{tab:notation}
\end{table}

Given $\state$ and $\param$,
the evaluation of $\fODE$ in \cref{eqn:ode} is typically done in one of two very different ways. 
Both involve the following steps, but differ in the particular combination of objective/constraints and the optimization approach used to enforce them. 

Given $\param$, and an initial set of values $\state_0$ for the concentrations, the time derivatives of the concentrations (e.g. RHS of Equation \cref{eqn:ode}) can be computed via the following steps.

\begin{enumerate}

    \item Compute preliminary updates of intracellular flux rates $\vIhat \in \mathbb{R}^{\numInt}\ (\numInt = 14)$ according to the concentrations $\state$ and given formulas of kinetic equations
    \begin{equation}
        \vIhat = \fkin(\state; \param) \label{eqn:kin_eq},
    \end{equation}
    where $\fkin: \mathbb{R}^{\numState \times \numParam} \mapsto \mathbb{R}^{\numInt}$ (see \secref{sec:fkin} for formulas of all kinetic expressions and \secref{sec:Nolan} for the changes of kinetic expressions we made based on the model in \cite{NOLAN2011108}). 
    
    \item The fluxes have to satisfy some constraints:
    \begin{itemize}
        \item Known kinetic expressions, i.e. Equation \cref{eqn:kin_eq}.
        
        \item The pseudo steady state assumption, which requires
        \begin{equation}
        \Smat \cdot \flux = 0, \label{eqn:PSS}
        \end{equation}
        involving the stoichiometric matrix 
        $\Smat \in \mathbb{R}^{\numMeta \times \numFlux}$
        ($\numMeta = 24$ is the number of metabolites at steady state, see \secref{sec:stoiMat} for all entries of $\Smat$).
        If we split the columns of $\Smat$ according to the $\numInt$ and $\numExt$ components (that is, $\SmatI = \Smat \cdot \BI, \SmatE = \Smat \cdot \BE$ where $\BI \in \mathbb{R}^{\numFlux \times \numInt}$ and $\BE \in \mathbb{R}^{\numFlux \times \numExt}$ 
        are two indicator matrices showing the intracellular and extracellular indices of all reactions), we have an equivalent form of \cref{eqn:PSS},
        \begin{equation*}
            \SmatI \cdot \vI + \SmatE \cdot \vE = 0
        \end{equation*}
        
        \item Among all 21 extracellular fluxes $\vE$, 14 of them are known to be irreversible, which requires
        \begin{equation}
            \vEir = \Bir \cdot \vE \geq 0, \label{eqn: bounds}
        \end{equation}
        where $\Bir \in \mathbb{R}^{\numExtir \times \numExt}$ is an indicator matrix containing the indices of irreversible fluxes among all extracellular ones.
    \end{itemize}
    Notice that the combination of Equations \cref{eqn:kin_eq} and \cref{eqn:PSS} consists of 38 independent linear equations,
    while the unknown variable $\flux$ is only 35-dimensional,
    leading to an overdetermined system.
    To address this issue,
    one can choose to satisfy some equations exactly, and others approximately (e.g. in a least squares sense).
    This leads to two substantially different approaches, the ``kinetic-based" and the ``stoichiometric-based",
    for computing intracellular flux rates $\vI$ and extracellullar $\vE$.
    It is important to state that these two approaches will, in general, lead to substantially different dynamic evolution for the same initial conditions of a metabolic kinetic scheme.
    
    \begin{enumerate}
        \item The kinetic-based approach: we satisfy the kinetic equations \cref{eqn:kin_eq}, and approximately satisfy the stoichiometric equations \cref{eqn:PSS}, which leads to
        \begin{equation}
        \begin{cases}
            \vI = \vIhat, \\
            \vE = \argmin_{\vE} ||\SmatI \cdot \vIhat + \SmatE \cdot \vE||_2^2, \text{ s.t. } \Bir \cdot \vE \geq 0.
        \end{cases} \label{eqn:kin_based}
        \end{equation}
        Here, we realize that the optimization problem is a linear least-squares problem with constraints (which implies it is actually a convex optimization problem). Moreover, if we ignored the constraints, we would be able to obtain an analytical solution for $\vE$ by computing the pseudo inverse of $\SmatE$:
        \begin{equation*}
            \vE = -(\SmatE)^+ \cdot \SmatI \cdot \vIhat. 
        \end{equation*}
       
       \item The stoichiometric-based approach \cite{NOLAN2011108}: we satisfy the stoichiometric equations exactly \cref{eqn:PSS}, and then approximately satisfy the kinetic equations \cref{eqn:kin_eq}, which leads to
       \begin{equation}
           (\vI, \vE) = \argmin_{\vI, \vE} ||\vI - \vIhat||_2^2, \text{ s.t. } \SmatI \cdot \vI + \SmatE \cdot \vE = 0, \Bir \cdot \vE \geq 0. \label{eqn:stoi_based}
       \end{equation}
       This is a least squares optimization problem with linear constraints. If we ignored the inequality constraints, we would obtain an analytical solution of $(\vI, \vE)$ by the Lagrange multiplier approach, see \secref{sec:stoi_sol} for details.
    \end{enumerate}
    
    To help with the numerics of the two embedded optimization problems,
    \newcommand{\fluxScaling}{1000}
    we in fact rescale  the supplied $\vIhat$ values: divide them by $\fluxScaling$ to
    adjust their numerical range to $\sim 1-10$
    upon entering either optization problem,
    and multiply the resulting fluxes $\vI$ and $\vE$
    by the same factor before exiting.
    This does not change the solution,
    but improves the numerical conditioning.
    
    After the optimization step we have
    \begin{equation}
        \flux = \BI \cdot \vI + \BE \cdot \vE = \foptim (\vIhat) \label{eqn:optim},
    \end{equation}
    where $\ell\in\{k,s\}$
    indicates whether we are using the \textcolor{red}{k}inetic or \textcolor{red}{s}toichiometric approach to finding fluxes.
    
    \item Compose \cref{eqn:ode}, \cref{eqn:optim}, and \cref{eqn:kin_eq}
    to create
    \begin{equation}
        \ddt{\state} = \fODE(\state; \foptim(\fkin(\state; \param))). \label{eqn:ode_2}
    \end{equation}
    As the fluxes $\flux$ are the reaction rates for each of the $\numExt + \numInt$ reactions,
    this follows directly from the stoichiometry of these reactions.
\end{enumerate}

Whether using the first or the second approach,
the resulting set of equations can subsequently be integrated
using an error-controlled integrator to obtain a full time series of all concentrations, for example,
\begin{equation}
\label{eqn:int}
    \state(t = t_{n + 1}) = \fint(\state(t = t_n); \fODE) = \state(t = t_n) + \int_{t_n}^{t_{n + 1}} \fODE(\state; \flux(\state; \param)) \ \mathrm{d}t,
\end{equation}
where $\{t_i: i = 0, 1, 2, \cdots \}$ is the set of equal-spaced timestamps.
It is important however to note that rate discontinuities potentially can (and actually do) arise at time instances when different constraints become active
(see \cref{fig:boundDisco}).
Note also that typical operating protocols of bioreactors
often call for the addition of species (e.g. nutrients) at particular time instances,
thus leading to temporal discontinuities in the system states.
We will illustrate both these types of discontinuities below in \secref{sec:wbData}.

Several important contributions
on which this paper is based
were established in previous work,
beginning with applications to \textit{E. coli} \cite{mahadevan2002dynamic}
and then proceeding to the more recent mammalian biomanufacturing targets 
\cite{NOLAN2011108}.
Beyond the constraints on our inner (optimization) problem
that we showed above,
additional constraints were imposed in \cite{mahadevan2002dynamic} on their outer (time-integration) problem,
such as non-negative metabolites
and limits on the rate-of-change of fluxes.
In their dynamic (resp. static) optimization approach (DOA, resp. SOA) they determined fluxes over an entire trajectory (resp. one trajectory segment, with constant fluxes).
Our simulations can be thought of as a form SOA, with the segment being a single integration step (as also in \cite{NOLAN2011108}). 

Before we start, a note on the computation of model gradients: 
many accurate integrators require the system Jacobian as well as sensitivities w.r.t. parameters.
This is also important for the integration
of differential-algebraic systems of equations
(differential equations with equality constraints).
Furthermore, these gradients
(w.r.t. state variables and/or parameters)
are crucial in identification tasks:
training neural networks to approximate the system equations
and/or estimate their parameters from data.
As we described above, our evolution equations
are not simple explicit ordinary differential equations,
but rather, their right-hand side arises as the result of solving an optimization problem,
depending on the current state.
This renders the accurate evaluation of these ODEs 
(as well as their sensitivity and variational computations) less straightforward than the explicit
right-hand-side case. 

\subsection{Black-box Model}
Our black-box model is a multi-layer perceptron (MLP) embedded within a numerical integrator scheme (e.g. the forward-Euler scheme or the Runge-Kutta template), where the MLP ($\mathrm{NN_b}(\cdot; \netParam)$) is used to learn the right-hand-side (RHS) of the ODE:
\begin{equation}
     \label{eqn:BlackBox}
     \statePred(t = t_{n + 1}) = \fint(\state(t = t_n); \mathrm{NN_b}) = \state(t = t_n) + \int_{t_n}^{t_{n + 1}} \mathrm{NN_b} (\state; \netParam) \ \mathrm{d}t.
\end{equation}
The details of generating the datasets can be found in \secref{sec:wbData}.
Note that here the right-hand-side depends only on the 
system state; it is also possible to make the Neural Network
 \cref{eqn:BlackBox}
dependent on physical input parameters
(such as feeding conditions
or basal gene expression rates),
by including these parameters as additional inputs to the NN function.
This will be important if, at a later stage, one wishes to optimize operating conditions towards some additional global objective (e.g.
maximal biomass production). 
This possibility has been demonstrated in older work
\cite{rico1992discrete}; it will not be repeated here.

\subsection{White-box and Gray-box Models}
\subsubsection{Model Structures}
In contrast with the black-box model which is purely data-driven, white-box and gray-box models benefit from existing physical knowledge,
leaving only the unknown parts of the model trainable.
In this paper, these two models have structure similar to that of what we deem the ground-truth biophysical model
(see \cref{fig:architecture}),
with changes limited to the computation of preliminary intracellular fluxes $\vIhat$.
While the white-box model assumes that some of the kinetic parameters $\param$
are unknown or need calibration,
the gray-box model suggests that part of the kinetic expressions $\fkin$ have no known functional form
and therefore replaces them with neural network approximations.
It is natural to also construct a mixed version of the hybrid model that contains
both unknown (``white-box'') kinetic parameters
and unknown (``black-box'') kinetic expressions,
resulting in an overall gray-box model.
\usetikzlibrary{fit}
\newcommand{\mainSpacing}{0.5cm}

\tikzstyle{operation} = [
    rectangle, rounded corners=0.3cm,
    inner sep=2mm,
    text centered, draw=black,
    align=center
]
\tikzstyle{branch} = [
    diamond, rounded corners=0.1cm,
    inner sep=-1mm,
    text centered, draw=black,
    align=center
]
\tikzstyle{variableQty} = [
    text centered,
    align=center
]
\tikzstyle{data} = [-stealth, line width=0.7mm]

\newcommand{\GBColor}{brown}
\newcommand{\WBColor}{lime}
\newcommand{\commonColor}{darkgray}
\newcommand{\BBColor}{purple}
\tikzstyle{PhysicsStyle} = [dash dot, data]
\tikzstyle{GBStyle} = [PhysicsStyle, draw=\GBColor, data]
\tikzstyle{WBStyle} = [PhysicsStyle, draw=\WBColor, data]
\tikzstyle{BBStyle} = [dotted, draw=\BBColor, data]
\tikzstyle{CommonStyle} = [draw=\commonColor, data]

\newcommand{\ySpacing}{1cm}

\newcommand{\numLinePattern}{dashed}
\newcommand{\numLineOpacity}{1}
\newcommand{\ddtsub}[1]{\frac{\mathrm{d} \vec C_{#1}}{\mathrm{d} t}}
\begin{figure}[!ht]
    \centering
\begin{tikzpicture}[node distance=1cm and 1cm,
]
\scalebox{.8}{ 
    \input{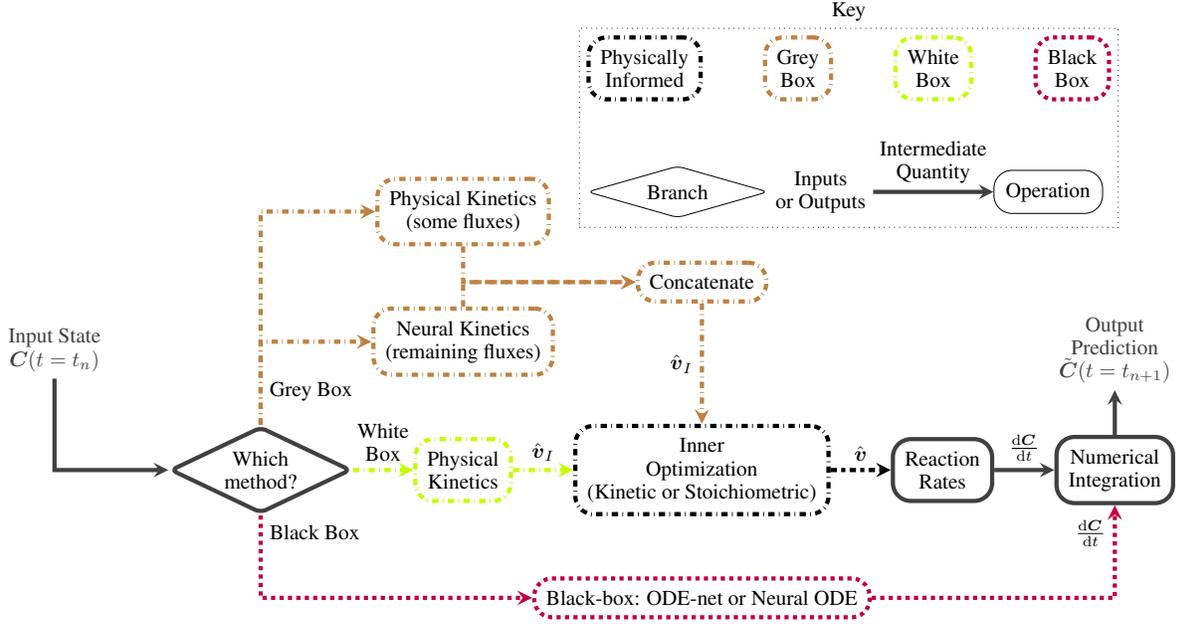}
}
\end{tikzpicture}

    \caption{
        \textbf{White, gray, and black inner architectures.}
        Operations are boxed,
        data or predictions are unboxed,
        and notable named intermediates are labeled on edges.
        Color and pattern are used to distinguish between
        model pathways that are distinct between the gray-, white-, and black-box approaches,
        or common to all three.
    }
    \label{fig:architecture}
\end{figure}

Though the white-box model superficially resembles typical parameter-fitting problems,
due to the presence of the inner optimization step in its evaluation,
traditional fitting approaches like general linear least-squares 
cannot be well-adapted into our framework.
Instead, a gradient-based fitting approach will be employed
using a differentiable convex optimization layer as described in \secref{sec:whiteBoxGradients}.
We remind the reader here that, since the original biophysical model 
includes two different approaches for computing fluxes (see \cref{eqn:optim}), 
we also make our white- or gray-box frameworks in two versions: 
that is, we use kinetic-version models to learn on the dataset generated from the kinetic approach, 
and stoichiometric-version models on the dataset that came from the stoichiometric approach.

\subsubsection{Computation of Gradients in Convex Optimization}
\label{sec:whiteBoxGradients}

\newcommand{\definedWord}[1]{\textbf{\emph{#1}}}
\newcommand*{\stext}[1]{\text{ #1 }} 
For this section in particular,
we need to define some terms.

The \definedWord{model} refers to the differentiable program used to make predictions: this
includes RHS evaluations (white-, gray-, or black-box),
perhaps necessitating an embedded convex optimization program
\cref{eqn:kin_based} or \cref{eqn:stoi_based}
\newcommand{\CP}{ECOP}
(\definedWord{\CP});
as well as the use of these RHS in numerical integration steps. 

The \definedWord{inputs} for this \CP{}
include both\begin{equation}
    \label{eqn:ListCPinputs}
    \begin{array}{rcl}
        \stext{constants} &
        \SmatI, \SmatE, \stext{and} \Bir;
        & \stext{and}
        \\
        \stext{outputs from upstream modules}
        &
        \vIhat
        &
        \stext{(function evaluations).}
    \end{array}
\end{equation}
All of these will be considered constant
for the purpose of solving the \CP{} for each call to the RHS.

\definedWord{Parameters} here refers to those quantities which could be modified by our outer training loop, including both
kinetic parameters $\param$ of the kinetic equations \cref{eqn:kin_eq}
(when we perform white box parameter estimation
or gray box parameter estimation); and
neural network parameters
$\netParam$,
i.e. trainable weights and biases
(when we train gray box networks
to recover unknown functional dependencies):
\begin{equation}
    \label{eqn:ListParams}
        \begin{array}{rcl}
        \param & \stext{and} & \netParam.
    \end{array}
\end{equation}
We will divide these
into trainable and untrainable (fixed) parameters
depending on the particular experiment.

The \definedWord{outputs} of the \CP{}
are the reported converged values of the \definedWord{variables} the problem solves for,
including both
\begin{equation}
    \label{eqn:ListCPVars}
    \begin{array}{rcl}
         \text{the fluxes} & \vE \stext{and} \vI;
         & \stext{and, possibly} 
         \\
         \stext{auxiliary variables (described below)} & \res.
    \end{array}
\end{equation}

For the purposes of training,
we would like our
loss function
(see \cref{eq: MSE} described for particular experiments
in later sections)
to be differentiable 
with respect to all of the trainable parameters \cref{eqn:ListParams}.
This requires that {\em the model predictions} be differentiable,
and therefore for each step in the model to be differentiable,
including the \CP{}, with respect to the same.

To enable this in a gradient-based computing framework such as PyTorch,
we turn to the package \texttt{cvxpylayers}
which was developed with such problems in mind
\cite{OptNetKolter}.
This package itself uses
\texttt{cvxpy} (a Python-embedded modeling language for convex optimization problems)
\cite{diamond2016cvxpy,agrawal2018rewriting,agrawal2019differentiable}
and \texttt{diffcp} (a Python package for computing the derivative of a cone program, which is a special case of convex programming)
\cite{diffcp2019,diffcp,amos2019differentiable}.

In order for these packages to correctly evaluate the gradient of the outputs
\cref{eqn:ListCPVars}
of the \CP{}
with respect to all of its inputs \cref{eqn:ListCPinputs},
certain structural characteristics must hold.
Specifically, the problem needs to be rewritten conforming with the rules of Disciplined Convex Programming (DCP) and of Disciplined Parametrized Programming (DPP).
DCP is a system for constructing convex programs
that combines common convex functions (e.g. $x^2, |x|$) with composition and combination rules
(e.g. $f \circ g$ is convex if both $f$ and $g$ are convex; nonnegative linear combination of convex functions is still convex).
If these rules are followed,
the library can automatically determine whether the full problem is indeed convex.

DPP is a subset of DCP, which further requires 
that all expressions
of the \CP{}
are affine with respect to the \CP{} inputs
\cref{eqn:ListCPinputs}.
It has been proved 
\cite{agrawal2019differentiable}
that a DPP-supportable convex program can be invertibly transformed into a cone program (and its derivative information can be obtained from \texttt{diffcp}). 
Therefore, DPP is mainly used in input-dependent convex programming, which allows the entire program to be differentiable without actually unrolling and back-propagating through the optimization loop.

Because DCP requires that the expressions in the \CP{} be affine w.r.t. the problem inputs \cref{eqn:ListCPinputs}, 
the product of two inputs is not an acceptable expression.
\newcommand{\resExp}{\SmatI \cdot \vIhat + \SmatE \cdot \vE}
This means e.g. that $(\resExp)^T \cdot (\resExp)$ in the objective function of \cref{eqn:kin_based} has to be reformulated.
This is resolved by the addition of another variable $\res$ in \cref{eqn:DCPkin} and \cref{eqn:DCPstoi},
and then equality constraints on this additional variable, such that the newly defined problems are equivalent to the old.
Further,
in the kinetic case,
to avoid the direct input product
$\SmatI \cdot \vIhat$,
we need to include $\vI$ as an optimization variable,
but then upgrade
what was a pre-optimization expression $\vI = \vIhat$ from \cref{eqn:kin_based}
to an actual equality constraint in \cref{eqn:DCPkin}.

In summary, for the kinetic approach,
we rewrite \cref{eqn:kin_based} as
\begin{equation}
\begin{aligned}
    \min_{\res, \vI, \vE} \ & ||\res||_2^2 \\
    \text{s.t. } & \Bir \cdot \vE \geq 0 \\
    & \res = \resExp \\
    & \vI - \vIhat = 0
\end{aligned}
\label{eqn:DCPkin}
\end{equation}
and for the stoichiometric approach,
we rewrite \cref{eqn:stoi_based} as
\begin{equation}
\begin{aligned}
    \min_{\res, \vI, \vE} \ & ||\res||_2^2 \\
    \text{s.t. } & \Bir \cdot \vE \geq 0 \\
    & \res = \vI - \vIhat \\
    & \resExp = 0
\end{aligned}
\label{eqn:DCPstoi}
\end{equation}
with inputs \cref{eqn:ListCPinputs}.

With these changes,
the two problems are DPP-compliant;
so,
we are able to evaluate derivatives of the problem outputs \cref{eqn:ListCPVars} (in particular, the argmins $\vI$ and $\vE$)
with respect to the problem inputs \cref{eqn:ListCPinputs}
and also evaluate vector-Jacobian products as needed in a larger 
PyTorch back propagation
to eventually get loss gradients w.r.t. parameters \cref{eqn:ListParams}.

\subsection{Auto-regressive Loss}

For supervised learning of the dynamics underlying time series data,
one approach is to use the ground truth prediction/output from a prior time step as the input for the current time step,
which leads to the teacher-forcing method (also known as professor-forcing in \cite{prof_force}).
Alternatively, we could use the model prediction from the prior time step as input,
which is called ``autoregressive training".
In fact, if contiguous data trajectories are divided into episodes of $M$ steps each,
and $M$ reduced to $2$, we see that the first is in fact a special case of the second.
So, in general, we use an autoregressive model structure
(however, see also \cref{eqn:BlackBox}),
which means the forward pass of the model can be written as
\begin{equation}
    \statePred(t = t_{i + 1}) = \mathrm{Model} (\statePred(t = t_i)), \statePred(t = t_0) = \state(t = t_0), \label{eq: MSE}
\end{equation}
where $\mathrm{Model}$ can represent the integration of the black-, white- or gray-box RHS.
The mean-squared loss (MSE) between the two time series can be therefore computed as
\begin{equation}
    \mathrm{MSE}(\{\statePred(t = t_i)\}, \{\state(t = t_i)\}) = \frac{1}{\numState \rolloutLength} \sum_{j = 1}^{\rolloutLength} ||\statePred(t = t_j) - \state(t = t_j)||_2^2,  
\end{equation}
where $\rolloutLength+1$ is the length of the dataset $\{\state(t = t_i): i = 0, 1, 2, \cdots, \rolloutLength \}$ and $\numState = 14$ is the dimension the state vector
as we have shown in \tabref{tab:notation}.

\section{Results}
\label{sec:experiments}

In this section, we will describe each of the several computational experiments tabulated in \cref{tab:experiments}
which we performed in this paper.
We first begin by describing the data generation procedure that was used for each of the parameter-identification
and neural-network training experiments that follow.
We include an analysis of the impact of the constraints
in the inner optimization problem,
considering events when constraints switch to (resp. from) active (resp. inactive).
We then begin our actual training experiments
with a black-box example.
All of our training experiments
include both kinetic and stoichiometric variants.
Subsequently, we perform white-box identification, in both two-free-parameter and five-free-parameter variants.
Finally, we will perform a mixture of these two tasks with gray-box modeling:
    First we will use a neural network to replace one of the kinetic expressions in $\fkin$;
    then we will repeat this, also allowing one of the physical parameters $\param$ to the trainable.

\begin{table}[h]
    \centering
    \begin{tabular}{rl}
        \textbf{Experiments} & \textbf{Section}
        \\ \hline
        Data Generation & \S\ref{sec:wbData}
        \\
        Black-box & \S\ref{sec:blackBox}
        \\
        White-box (2 Parameters Unknown) & \S\ref{sec:wb2p}
        \\
        White-box (5 Parameters Unknown) & \S\ref{sec:wb5p}
        \\
        Gray-box (1 Expression Unknown) & \S\ref{sec:grayBoxG}
        \\
        Gray-box (1 Expression Unknown + 1 Parameter Unknown) & \S\ref{sec:grayBoxW}
        \\
        \hline
    \end{tabular}
    \caption{
    Summaries of computational experiments in this paper.
    }
    \label{tab:experiments}
\end{table}

\subsection{Data Generation}
\label{sec:wbData}

We begin by
simulating short trajectories for a variety of initial conditions,
and collecting these flows as a dataset {\textit for a single set of parameter values}.
We then implement the Neural Network model
in PyTorch exactly as described in \secref{sec:whiteboxStructure}
and trained to match these flows.

\begin{figure}[!ht]
    \centering
        \includegraphics[width=0.96\textwidth]{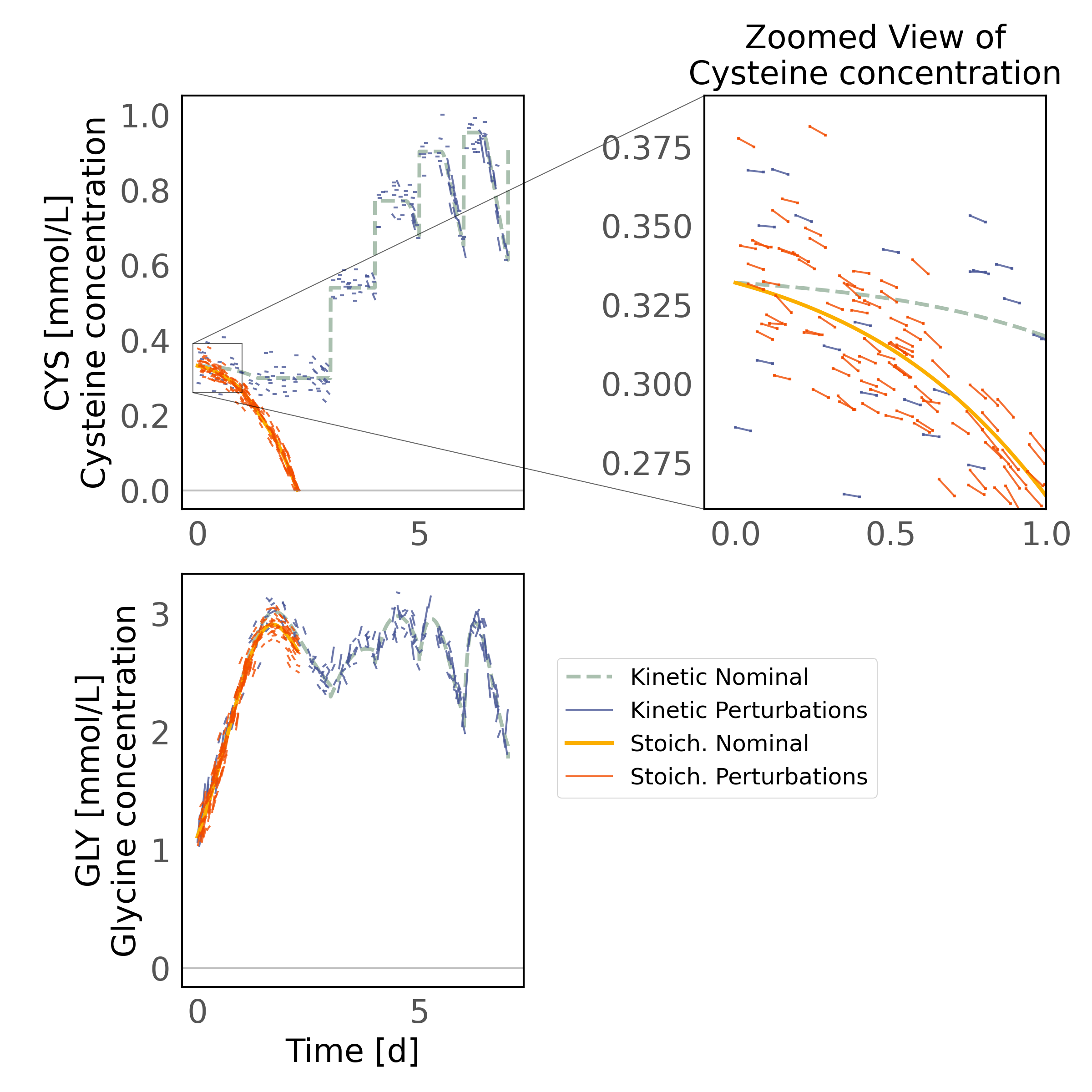}
        \caption{Nominal trajectories (kinetic and stoichiometric) overlaying sampled short-time flows from perturbed initial conditions. Trajectories for only a few of the $\numState$ variables are shown.
        Trajectories for all variables are shown in \figref{fig:nominalAll}.
        Color is used to distinguish between different curves described in the legend.
        }
    \label{fig:nominal}
\end{figure}

The dataset consists of $\numRollouts$ transients of the full model \cref{eqn:ode}, or equivalently, \cref{eqn:ode_2},
from initial conditions (ICs) taken as Gaussian perturbations around means; the means themselves are sampled uniformly (in time) at random along a central nominal trajectory (NT).
The per-variable standard deviations are proportional to the extent of variation of that variable in the NT.
That is, the sample of ICs is given by
\begin{equation}
    \label{eqn:ICs}
    \left\{
    \state^{(i)}(t = 0)
    =
    \left[
    \begin{array}{rcl}
         C_1^{(i)}(t = 0) & \sim & \mathcal{N}(\bar{C}_1(t = t_i), \sigma_1)
         \\
         &\vdots
         \\
         C_\numState^{(i)}(t = 0) & \sim & \mathcal{N}(\bar{C}_{\numState}(t = t_i), \sigma_{\numState})
    \end{array} 
    \right] \Bigg| i = 1, 2, \cdots, \numRollouts
    \right\},
\end{equation}
where the nominal trajectory $\bar{\vec{C}}(t) = (\bar{C}_1(t), \bar{C}_2(t), \cdots, \bar{C}_{\numState}(t))^T$ starts from a particular set of initial conditions that were measured during a laboratory experiment. 
The feeding events were implemented as state discontinuities.
This nominal trajectory (in both its ``kinetic" and its ``stoichiometric" integrations) appears in \figref{fig:nominal};
the ``stoichiometric" NT  stops just before the state goes negative
(and thus before any feeding events have occurred).

Each initial condition from \cref{eqn:ICs}
is then accurately simulated
(with an
order $8(5,3)$
explicit Runge-Kutta method
\cite{dop853}
with absolute tolerance $10^{-8}$ and relative tolerance $10^{-7}$)
to a time horizon 
generally significantly shorter than the entire nominal trajectory,
(circular dots in \figref{fig:bb_results_kin} and \figref{fig:bb_results_stoi})
giving us data in the form of several trajectory ``windows", constituting one ``episode".

\subsubsection{Detecting transitions in constraint activity during
simulation}
The inner optimization described in \cref{sec:whiteboxStructure} includes bounds on some of the fluxes computed (lower bounds indicating irreversible reactions). 
Depending on the current state of the simulated variables $\state$, the optimal \textit{unconstrained} fluxes may not lie inside the bounded domain. 
The \textit{constrained} optimum will then instead lie 
on constraint boundaries (or even possibly intersections of them).

\begin{figure}
    \centering
    \begin{subfigure}[b]{0.45\textwidth}
        \includegraphics[width=\textwidth]{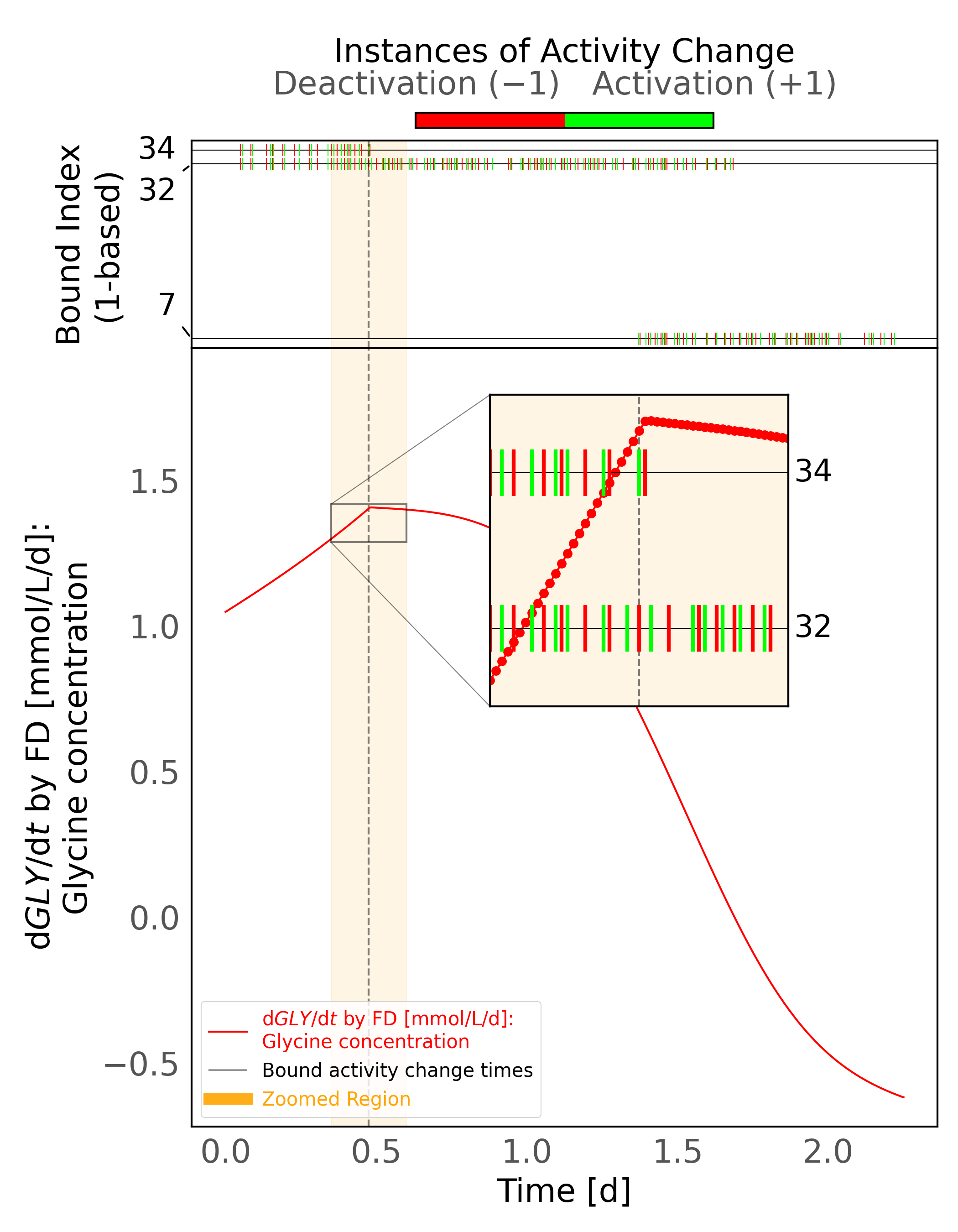}
        \caption{Glycine; first derivative}
        \label{fig:boundDisco1DerGLY}
    \end{subfigure}
    \begin{subfigure}[b]{0.45\textwidth}
        \includegraphics[width=\textwidth]{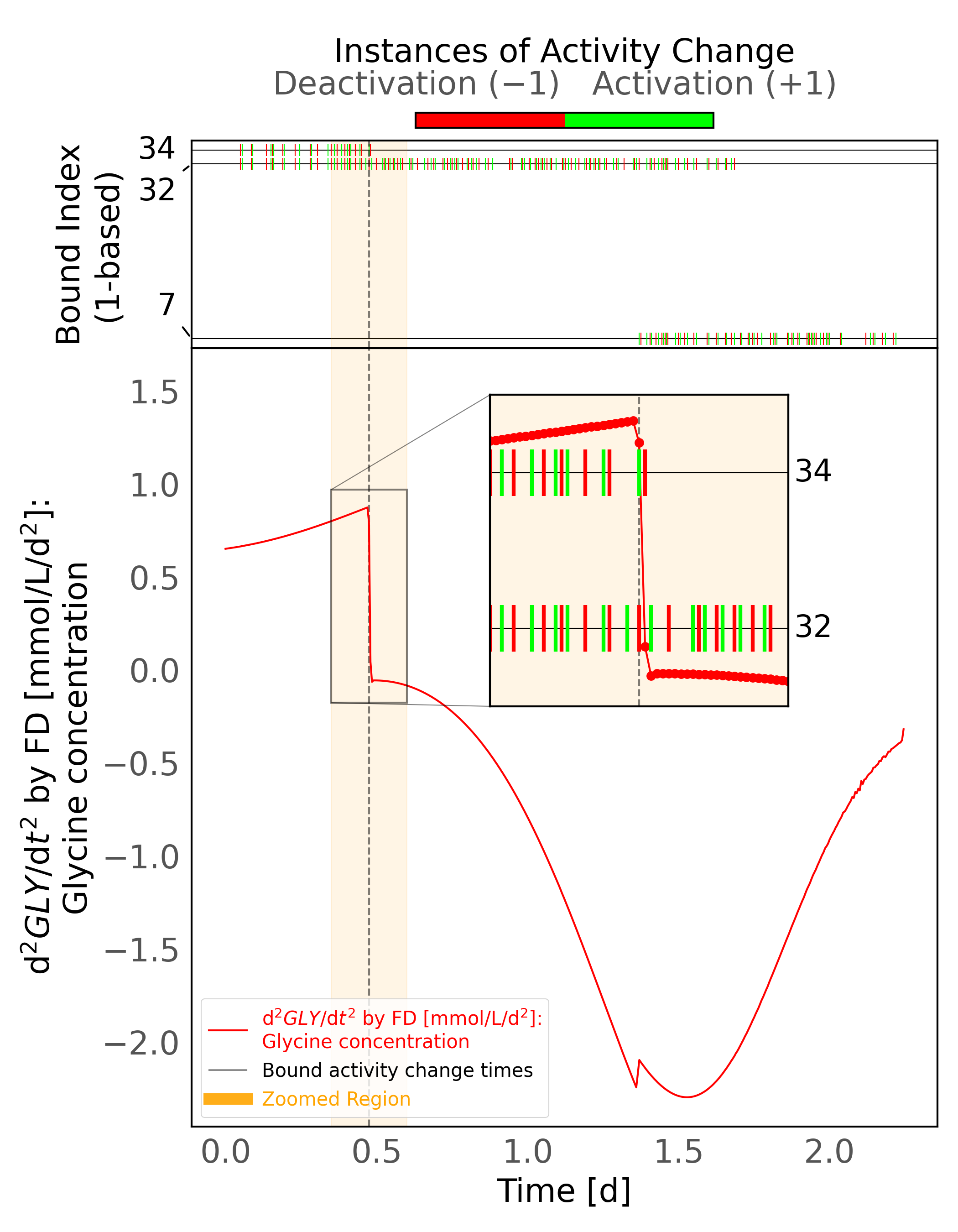}
        \caption{Glycine; second derivative}
        \label{fig:boundDisco2DerGLY}
    \end{subfigure}
    \begin{subfigure}[b]{0.45\textwidth}
        \includegraphics[width=\textwidth]{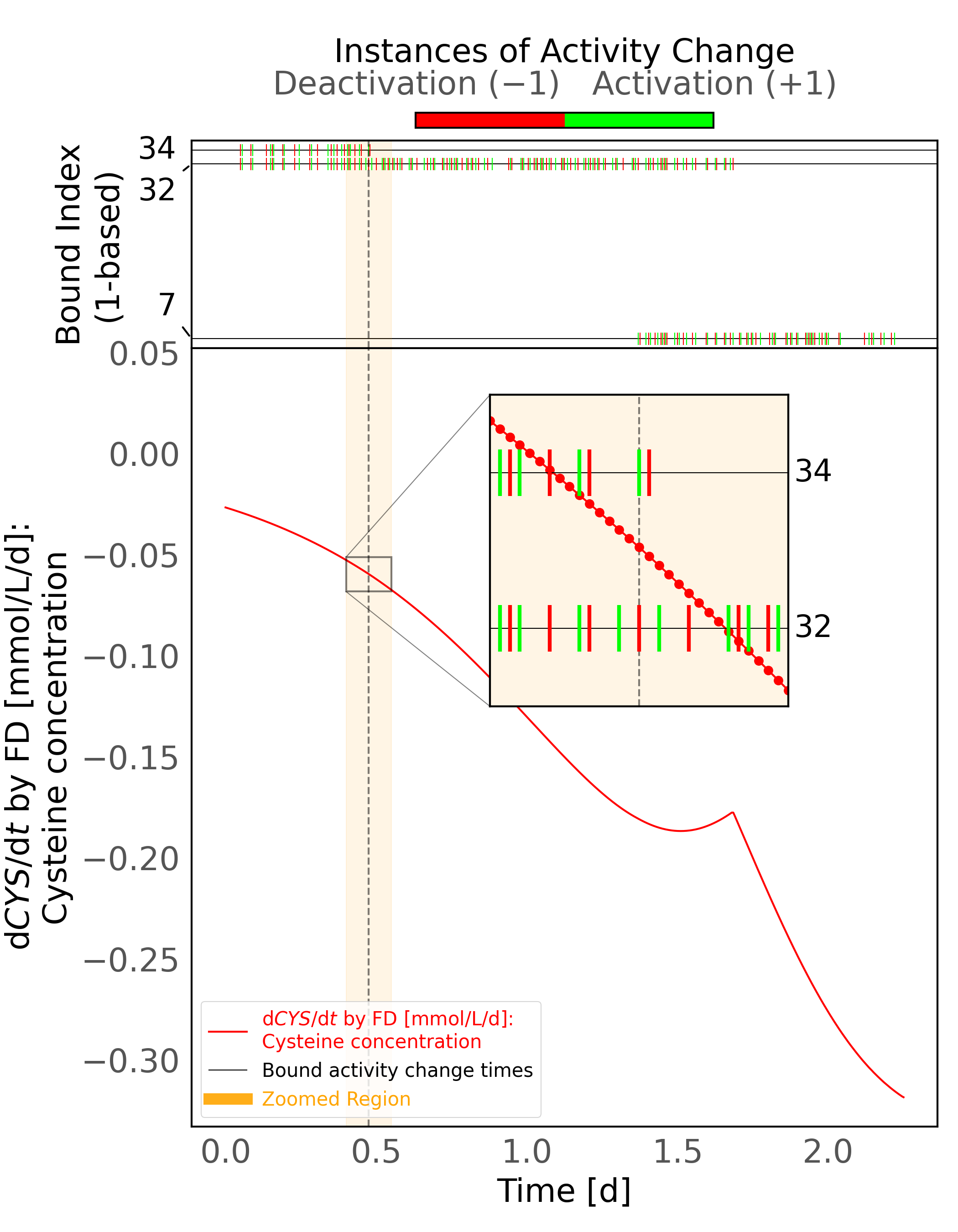}
        \caption{Cysteine; first derivative}
        \label{fig:boundDisco1DerCYS}
    \end{subfigure}
    \begin{subfigure}[b]{0.45\textwidth}
        \includegraphics[width=\textwidth]{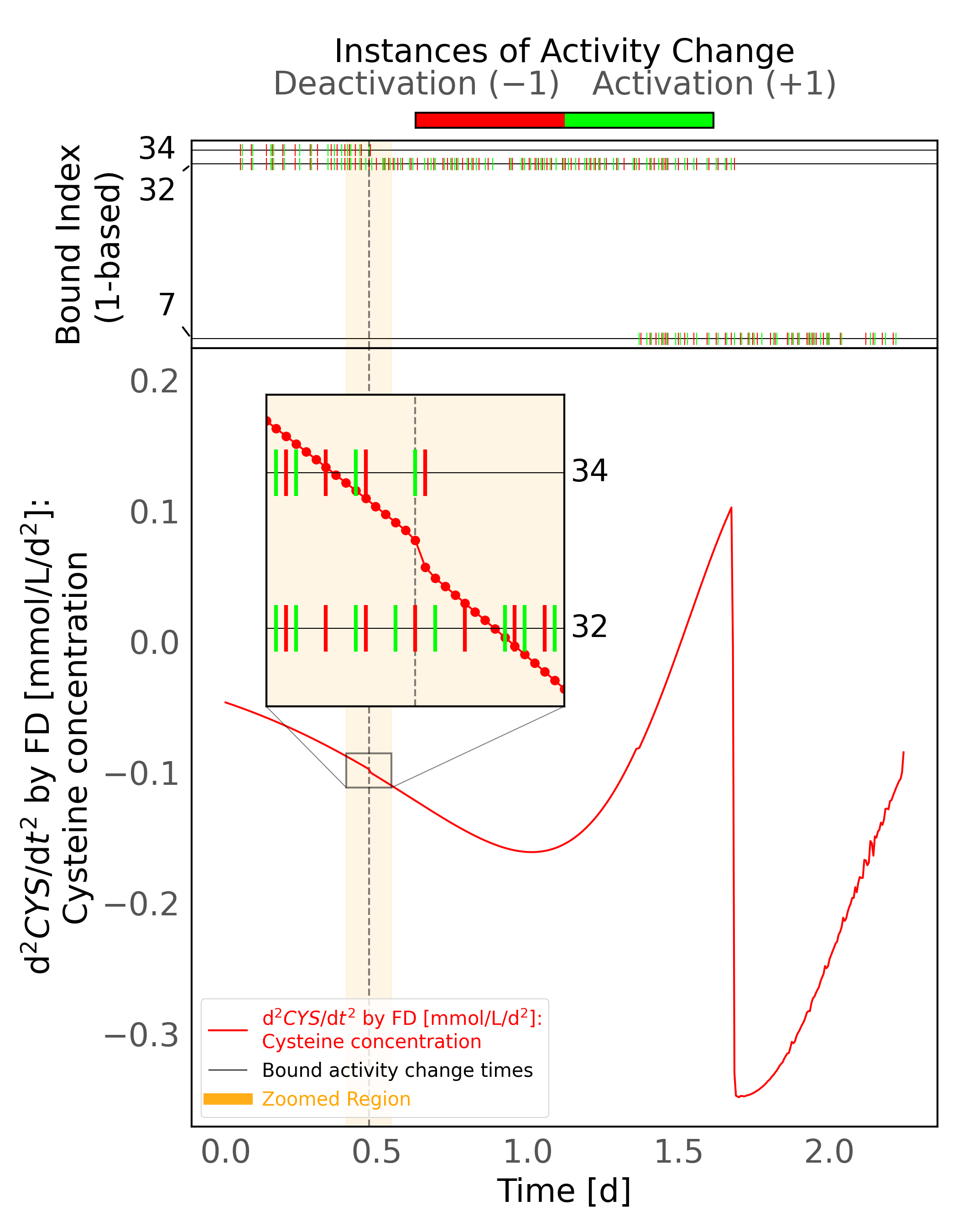}
        \caption{Cysteine; second derivative}
        \label{fig:boundDisco2DerCYS}
    \end{subfigure}
    \caption{
        \textbf{Activity of bound constraints along a sample run for the stoichiometric case},
        Observe the discontinuities arising in the second derivative
        (\ref{fig:boundDisco2DerGLY} and \ref{fig:boundDisco2DerCYS}) of the concentration evolution.
        Time derivatives for plotting were estimated by local forward finite differences (FD).
     Color is used to distinguish between different curves described in the legend.
    }
    \label{fig:boundDisco}
\end{figure}

At the onset---or the end---of occurrence of such events, the trajectory 
of $\flux$ may/will develop sharp corners.
To explore the impact of this phenomenon on the system dynamics we report,
at each timestep of a simulation, which flux bounds are active and which are not.
For an event-driven version of such simulations we
refer the reader to
the package in 
\cite{barton2014},
in which
several failure modes are considered
(beyond the ones arising in our work),
including both an infeasible inner optimization problem,
and a problem with multiple solutions (leading to a set-valued differential equation).
In general, events can be either ``time events" or ``state events";
the first require only accurately stopping the integration at a particular time.
State events, on the other hand, occur when some condition(s) of the continuous state become satisfied.
In some cases, this can be detected by locating zeros
of some interpolating polynomial(s)
\cite{barton2014,barton2000};
further difficulty arises
when the event condition cannot be described by the root of a continuous function
(e.g., the case of \cref{fig:boundDisco}, in which a Boolean quantity changes at the event).
We intend to explore the proper analysis of such event detection in future work (as well as the integration between events, possibly modifying the RHS evaluation between each pair of events to satisfy the active bounds by construction).
We employ instead a less sophisticated visualization-based method, shown in \cref{fig:boundDisco}:
We depict, in \cref{fig:boundDisco}, such changes in constraint activity status alongside the first 
(\ref{fig:boundDisco1DerCYS}, \ref{fig:boundDisco1DerGLY}) 
and second 
(\ref{fig:boundDisco2DerCYS}, \ref{fig:boundDisco2DerGLY}) 
time derivatives of two key simulated variables in a temporally aligned fashion. 
This is shown here along a short run of the stoichiometric model; a constraint turning active is marked by a short green tick, and its turning inactive by a brief red tick). Notice the jumps in the second derivative, and the sharp corners in the first derivative of the concentration evolution.

During this run we observe that three fluxes (7, 32 and 34) had encounters with their corresponding lower bounds. 
More specifically, around $t^* \approx 0.5$, when the bound for reaction 34 becomes {\em persistently inactive} (thus the corresponding flux moves well away from zero), we see sharp downward discontinuities \textit{in the second time derivative} of the evolutions of both cysteine and glycine.
Note that reaction 34 is the (irreversible) breakdown of NADH (see full stoichiometry matrix in \cref{sec:stoiMat}, or the relevant parts of the reaction network diagram in \cite{NOLAN2011108}); and that, at this time, its flux continuously changes from $0$ to positive. 
We therefore expect, $\mathrm{d} v_{34} / \mathrm{d}t$ may experience a discontinuity at $t^*$.
Furthermore, one of the (reversible) reactions that makes GLY takes NADH as an input. If that reaction rate is positive at that given moment, one of its inputs suddenly becomes less available. Therefore, because $\mathrm{d} \mathrm{GLY}/\mathrm{d}t \sim -v_{34}$, and $\mathrm{d}v_{34}/\mathrm{d}t$ is discontinuous, we can expect that  $\mathrm{d}(\mathrm{d} \mathrm{GLY} /\mathrm{d}t)/\mathrm{d}t$ will also be discontinuous, and that is clearly visible in \cref{fig:boundDisco2DerGLY}.

Such a rationalization can be repeated for CYS, which also relies on NADH as an input, and which also experiences a discontinuity in its second derivative (\cref{fig:boundDisco2DerCYS}). 
Note that CYS has a much larger discontinuity associated with the activation of the lower bound on flux 32.
Note also that GLY has a second discontinuity occurring later (around $t ~ 1.4$); this is related to flux 7, which
however involves different pathways.

\subsection{Black-Box Neural Network Identification}
\label{sec:blackBox}
\newcommand{\BBKINETICArchitecture}{20, 64, 32}
\newcommand{\BBKINETICActivation}{Tanh}
\newcommand{\BBKINETICNepochs}{8000}
\newcommand{\BBKINETICBatchesPerEpoch}{12}
\newcommand{\BBKINETICBatchSize}{64}
\newcommand{\BBKINETICLearningRate}{0.0004}
\newcommand{\BBKINETICEvaluationHorizon}{1.2}
\newcommand{\BBKINETICEvaluationDt}{0.1}
\newcommand{\BBKINETICnData}{768}
\newcommand{\BBKINETICnTimes}{13}

\newcommand{\BBSTOICHIOMETRICArchitecture}{20, 64, 32}
\newcommand{\BBSTOICHIOMETRICActivation}{Tanh}
\newcommand{\BBSTOICHIOMETRICNepochs}{8000}
\newcommand{\BBSTOICHIOMETRICBatchesPerEpoch}{12}
\newcommand{\BBSTOICHIOMETRICBatchSize}{64}
\newcommand{\BBSTOICHIOMETRICLearningRate}{0.0004}
\newcommand{\BBSTOICHIOMETRICEvaluationHorizon}{1.2}
\newcommand{\BBSTOICHIOMETRICEvaluationDt}{0.1}
\newcommand{\BBSTOICHIOMETRICnData}{768}
\newcommand{\BBSTOICHIOMETRICnTimes}{13}

\newcommand{\BBPredictorDt}{0.01}
\newcommand{\BBNetworkMethod}{Runge-Kutta 4}

To demonstrate system identification
with no assumed prior knowledge of the system mechanisms,
we performed black-box RHS learning,
in which we represent the entire system of ODEs as an end-to-end neural network.

Since there are two approaches in evaluating \cref{eqn:ode_2} as mentioned in \secref{sec:whiteboxStructure}, 
we also performed two experiments: one of them used
data generated from the ground-truth kinetic system
(sampling every $\Delta t=\BBKINETICEvaluationDt$ hours over a $t_\mathrm{max}=\BBKINETICEvaluationHorizon$-hour horizon, producing $\BBKINETICnTimes$ steps for each of the $\BBKINETICnData$ data trajectories),
and the other used data generated from the ground-truth stoichiometric one 
(with $\Delta t=\BBSTOICHIOMETRICEvaluationDt$, $t_\mathrm{max}=\BBSTOICHIOMETRICEvaluationHorizon$, $\BBSTOICHIOMETRICnTimes$ steps, and $\BBSTOICHIOMETRICnData$ data trajectories).

In both cases,
the black-box ODE was trained by taking steps of fixed size $\BBPredictorDt$ between the data samples
with a \BBNetworkMethod{} integrator (the black box identification ``does not know" about discontinuities in the model - it smoothly interpolates between data points in time). 

\begin{figure}[!ht]
    \centering
    \begin{subfigure}[b]{0.45\textwidth}
        \includegraphics[width=\textwidth]{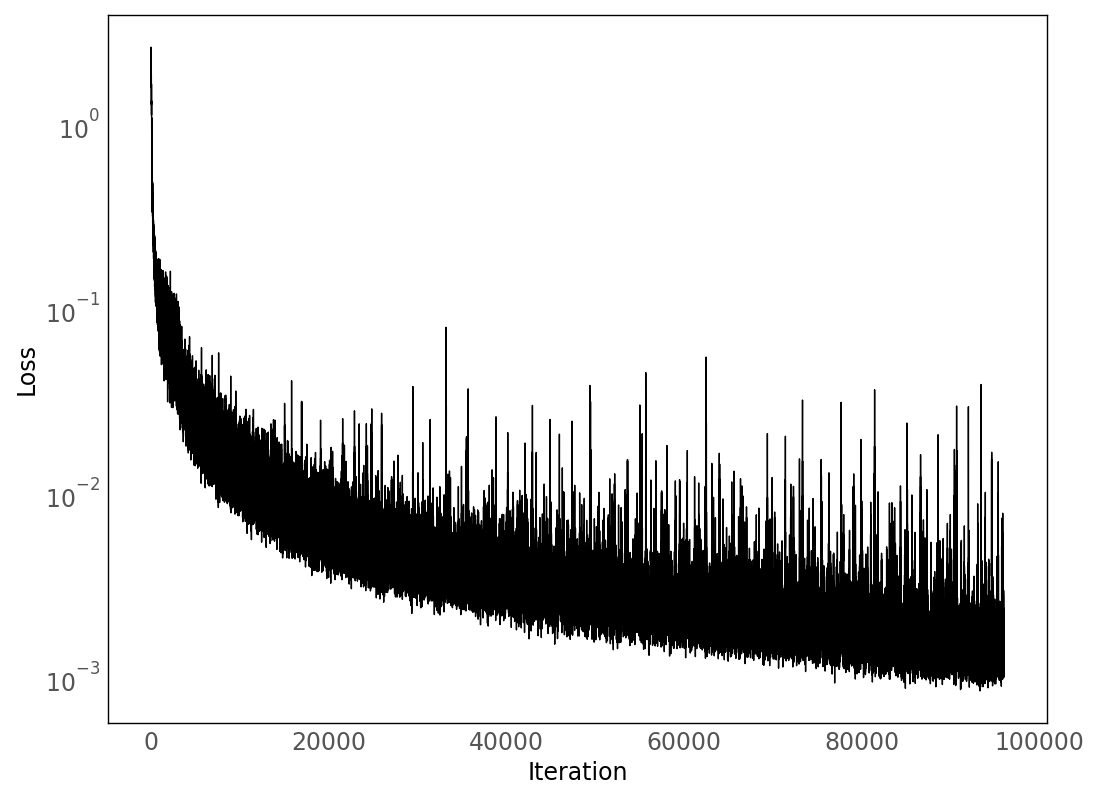}
        \caption{}
    \end{subfigure}
    \begin{subfigure}[b]{0.45\textwidth}
        \includegraphics[width=\textwidth]{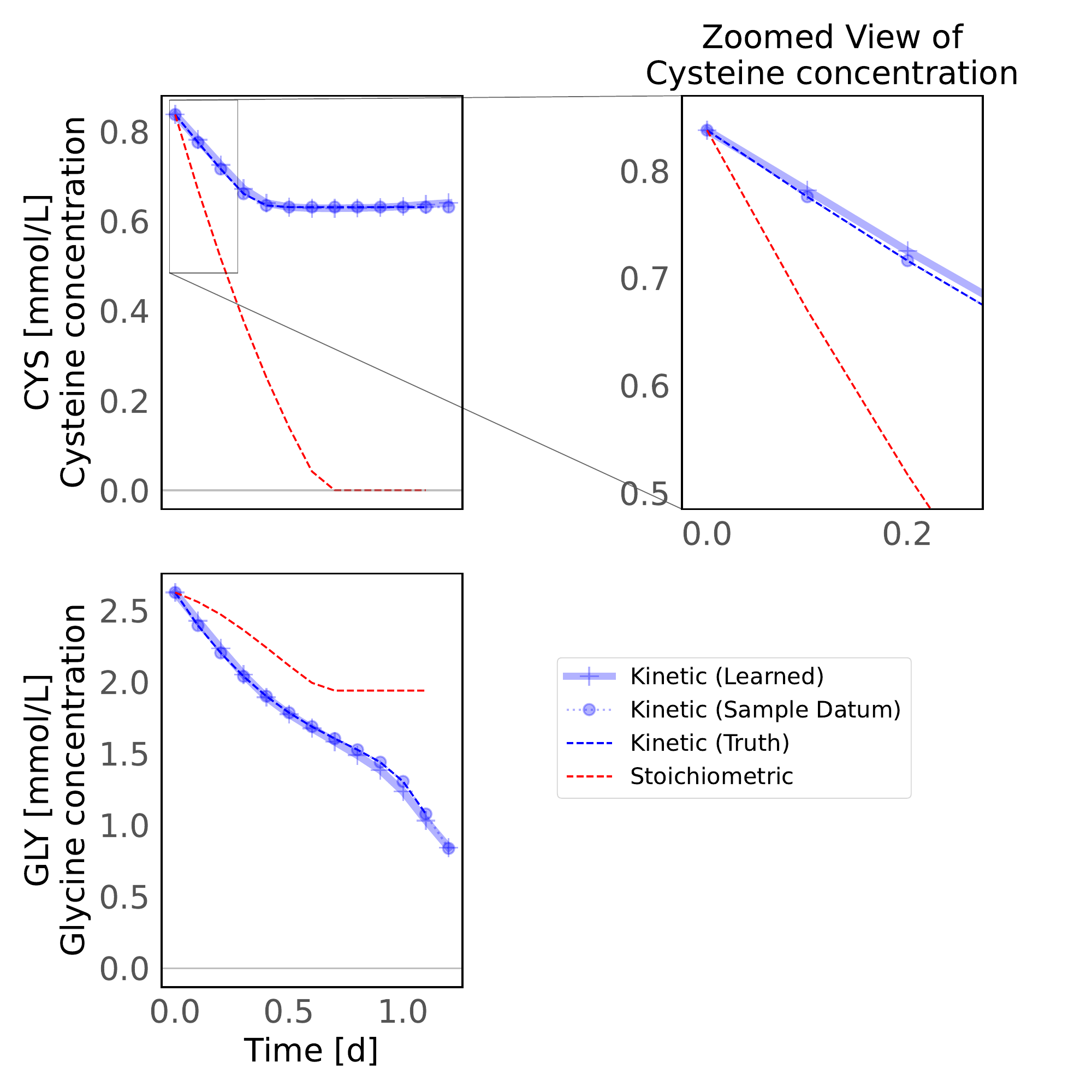}
        \caption{}
    \end{subfigure}
    \caption{Black-box training results (kinetic).
        See also \cref{fig:bb_traj_comparison-KINETIC-14} for more detailed results.
        Color is used to distinguish between different curves described in the legend.
    }
    \label{fig:bb_results_kin}
\end{figure}

\begin{figure}[!ht]
    \centering
    \begin{subfigure}[b]{0.45\textwidth}
        \includegraphics[width=\textwidth]{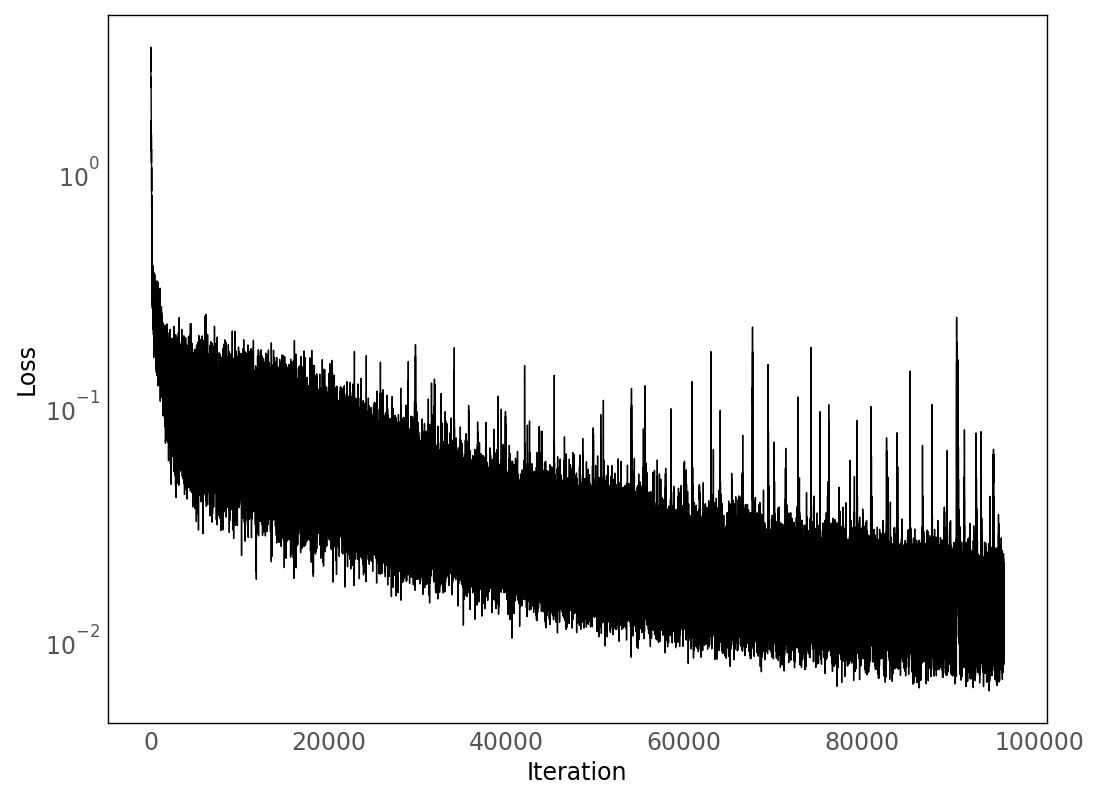}
        \caption{}
    \end{subfigure}
    \begin{subfigure}[b]{0.45\textwidth}
        \includegraphics[width=\textwidth]{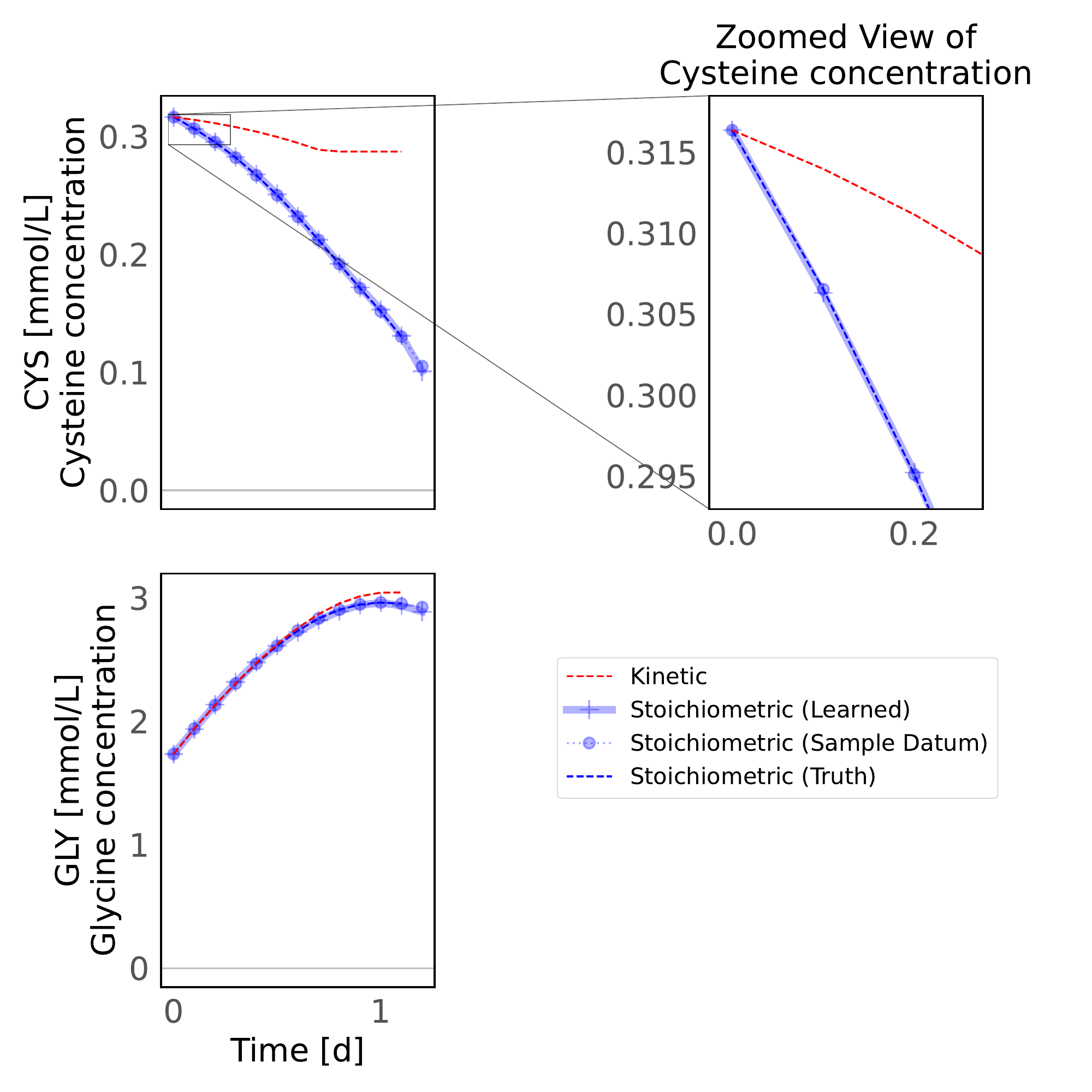}
        \caption{}
    \end{subfigure}
    \caption{Black-box training results (stoichiometric).
        See also \cref{fig:bb_traj_comparison-STOICHIOMETRIC-14} for more detailed results.
        Color is used to distinguish between different curves described in the legend.
    }
    \label{fig:bb_results_stoi}
\end{figure}

As can be seen in \cref{fig:bb_results_kin} and \cref{fig:bb_results_stoi},
our black-box neural ODE was able to fit the data trajectories quite tightly.
This validates the underlying approach, and suggests that such ODEs with inner optimization steps can be successfully approximated as (possibly slightly ``smoothened") closed-form functions.

\subsection{White-Box Neural Network: Parameter Estimation}
\label{sec:whiteBox}
\newcommand{\WBKINETICTwoParTrainableIndices}{0, 1}
\newcommand{\WBKINETICTwoParNepochs}{4000}
\newcommand{\WBKINETICTwoParBatchSize}{10}
\newcommand{\WBKINETICTwoParBatchesPerEpoch}{1}
\newcommand{\WBKINETICTwoParLearningRate}{0.001}
\newcommand{\WBKINETICTwoParEvaluationHorizon}{0.05}
\newcommand{\WBKINETICTwoParMethod}{Euler}
\newcommand{\WBKINETICTwoParEvaluationDt}{0.05}
\newcommand{\WBKINETICTwoParDataDt}{0.05}
\newcommand{\WBKINETICTwoParnData}{10}
\newcommand{\WBKINETICTwoParnTimes}{2}
\newcommand{\WBKINETICTwoParOptimizer}{RMSprop}

\newcommand{\WBKINETICFiveParTrainableIndices}{0, 1, 2, 3, 4}
\newcommand{\WBKINETICFiveParNepochs}{4000}
\newcommand{\WBKINETICFiveParBatchSize}{10}
\newcommand{\WBKINETICFiveParBatchesPerEpoch}{1}
\newcommand{\WBKINETICFiveParLearningRate}{0.001}
\newcommand{\WBKINETICFiveParEvaluationHorizon}{0.05}
\newcommand{\WBKINETICFiveParMethod}{Euler}
\newcommand{\WBKINETICFiveParEvaluationDt}{0.05}
\newcommand{\WBKINETICFiveParDataDt}{0.05}
\newcommand{\WBKINETICFiveParnData}{10}
\newcommand{\WBKINETICFiveParnTimes}{2}
\newcommand{\WBKINETICFiveParOptimizer}{RMSprop}

\newcommand{\WBSTOICHIOMETRICTwoParTrainableIndices}{0, 1}
\newcommand{\WBSTOICHIOMETRICTwoParNepochs}{4000}
\newcommand{\WBSTOICHIOMETRICTwoParBatchSize}{10}
\newcommand{\WBSTOICHIOMETRICTwoParBatchesPerEpoch}{1}
\newcommand{\WBSTOICHIOMETRICTwoParLearningRate}{0.001}
\newcommand{\WBSTOICHIOMETRICTwoParEvaluationHorizon}{0.05}
\newcommand{\WBSTOICHIOMETRICTwoParMethod}{Euler}
\newcommand{\WBSTOICHIOMETRICTwoParEvaluationDt}{0.05}
\newcommand{\WBSTOICHIOMETRICTwoParDataDt}{0.05}
\newcommand{\WBSTOICHIOMETRICTwoParnData}{10}
\newcommand{\WBSTOICHIOMETRICTwoParnTimes}{2}
\newcommand{\WBSTOICHIOMETRICTwoParOptimizer}{RMSprop}

\newcommand{\WBSTOICHIOMETRICFiveParTrainableIndices}{0, 1, 2, 3, 4}
\newcommand{\WBSTOICHIOMETRICFiveParNepochs}{4000}
\newcommand{\WBSTOICHIOMETRICFiveParBatchSize}{10}
\newcommand{\WBSTOICHIOMETRICFiveParBatchesPerEpoch}{1}
\newcommand{\WBSTOICHIOMETRICFiveParLearningRate}{0.001}
\newcommand{\WBSTOICHIOMETRICFiveParEvaluationHorizon}{0.05}
\newcommand{\WBSTOICHIOMETRICFiveParMethod}{Euler}
\newcommand{\WBSTOICHIOMETRICFiveParEvaluationDt}{0.05}
\newcommand{\WBSTOICHIOMETRICFiveParDataDt}{0.05}
\newcommand{\WBSTOICHIOMETRICFiveParnData}{10}
\newcommand{\WBSTOICHIOMETRICFiveParnTimes}{2}
\newcommand{\WBSTOICHIOMETRICFiveParOptimizer}{RMSprop}

To demonstrate the full-structure physical-parameter estimation setting that we term ``white-box'' learning,
we tried to recover (a) two or (b) five of the nominal parameter values.
Specifically,
we performed simulations at the nominal parameter values, collected the transient data, considered forty three (resp. forty) of them known, 
and then 
used a gradient-based training method to estimate the values of the remaining two (resp. five) from the data.
Our initial guess (a perturbation of the truth) is marked in \cref{fig:wb2pGrad}.
For all four of these numerical experiments,
the dataset consisted of {\WBSTOICHIOMETRICTwoParnData} short single-Euler-step ``trajectories''
each {\WBSTOICHIOMETRICTwoParEvaluationHorizon} hours in length (Runge-Kutta integration gave comparable results, not shown); the network ansatz was also {\WBKINETICTwoParMethod}
with a step size of $\WBKINETICTwoParEvaluationDt$.
Training was {\WBKINETICTwoParNepochs} epochs
of {\WBKINETICTwoParOptimizer}
with {\WBKINETICTwoParBatchesPerEpoch} batch per epoch.
This demonstrates the use of the algorithms in \cite{OptNetKolter} 
to carry out differentiation through the inner optimization problem of evaluating the equation right-hand-side,  discussed in \cref{sec:whiteboxStructure},
enabling gradient-based learning for this experiment, and
serving as an initial validation of the algorithms before the gray-box methods of \cref{sec:grayBox} that follow below.

\subsubsection{Known Model, Two Unknown Parameters}
\label{sec:wb2p}

\begin{figure}[!ht]
    \centering
    \includegraphics[width=0.5\textwidth]{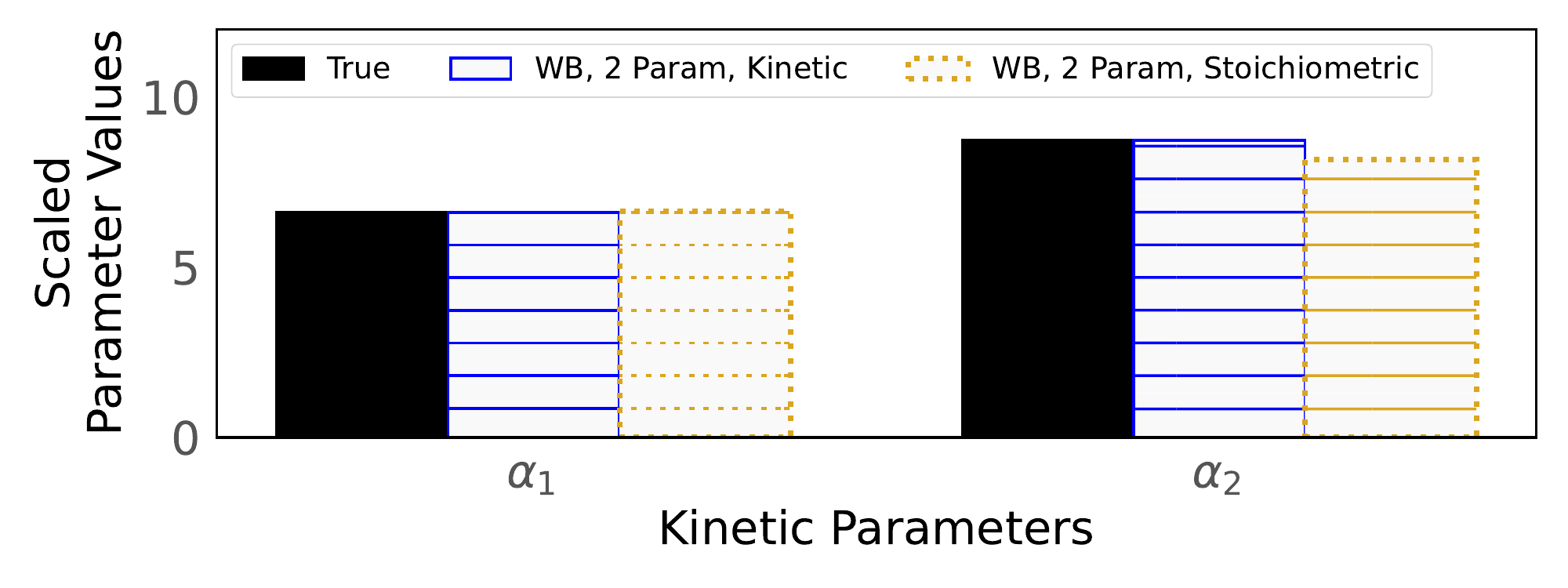}
    \caption{Parameter comparison for white-box two-parameter.
    \label{fig:parameter_compare_2P}
    }
\end{figure}

\begin{figure}[!ht]
    \centering
    \begin{subfigure}[b]{0.49\textwidth}
        \includegraphics[width=\textwidth]{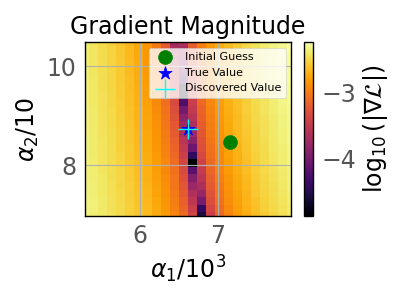}
        \caption{\textbf{\typeOne:} data is from ground-truth kinetic simulation; WB ansatz is also kinetic simulation.}
        \label{fig:wb2pGradMagKin}
    \end{subfigure}
    \begin{subfigure}[b]{0.49\textwidth}
        \includegraphics[width=\textwidth]{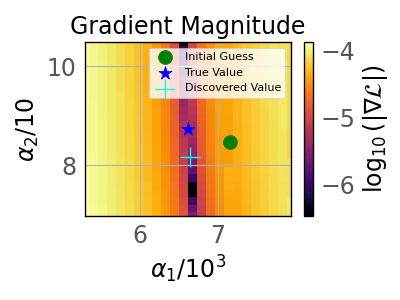}
        \caption{\textbf{\typeTwo:} data is from ground-truth stoichiometric simulation; WB ansatz is also stoichiometric simulation.}
        \label{fig:wb2pGradMagStoi}
    \end{subfigure}
    \caption{
        \textbf{Gradient
        landscape
        for the white-box, two-parameter case.
        }
        Kinetic (\ref{fig:wb2pGradMagKin})
        vs
        stoichiometric (\ref{fig:wb2pGradMagStoi})
        implementation from \secref{sec:whiteboxStructure}.
        The stiff gradient vectorfield leads to some degree of parameter nonidentifibility.
    }
    \label{fig:wb2pGrad}
\end{figure}

\begin{figure}[!ht]
    \centering
    \begin{subfigure}[b]{0.45\textwidth}
        \includegraphics[width=\textwidth]{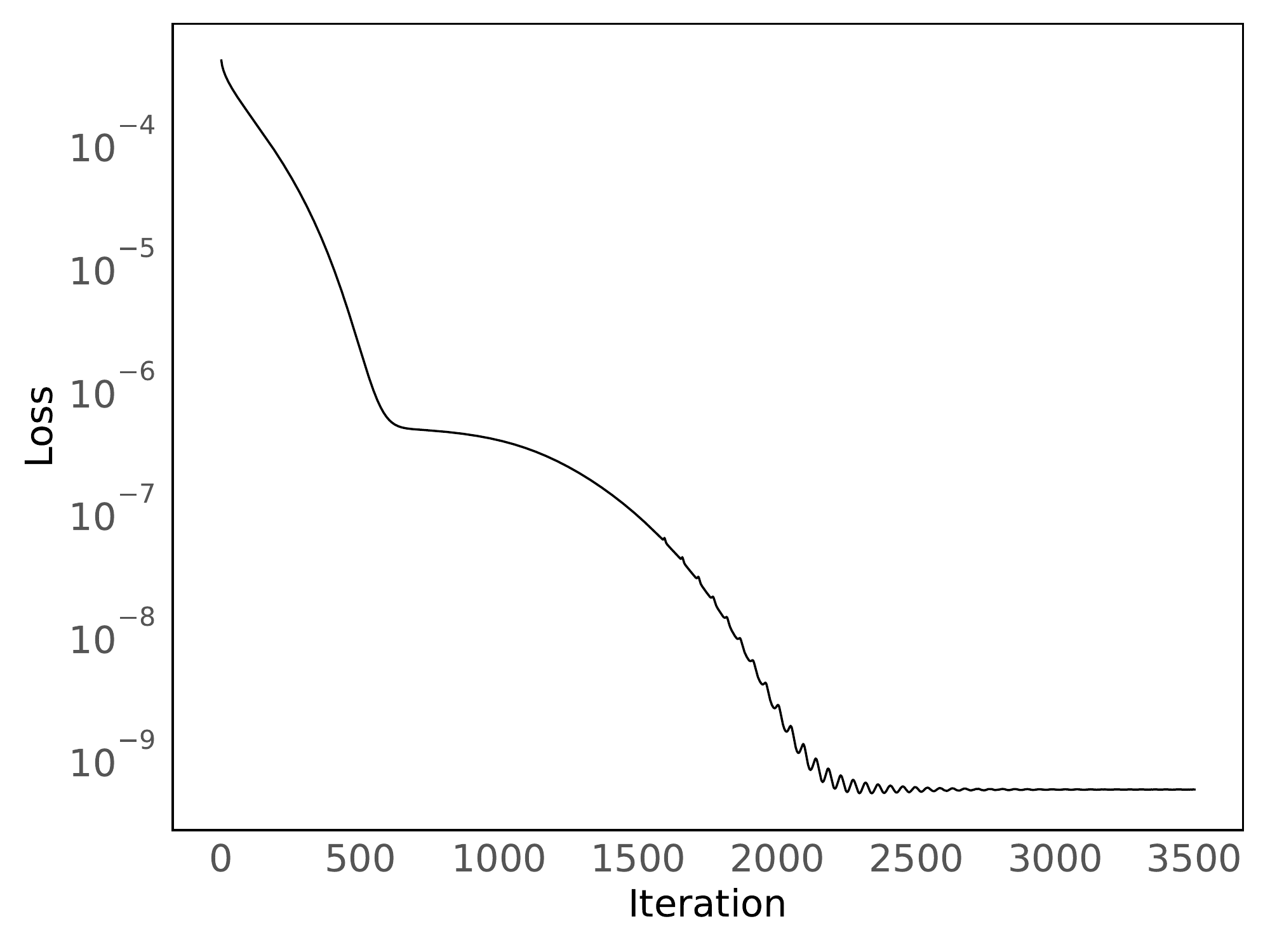}
        \caption{\typeOne}
        \label{fig:wb2pLossKin}
    \end{subfigure}
    \begin{subfigure}[b]{0.45\textwidth}
        \includegraphics[width=\textwidth]{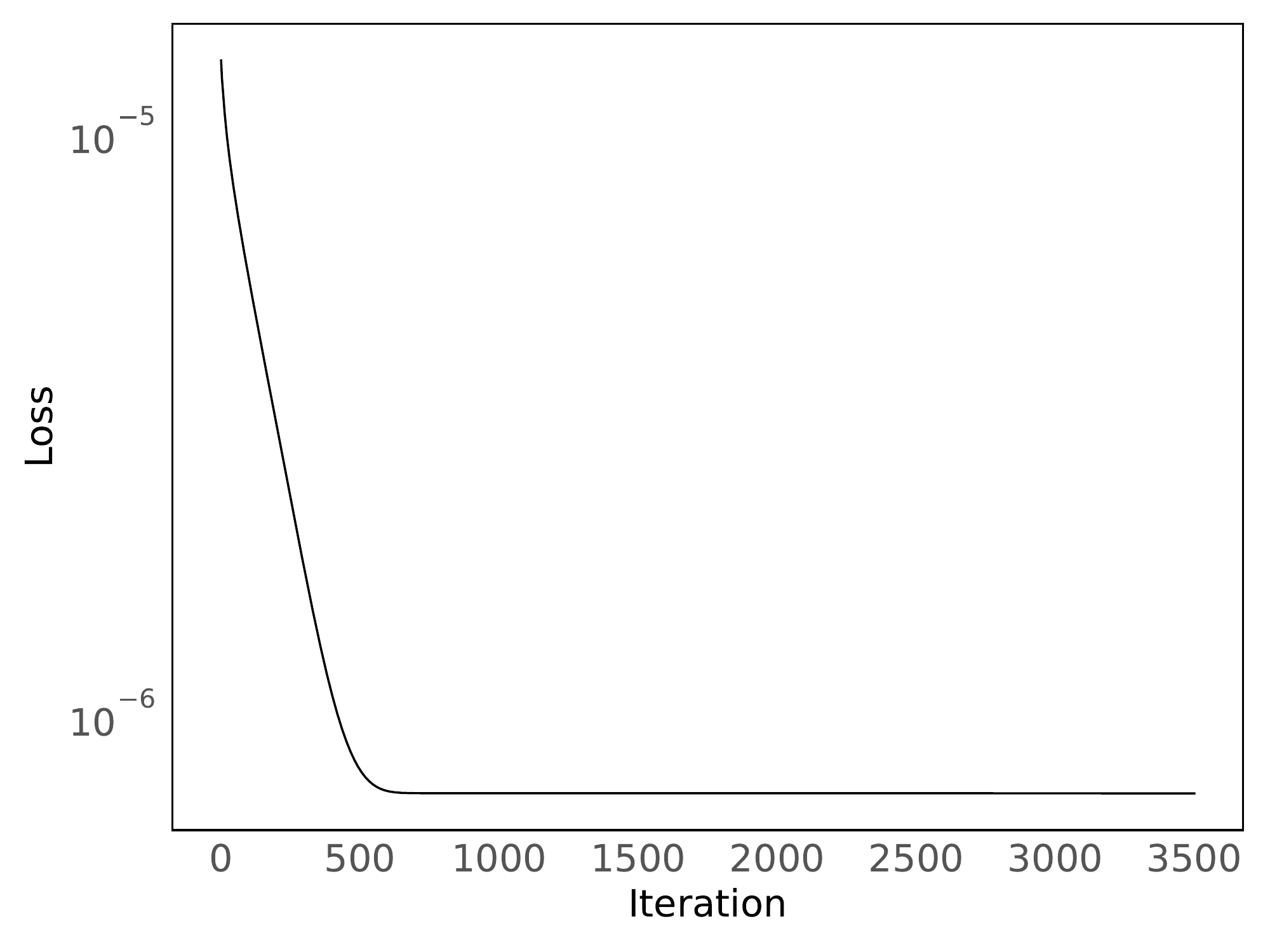}
        \caption{\typeTwo}
        \label{fig:wb2pLossSto}
    \end{subfigure}
    \caption{
        \textbf{Convergence of the training to the final parameter estimates for the white-box, two-parameter case.}
        Kinetic (\ref{fig:wb2pLossKin})
        vs
        stoichiometric (\ref{fig:wb2pLossSto})
        implementation from \secref{sec:whiteboxStructure}.
        See also \figref{fig:absrelTwoParam}.
    }
    \label{fig:wb2pLoss}
\end{figure}

In our first such white-box learning experiment,
we trained with only two unknown parameters from \cref{tab:parameters}; we find (\cref{fig:parameter_compare_2P}) that we can recover the two parameter values
reasonably well.
A motivation for this initial experiment
is to help visualize the gradient landscape in \cref{fig:wb2pGrad}.
We see in \cref{fig:wb2pLoss} that the induced gradient dynamics of the learning problem are highly stiff, making adaptive training methods such as Adam \cite{Kingma2014AdamAM} an absolute necessity.
The training exhibits a two-staged descent,
consisting of (a) first, a fast approach to a deep trough in the
parameter space, and then (b) a slower motion 
within the trough,
with some oscillations induced by the finite learning step size.
Furthermore, in the \typeTwo{} case (\cref{fig:wb2pGradMagStoi} and \cref{fig:wb2pLossSto}),
we observe that the final gradient is so shallow
that even Adam takes prohibitively long to move
any appreciable distance within the loss trough.
Note that, although the true $\vec{\param}$ values
indeed mark a minimum for the optimization problem posed,
as seen in \cref{fig:wb2pGradMagStoi}, this minimum is
extremely shallow, making the discovery of the 
true value for $\param_2$ imperfect - for all practical
purposes, the entire ``bottom of the trough" leads to a
good fit.
This is an instance of what is termed ``model sloppiness''
\cite{DANIELS2008389, HOLIDAY2019419}; along the bottom of
the trough the loss function posed is not strongly sensitive to the parameters, leading to parameter nonidentifiability.

\subsubsection{Known Model, Five Unknown Parameters}
\label{sec:wb5p}

\begin{figure}[!ht]
    \centering
    \includegraphics[width=0.5\textwidth]{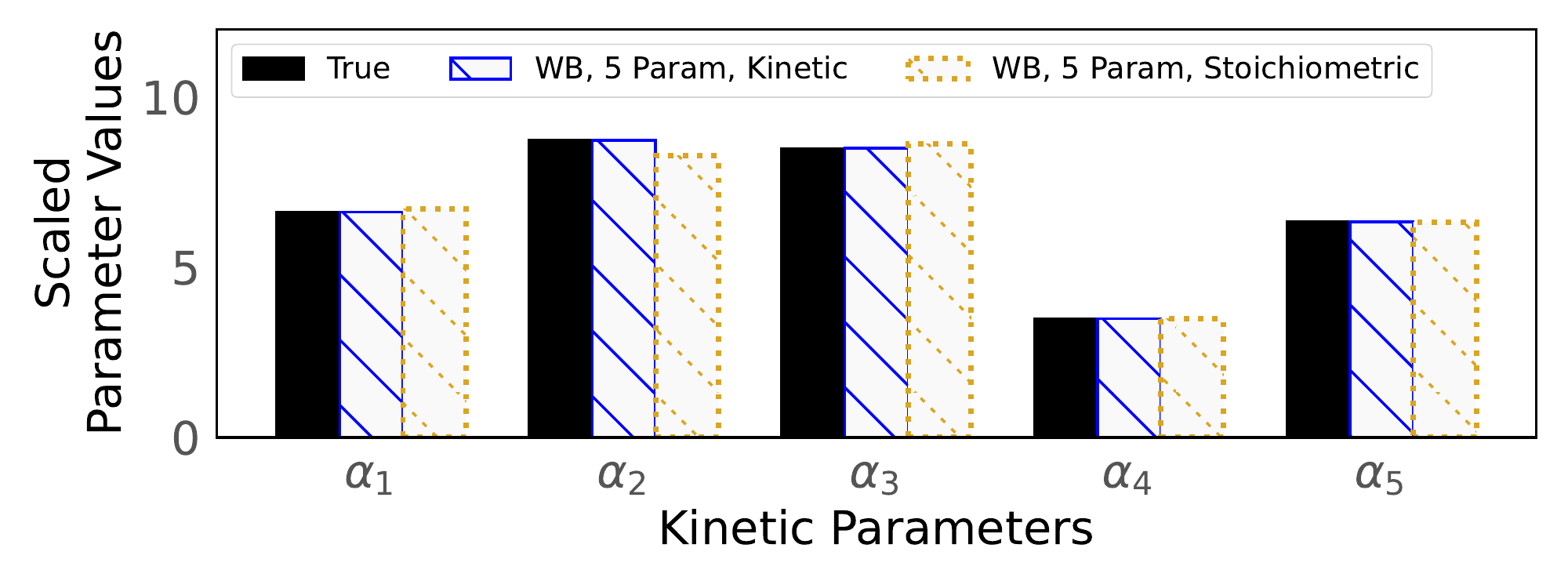}
    \caption{Parameter comparison for the white-box, five-parameter case.
    \label{fig:parameter_compare_5P}
    }
\end{figure}

\begin{figure}[!ht]
    \centering
    \begin{subfigure}[b]{0.45\textwidth}
        \includegraphics[width=\textwidth]{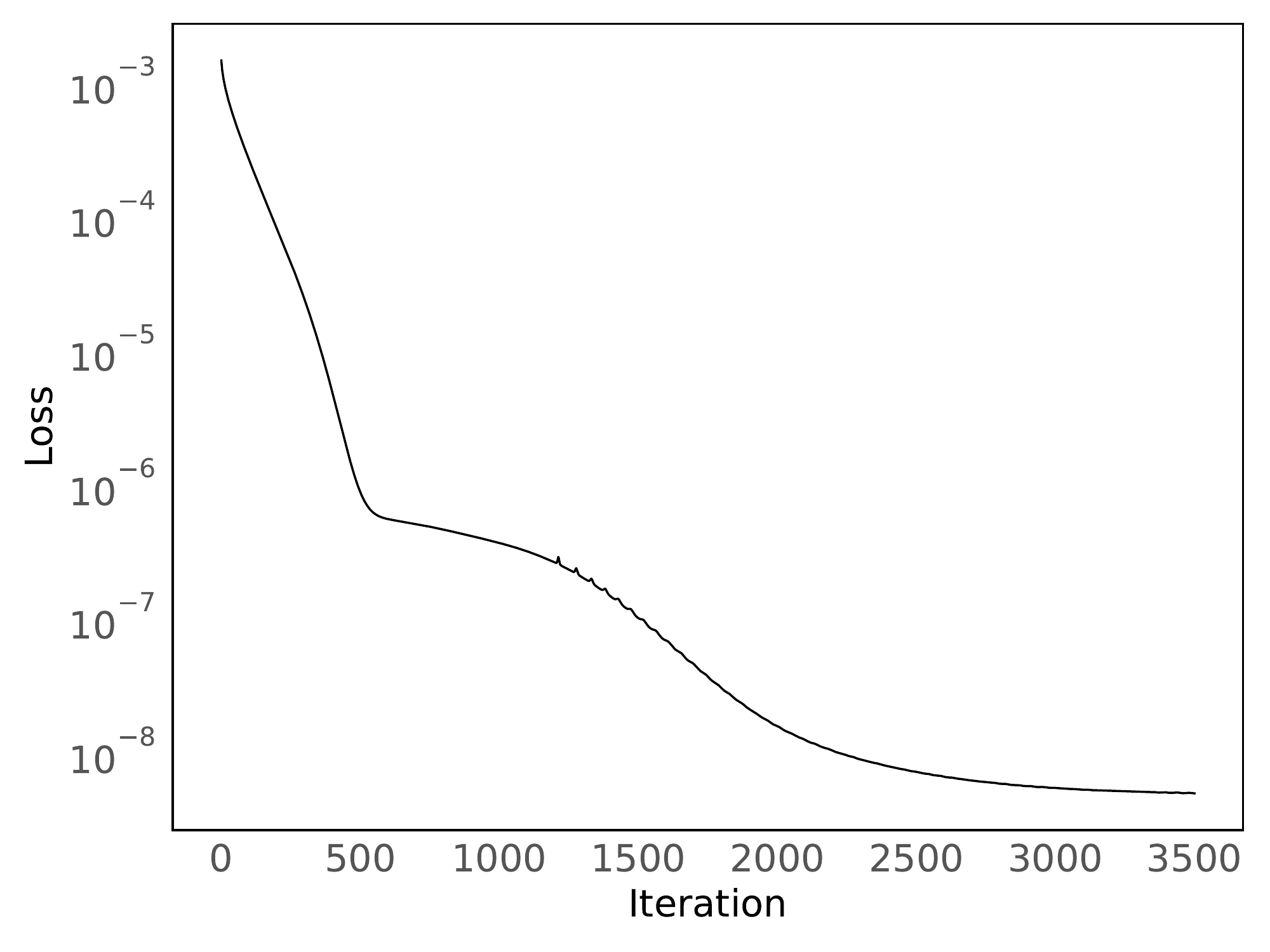}
        \caption{\typeOne}
        \label{fig:wb5pLossKin}
    \end{subfigure}
    \begin{subfigure}[b]{0.45\textwidth}
        \includegraphics[width=\textwidth]{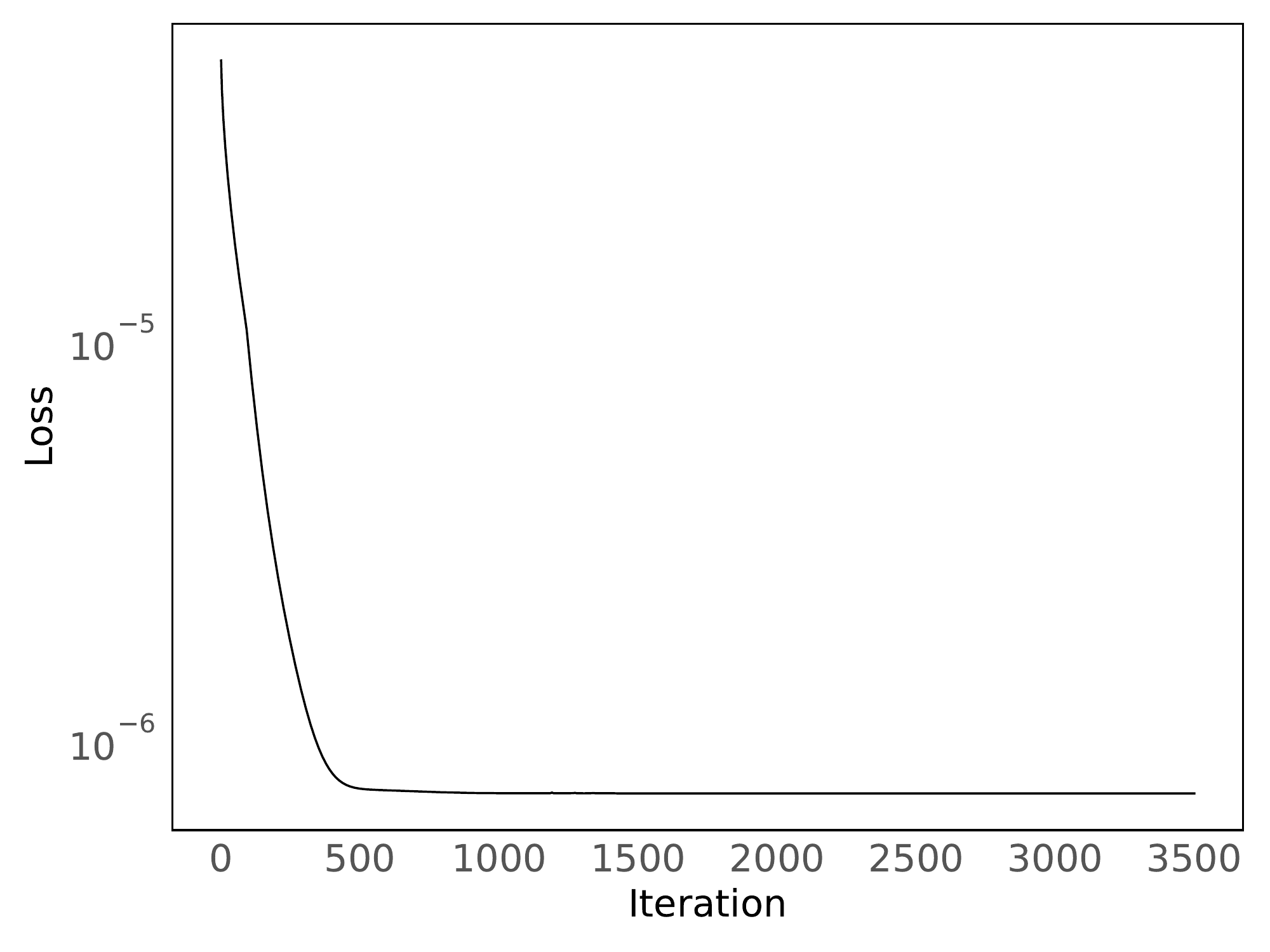}
        \caption{\typeTwo}
        \label{fig:wb5pLossSto}
    \end{subfigure}
    \caption{
        \textbf{Convergence of the training to the final parameter estimates for the white-box, five-parameter case.}
        Kinetic (\ref{fig:wb5pLossKin})
        vs
        stoichiometric (\ref{fig:wb5pLossSto})
        implementation from \secref{sec:whiteboxStructure}.
        See also \figref{fig:absrelFiveParam}.
        \label{fig:wb5pLoss}
    }
\end{figure}

Next, we repeated the parameter estimation experiment
of the previous section but now with five, rather than two, unknown  $\param$ values.
We find that the numerical values are again recovered
reasonably well (\cref{fig:parameter_compare_5P}),
and with loss dynamics (\cref{fig:wb5pLoss}) similar to the two-parameter case.

\subsection{Partially Known Model: Gray-box Identification}
\label{sec:grayBox}

For the work in this section, 
we assumed the expression of $\hat{v}_{I, 2}$ in $\fkin$ was not known: instead, we only knew that it is a function of GLC and LAC, and we replaced it by a 2-8-8-1 multi-layer perceptron (MLP) with trainable weights and biases.
We embedded this MLP into our gray-box computation graph
visualized in \cref{fig:architecture}
to make predictions for loss evaluation.

For all four of these experiments,
the dataset consisted of 800 single-Euler-step ``trajectories''
each 0.05 hours in length,
The network ansatz was also Euler with a step size of 0.05.
Training was 500 epochs
of RMSProp
with 20 batches per epoch.

\subsubsection{Partially Known Model, All Parameters Known}
\label{sec:grayBoxG}
For our first gray-box experiments
we further assumed that the values of all kinetic parameters
used in expressions beyond that for $\hat{v}_{I,2}$
are correct and do not need to be calibrated.
We performed the experiment twice:
once with kinetic-based data
(data from type-1 simulations)
and with the white portion of our gray-box also based on the kinetic formulation; and once 
with stoichiometric data/version respectively.

\begin{figure}[!ht]
    \centering

    \begin{subfigure}[b]{0.64\textwidth}
        \includegraphics[width=\textwidth]{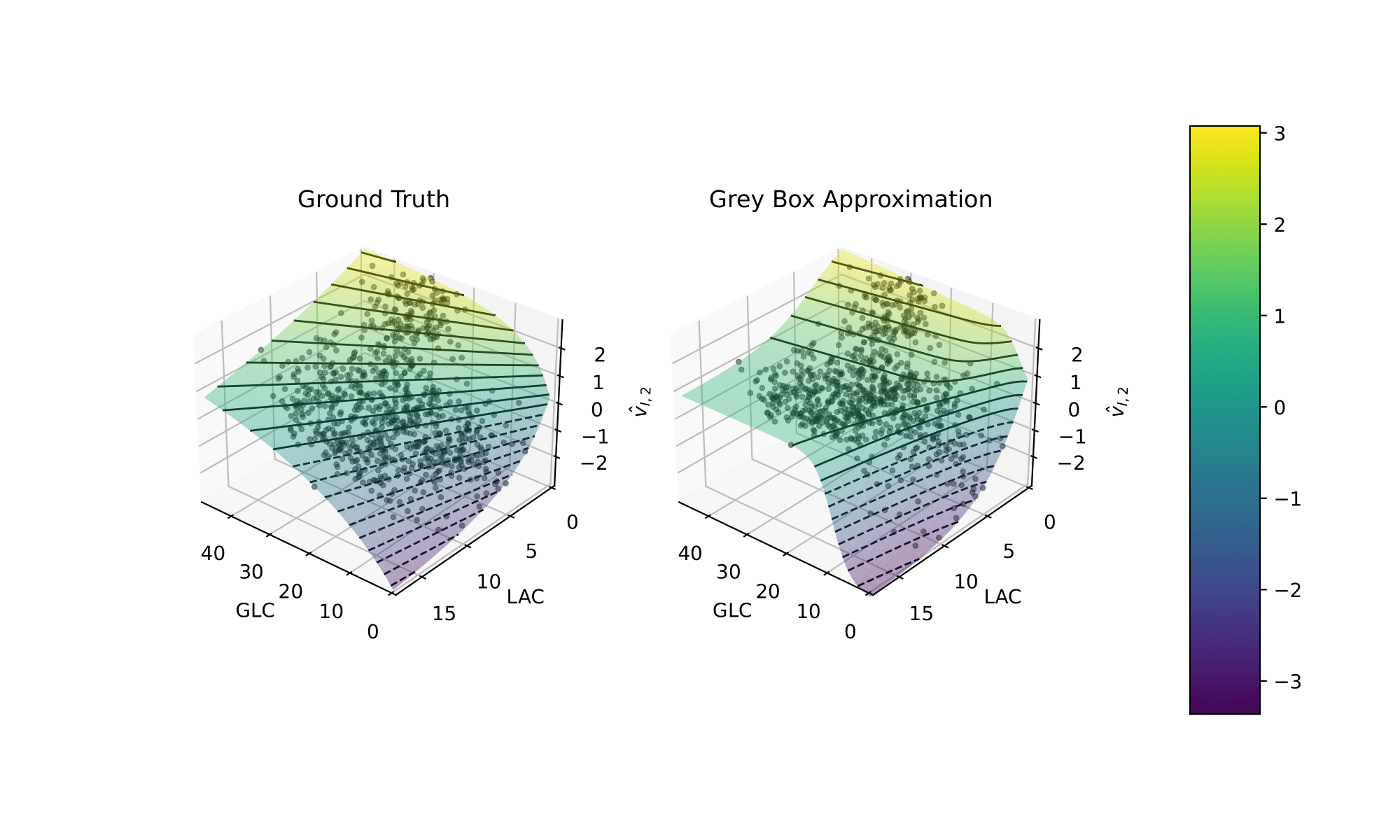}
        \caption{\label{fig:gbAFluxFuncs}}
    \end{subfigure}
    \begin{subfigure}[b]{0.32\textwidth}
        \includegraphics[width=\textwidth]{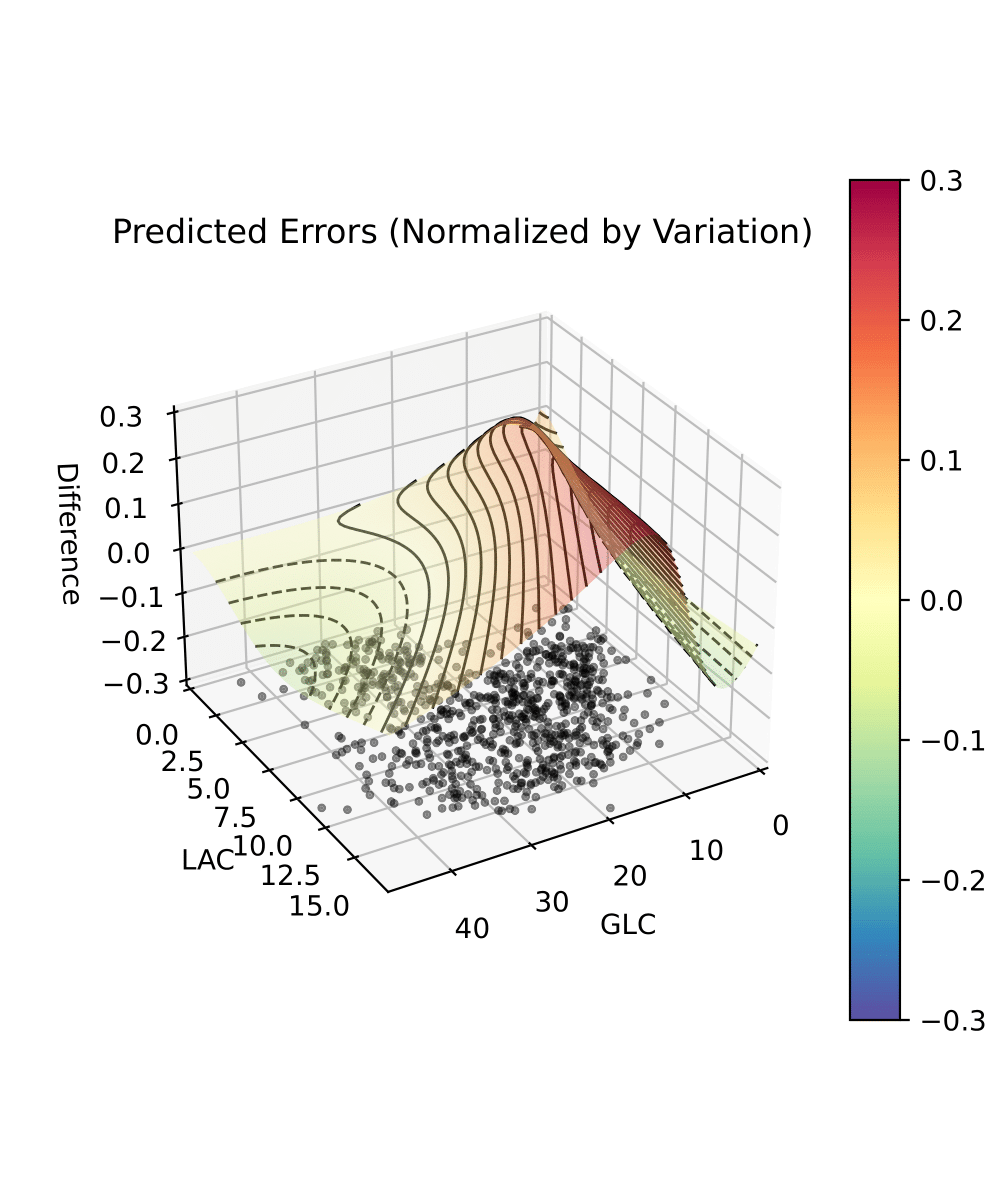}
        \caption{\label{fig:gbAErr}}
    \end{subfigure}
    
    \caption{
        \textbf{Comparison of fluxes (ground-truth vs kinetic gray-box model).}
        Note that the (GLC, LAC) points visited in training are scattered on the surfaces (\ref{fig:gbAFluxFuncs}) or on the base plane (\ref{fig:gbAErr}).
        \textbf{\ref{fig:gbAFluxFuncs}:} ground-truth and neural net approximations of the fluxes given the inputs of the neural net (GLC and LAC).
        \textbf{\ref{fig:gbAErr}:} normalized prediction errors (fraction of max-min of the true function, across the data).
        Note the relative rotation (for visual clarity) between (\ref{fig:gbAFluxFuncs}) and (\ref{fig:gbAErr}).
    \label{fig:gbA}
    }
\end{figure}

\begin{figure}[!ht]
    \centering

    \begin{subfigure}[b]{0.64\textwidth}
        \includegraphics[width=\textwidth]{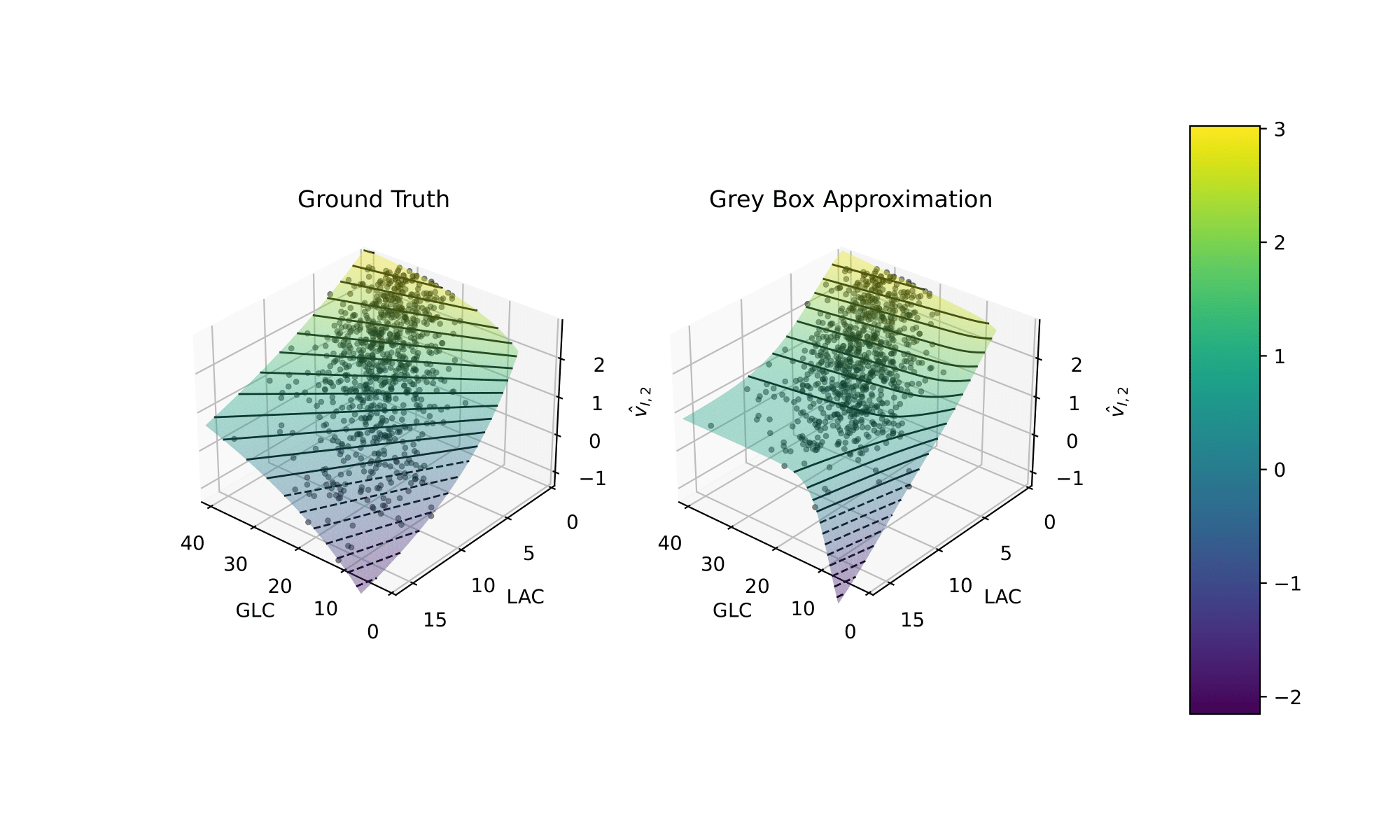}
        \caption{\label{fig:gbBFluxFuncs}}
    \end{subfigure}
    \begin{subfigure}[b]{0.32\textwidth}
        \includegraphics[width=\textwidth]{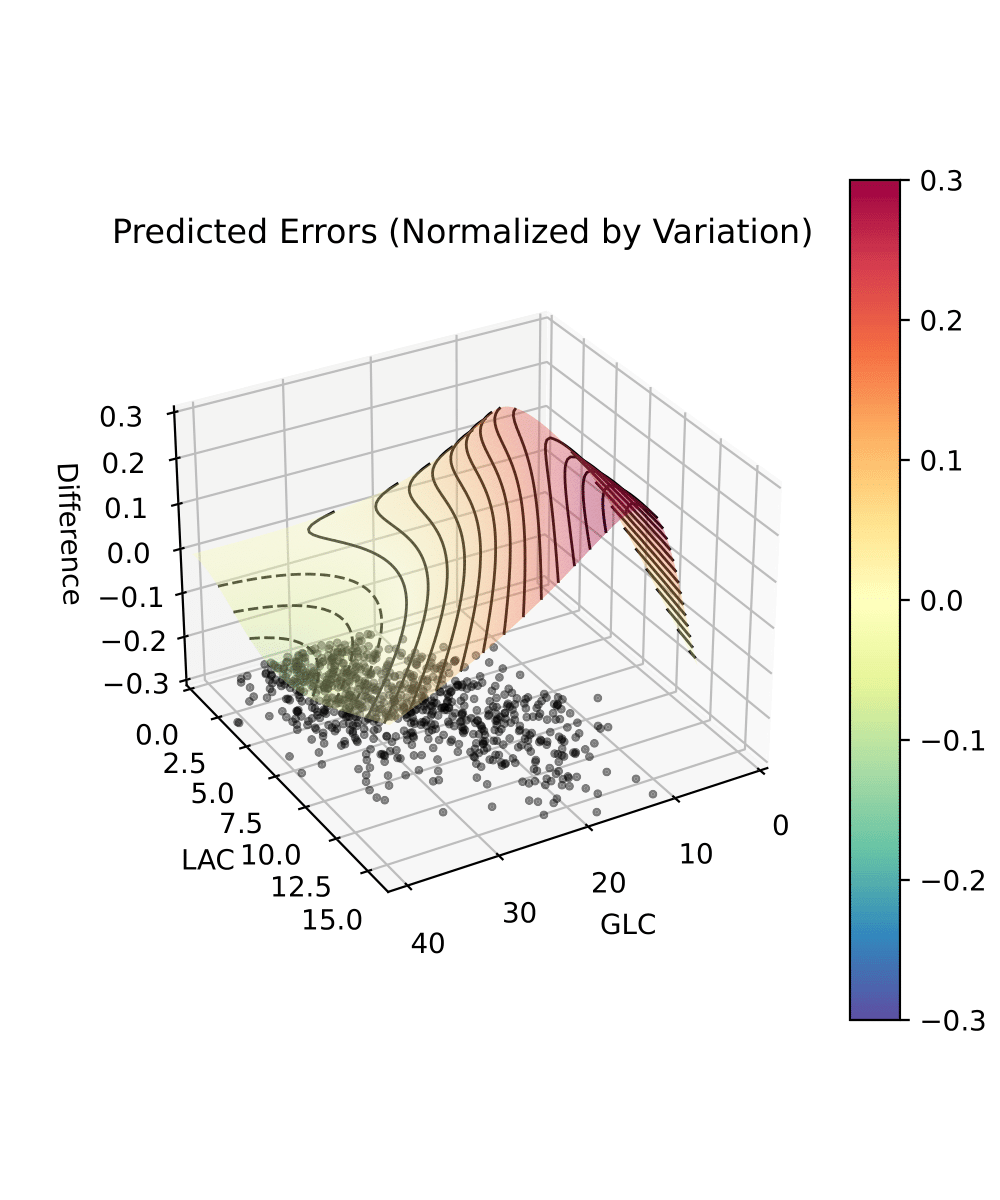}
        \caption{\label{fig:gbBErr}}
    \end{subfigure}
    
    \caption{
        \textbf{Comparison of fluxes (ground-truth vs stoichiometric gray-box model).}
        \ref{fig:gbBFluxFuncs}: ground-truth and neural net approximations of the fluxes given the neural net inputs, GLC and LAC.
        \ref{fig:gbBErr}: normalized prediction errors (fraction of max-min of the true function, across the data.)
        Note that the (GLC, LAC) points visited in training are scattered on the surfaces (\ref{fig:gbBFluxFuncs}) or on the base plane (\ref{fig:gbBErr}).
        Note again the relative rotation (for visual clarity) between (\ref{fig:gbBFluxFuncs}) and (\ref{fig:gbBErr}).
    \label{fig:gbB}
    }
\end{figure}

For each experiment,
we find (\figref{fig:gbAFluxFuncs} and \figref{fig:gbBFluxFuncs}, resp.)
that the learned flux functions are largely reproduced correctly, but there are discrepancies (relatively flat network predictions) over some parts of their domain.
The greatest percent discrepancy
(given in \figref{fig:gbAErr} and \figref{fig:gbBErr},
scaled by the spread of true values across the training data)
arises, as one might expect,
at locations where the flux is approximately zero.
The most obvious explanation for this error is that this region (small GLC) is not frequently visited by the ground-truth dynamics used for training data, especially in the stoichiometric case.
Finally, errors in small fluxes lead have less egregious consequences in what we actually minimize in the network: the prediction error for the concentration evolution.

\subsubsection{Partially Known Model, Partially Known Parameters}
\label{sec:grayBoxW}

Finally,
we combine the physical-parameter (white-box) fitting of \cref{sec:whiteBox}
with the gray-box fitting of \cref{sec:grayBoxG}
to produce a model in which we train both neural and physical components jointly.
For this experiment, we still assume that the expression for $\hat{v}_{I, 2}$ in $\fkin$ is unknown;
but we also additionally assume the value of one kinetic parameter, $\alpha_1$, also needs to be calibrated. 
As before, we also studied kinetic and stoichiometric versions of the experiment.

\begin{figure}[!ht]
    \centering
    \begin{subfigure}[b]{0.55\textwidth}
        \includegraphics[width=\textwidth]{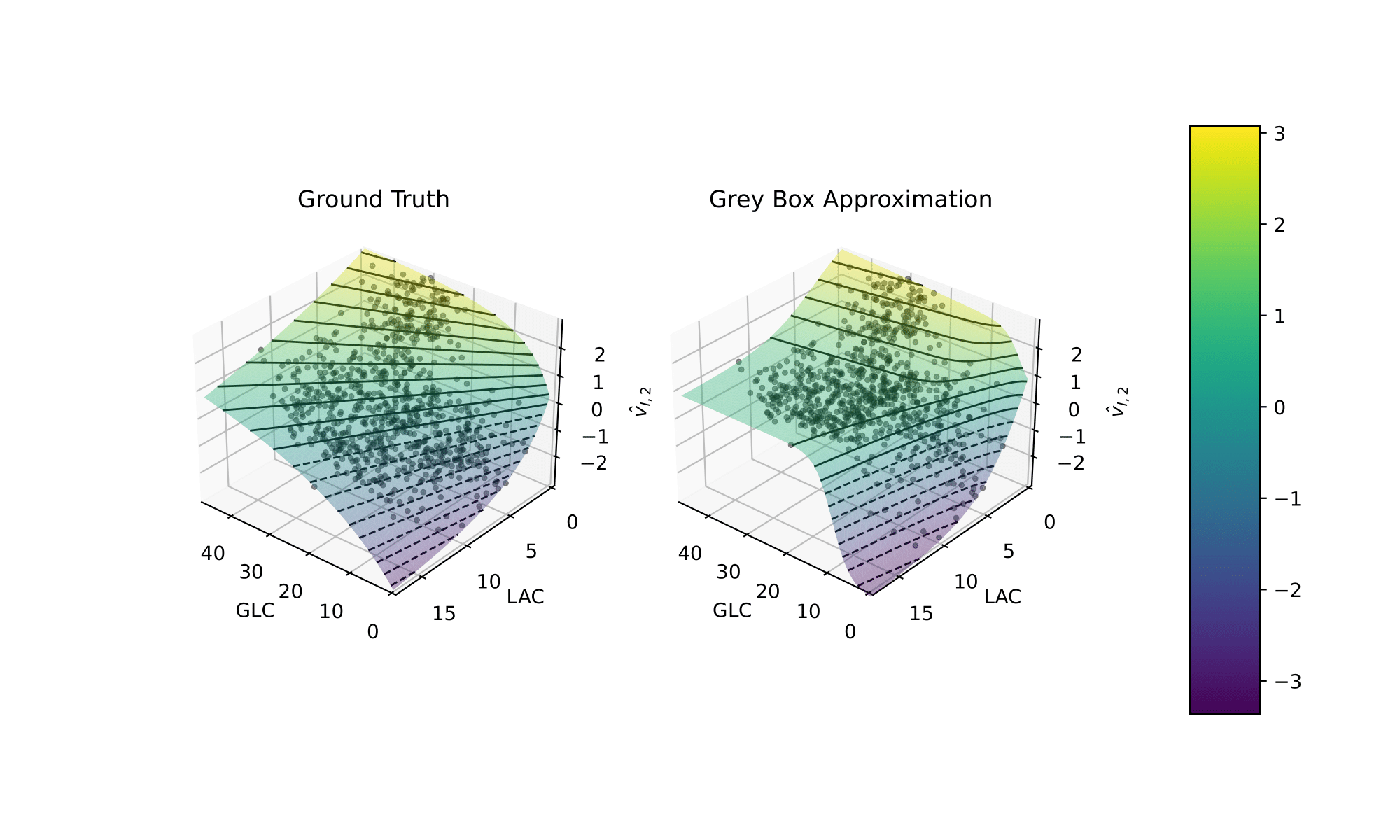}
        \caption{\label{fig:gbwbAFuncs}}
    \end{subfigure}
    \begin{subfigure}[b]{0.27\textwidth}
        \includegraphics[width=\textwidth]{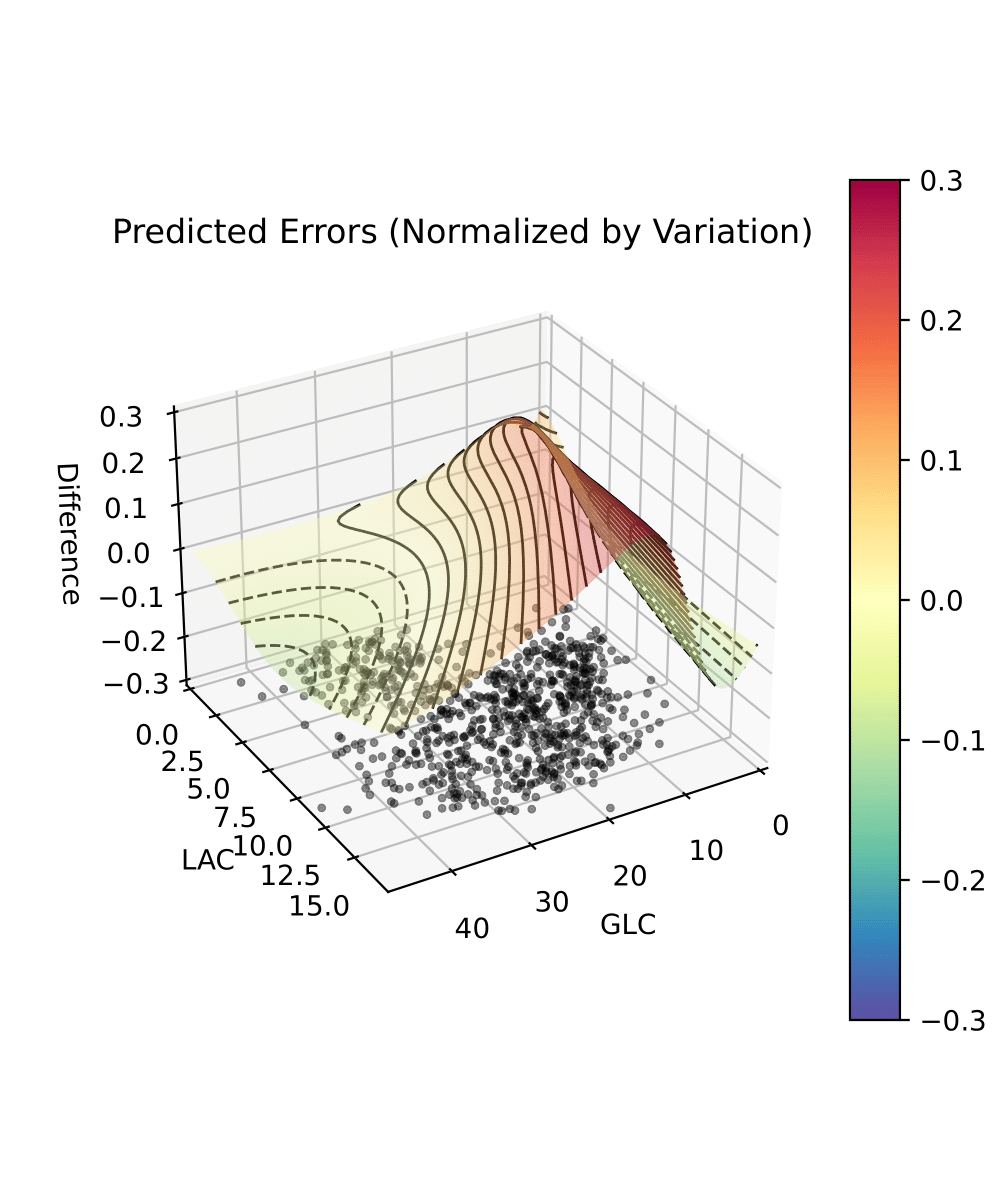}
        \caption{\label{fig:gbwbAErr}}
    \end{subfigure}
    \begin{subfigure}[b]{0.14\textwidth}
        \includegraphics[width=\textwidth]{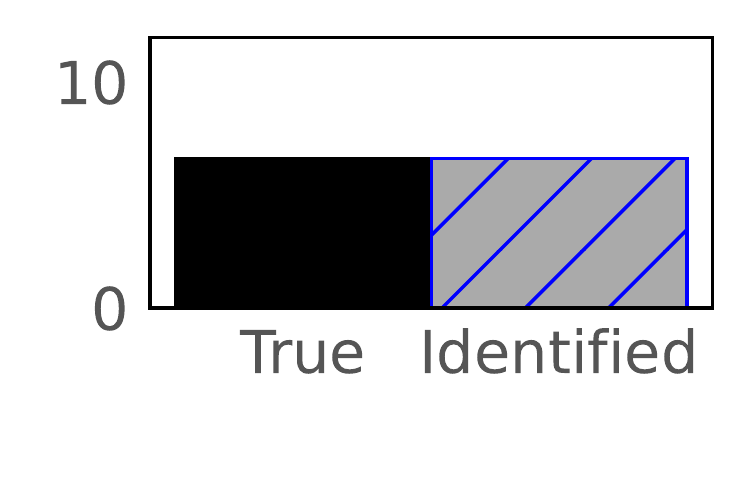}
        \caption{\label{fig:parameter_compare_EP}}
    \end{subfigure}
    \caption{
        \textbf{Comparison of fluxes (kinetic ground-truth vs gray-box model).} 
        \textbf{\ref{fig:gbwbAFuncs}:} ground-truth and neural net approximations of the fluxes as functions of the learned kinetic expression inputs, GLC and LAC.
        \textbf{\ref{fig:gbwbAErr}:} normalized (as in \figref{fig:gbAErr} and \ref{fig:gbBErr}) predicted errors.
        Black data points plotted as in \figref{fig:gbA} and \ref{fig:gbB}.
        \textbf{\ref{fig:parameter_compare_EP}:} Ground-truth value and the recovered value of the kinetic parameter $\alpha_1$.
    \label{fig:gbwbA}
    }
\end{figure}

\begin{figure}[!ht]
    \centering
    \begin{subfigure}[b]{0.55\textwidth}
        \includegraphics[width=\textwidth]{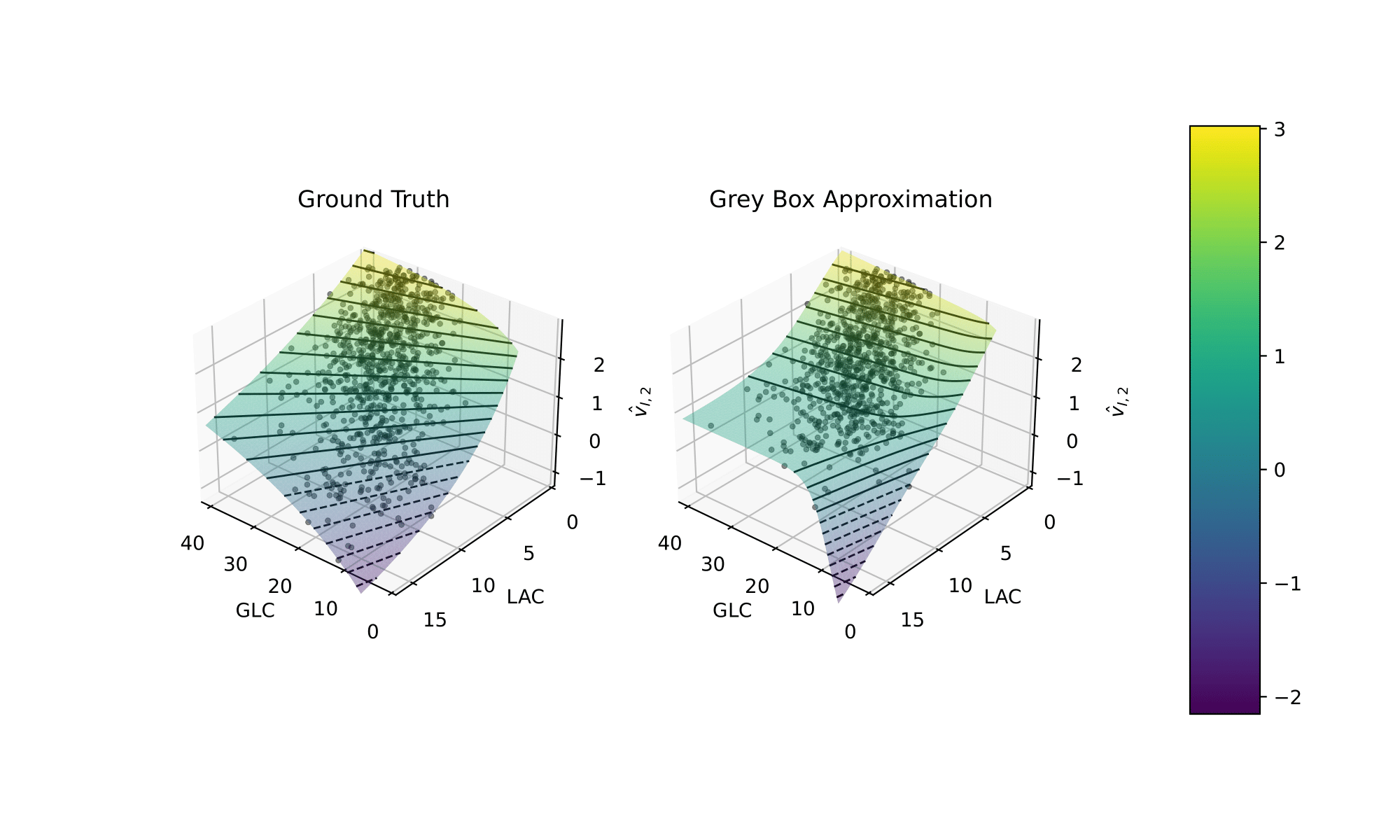}
        \caption{\label{fig:gbwbBFuncs}}
    \end{subfigure}
    \begin{subfigure}[b]{0.27\textwidth}
        \includegraphics[width=\textwidth]{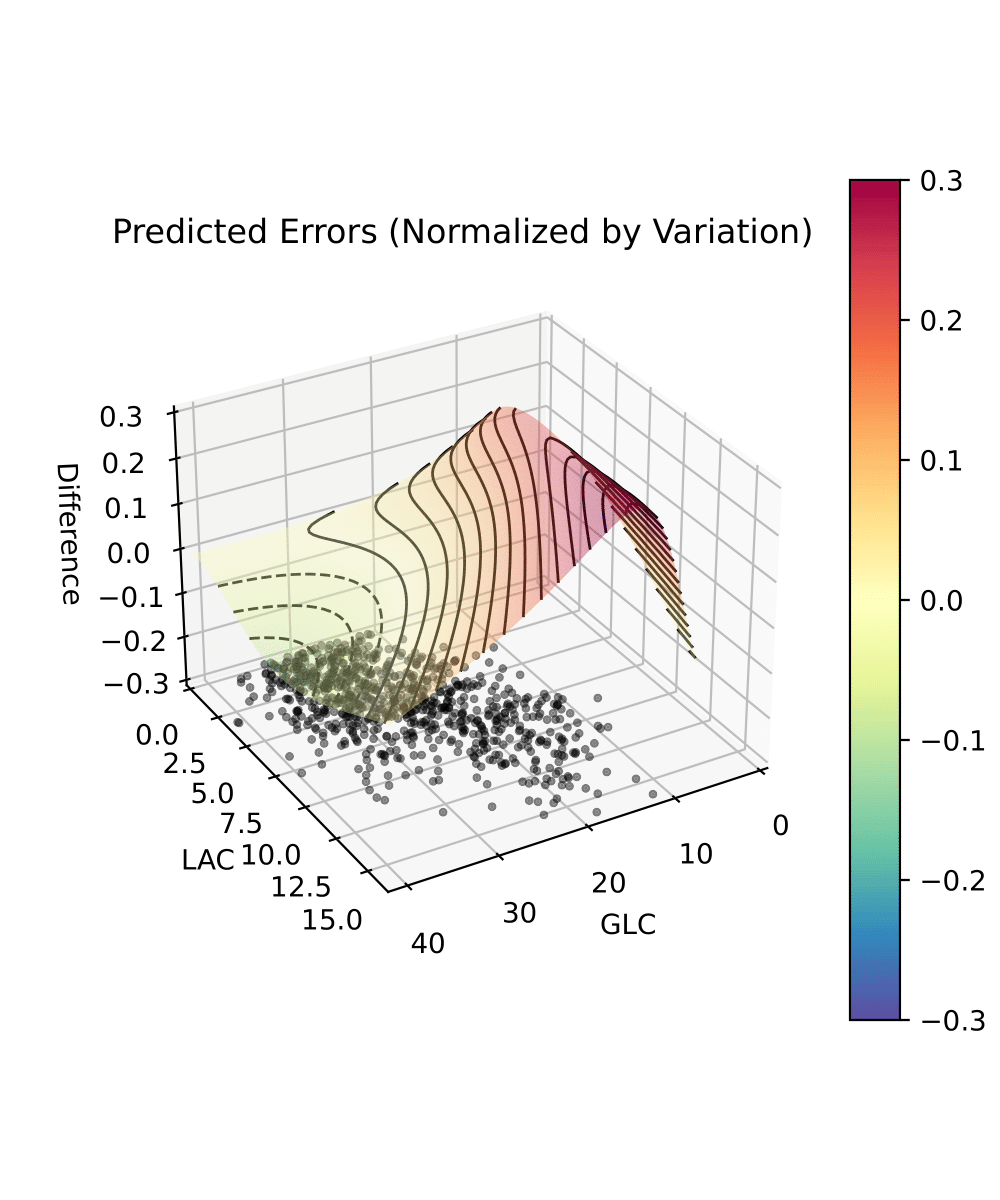}
        \caption{\label{fig:gbwbBErr}}
    \end{subfigure}
    \begin{subfigure}[b]{0.14\textwidth}
        \includegraphics[width=\textwidth]{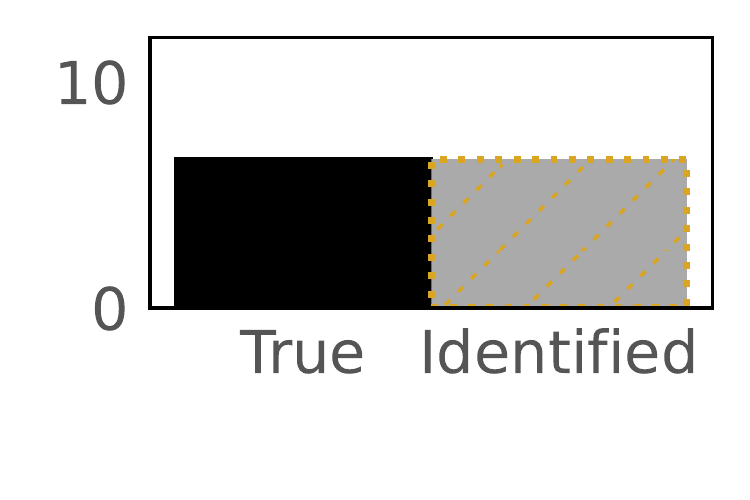}
        \caption{\label{fig:parameter_compare_EPS}}
    \end{subfigure}
    \caption{
        {
        \textbf{Comparison of fluxes (stoichiometric ground-truth vs gray-box model).} 
        \textbf{\ref{fig:gbwbBFuncs}:} ground-truth and neural net approximations of the fluxes as functions of the learned kinetic expression inputs GLC and LAC.
        \textbf{\ref{fig:gbwbBErr}:} normalized (as in \figref{fig:gbAErr} and \ref{fig:gbBErr}) predicted errors.
        Black data points plotted as in \figref{fig:gbA} and \ref{fig:gbB}.
        \textbf{\ref{fig:parameter_compare_EPS}:} Ground-truth value and the recovered value of the kinetic parameter $\alpha_1$.
        }
    \label{fig:gbwbB}
    }
\end{figure}

As we can see in 
\figref{fig:gbwbA}, \figref{fig:gbwbB} and \tabref{tab:parameters}
the recovery of the shape of the flux function
has similar characteristics to \cref{sec:grayBoxG};
yet we are also able to rediscover the parameter value accurately,
demonstrating the method's potential in joint learning 
with such mixed physical prior information.

\section{Conclusions and Future Directions}

In this paper, we revisited a mechanistic model of the biochemical reactions arising in Chinese Hamster Ovary (CHO) cell cultures. 
When simulating the dynamics of the model, evaluation of the temporal derivative of this system of equations practically necessitates the solution of a constrained convex problem
at each time step.
This ``inner optimization" can lead to: (a) discontinuities
in the second time derivatives of the evolving concentrations (that is, the solution itself is C1 as shown in \cref{fig:boundDisco});
and (b) difficulties in computing the system Jacobian, or sensitivity gradients of evolving states with respect to system parameters. 

We then demonstrated how to incorporate such mechanistic physical knowledge of the model along with data-driven approaches, so as to identify or calibrate this type of systems.
Our hybrid model can be black-, gray-, or white-box, depending on the portion of physical laws one is confident about {\em a priori}.
Importantly, we implemented a modification of traditional neural network/ODE-net architecture in our white- and gray-box models based on \cite{OptNetKolter}: this approach can encode the differentiable convex optimization layer within a numerical integrator (which can be considered as an unrolled recurrent neural network) so
as to overcome the obstacles of computing model gradients.
The potential of this type of data-driven models to
identify metabolic network dynamics from data, and perform regression tasks, was illustrated.

The approaches and model architectures that we designed an implemented in this paper should be of broad applicability
in fields of engineering where the right-hand-side of the evolution equations intrinsically involves an optimization
problem; robotics control, or differentiable Model Predictive Control \cite{BootsKolterMPC2018} come to mind.
In the metabolic engineering domain, such algorithms can be usefully combined with downstream optimization problems, for the design of experiments, the optimization of feed media composition, or the design of optimal feeding/harvesting policies in bioreactor operation. 

{\bf Acknowledgements:} This work was partially supported by AMBIC. The work of T.B., T.C. and I.G.K. was also partially supported by an AFOSR MURI. 
The work of C. M. was partially enabled by the DOE Center for Advanced Bioenergy and Bioproducts Science
(U.S. Department of Energy, Office of Science,
Office of Biological and Environmental Research
under Award Number DE-SC0018420)
and by the DOE Office of Science,
Office of Biological and Environmental Research
under Award Number DE-SC0018260).

\bibliographystyle{unsrt}
\bibliography{bi.bib}

\newcommand{\pDiscoveredKINETICKINETICbjj}{6617.2}
\newcommand{\pDiscoveredKINETICKINETICbja}{87.354}

\newcommand{\pDiscoveredKINETICKINETICejj}{6621}
\newcommand{\pDiscoveredKINETICKINETICeja}{87.371}
\newcommand{\pDiscoveredKINETICKINETICejb}{85.025}
\newcommand{\pDiscoveredKINETICKINETICejc}{3489.9}
\newcommand{\pDiscoveredKINETICKINETICejd}{6.3337}

\newcommand{\pDiscoveredSTOICHSTOICHbjj}{6643.8}
\newcommand{\pDiscoveredSTOICHSTOICHbja}{81.666}

\newcommand{\pDiscoveredSTOICHSTOICHejj}{6720.2}
\newcommand{\pDiscoveredSTOICHSTOICHeja}{82.774}
\newcommand{\pDiscoveredSTOICHSTOICHejb}{86.327}
\newcommand{\pDiscoveredSTOICHSTOICHejc}{3487}
\newcommand{\pDiscoveredSTOICHSTOICHejd}{6.3236}

\newpage
\appendix

\section{Physical Variables and Kinetic Parameters}

\begingroup
\setlength{\tabcolsep}{2pt}
\renewcommand{\arraystretch}{0.8}
\begin{table}[!ht]
\centering
\begin{tabular}{c|ccccccc}
    \toprule
    Variables & $C_1$ & $C_2$ & $C_3$ & $C_4$ & $C_5$ & $C_6$ & $C_7$ \\
    Abbreviations & BIOM & ANTI & GLC & LAC & ALA & ASN & ASP \\
    Names & Biomass & Antibody & Glucose & Lactate & Alanine & Asparagine & Aspartate \\
    Units & mmol/L & mmol/L & mmol/L & mmol/L & mmol/L & mmol/L & mmol/L \\
    Sample Values   & $1.18$ & $0.0801$ & $30.9$ & $1.36$ & $0.354$ & $5.23$ & $1.21$\\
    Kinetic $\mathrm{d}C_i/\mathrm{d}t$ & $2.02$ & $0.0361$ & $-1.76$ & $1.23$ & $0.179$ & $-0.152$ & $-0.097$\\
    Stoichiometric $\mathrm{d}C_i/\mathrm{d}t$ & $2.04$ & $0.0668$ & $-1.83$ & $1.2$ & $0.142$ & $-0.153$ & $-0.151$\\
    Kinetic - Stoichiometric& $-0.0116$& $-0.0307$& $0.07$& $0.0336$& $0.0372$& $0.00153$& $0.0536$\\
\midrule
    Sample Values   & $0.332$ & $4.34$ & $1.14$ & $7.74$ & $3.17$ & $2.68$ & $0.483$\\
    Kinetic $\mathrm{d}C_i/\mathrm{d}t$ & $-0.00657$ & $-0.997$ & $1.3$ & $-0.904$ & $0.181$ & $0.482$ & $0.875$\\
    Stoichiometric $\mathrm{d}C_i/\mathrm{d}t$ & $-0.0274$ & $-0.999$ & $1.05$ & $-0.764$ & $0.163$ & $0.679$ & $0.88$\\
    Kinetic - Stoichiometric& $0.0208$& $0.00239$& $0.242$& $-0.14$& $0.0179$& $-0.197$& $-0.00501$\\
\bottomrule
\end{tabular}
\caption{Names and units for all variables $\state = (C_1, C_2, \cdots, C_{14})^T$.}
\label{tab:concentrations}
\end{table}
\endgroup

\begingroup
\setlength{\tabcolsep}{2pt}

\begin{table}[!ht]
\centering
\begin{tabular}{rr|ccccccccc}
    \toprule
&    \changedText{}{Numbered Symbol} & $\alpha_1$ & $\alpha_2$ & $\alpha_3$ & $\alpha_4$ & $\alpha_5$ & $\alpha_6$ & $\alpha_7$ & $\alpha_8$ & $\alpha_9$ \\
&    \changedText{}{Physical Symbol} & $v_{max1}$ & $K_{i1}$ & $K_{m1}$ & $v_{max2f}$ & $K_{m2glc}$ & $v_{max3f}$ & $v_{max3r}$ & $K_{m3ala}$ & $K_{m3glc}$ \\
&    Values (True) & 6617.8 & 87.349 & 84.982 & 3490.4 & 6.3331 & 950.80 & 949.28 & 0.2165 & 2.0026 \\
\smash{\raisebox{-2mm}{\S\ref{sec:wb2p}}} &    Values (WB 2P Kin)   & $\pDiscoveredKINETICKINETICbjj$    & $\pDiscoveredKINETICKINETICbja$ \\
&    Values (WB 2P Sto)   & $\pDiscoveredSTOICHSTOICHbjj$    & $\pDiscoveredSTOICHSTOICHbja$ \\
\smash{\raisebox{-2mm}{\S\ref{sec:wb5p}}} &    Values (WB 5P Kin)   & $\pDiscoveredKINETICKINETICejj$ & $\pDiscoveredKINETICKINETICeja$ & $\pDiscoveredKINETICKINETICejb$ & $\pDiscoveredKINETICKINETICejc$ & $\pDiscoveredKINETICKINETICejd$ \\
&    Values (WB 5P Sto)   & $\pDiscoveredSTOICHSTOICHejj$ & $\pDiscoveredSTOICHSTOICHeja$ & $\pDiscoveredSTOICHSTOICHejb$ & $\pDiscoveredSTOICHSTOICHejc$ & $\pDiscoveredSTOICHSTOICHejd$ \\
\smash{\raisebox{-2mm}{\S\ref{sec:grayBoxW}}} &    Values (GB Kin)   & 6617.9 \\
&    Values (GB Sto)   & 6594.1 \\
&    Units & mmol/d & mmol/L & mmol/L & mmol/d & mmol/L & mmol/d & mmol/d & mmol/L & mmol/L \\
    \midrule
&    Numbered Symbol & $\alpha_{10}$ & $\alpha_{11}$ & $\alpha_{12}$ & $\alpha_{13}$ & $\alpha_{14}$ & $\alpha_{15}$ & $\alpha_{16}$ & $\alpha_{17}$ & $\alpha_{18}$ \\
&    Physical Symbol & $v_{max8f}$ & $v_{max8r}$ & $K_{m8gln}$ & $K_{m8glu}$ & $K_{m8nh3}$ & $v_{max9f}$ & $v_{max9r}$ & $K_{m9nh3}$ & $v_{max10f}$ \\
&    Values (True) & 9568.5 & 415.83 & 5.9198 & 2.0582 & 7.5053 & 3.3291 & 7.8132 & 0.6866 & 143.32 \\
&    Units & mmol/d & mmol/d & mmol/L & mmol/L & mmol/L & mmol/d & mmol/d & mmol/L & mmol/d \\
    \midrule
&    Numbered Symbol & $\alpha_{19}$ & $\alpha_{20}$ & $\alpha_{21}$ & $\alpha_{22}$ & $\alpha_{23}$ & $\alpha_{24}$ & $\alpha_{25}$ & $\alpha_{26}$ & $\alpha_{27}$ \\
&    Physical Symbol & $v_{max10r}$ & $K_{m10asn}$ & $K_{m10asp}$ & $K_{m10nh3}$ & $v_{max11}$ & $v_{max12f}$ & $v_{max12r}$ & $K_{m12ser}$ & $K_{m12gly}$ \\
&    Values (True) & 95.194 & 0.0157 & 3.5060 & 0.6301 & 0.5465 & 0.6330 & 92.978 & 3.0862 & 0.2020 \\
&    Units & mmol/d & mmol/L & mmol/L & mmol/L & mmol/d & 1 & mmol/d & mmol/L & mmol/L \\
    \midrule
&    Numbered Symbol & $\alpha_{28}$ & $\alpha_{29}$ & $\alpha_{30}$ & $\alpha_{31}$ & $\alpha_{32}$ & $\alpha_{33}$ & $\alpha_{34}$ & $\alpha_{35}$ & $\alpha_{36}$ \\
&    Physical Symbol & $v_{max13}$ & $K_{m13}$ & $v_{max16}$ & $K_{m16a}$ & $K_{m16b}$ & $v_{max17}$ & $K_{i17}$ & $v_{max2r}$ & $K_{m2lac}$ \\
&    Values (True) & 72.593 & 1.4396 & 6316.7 & 0.9967 & 0.9901 & 76.88 & 46.045 & 3996.2 & 4.3040 \\
&    Units & mmol/d & mmol/L & mmol/d & mmol/L & mmol/L & mmol/d & mmol/L & mmol/d & mmol/L \\
    \midrule
&    Numbered Symbol & $\alpha_{37}$ & $\alpha_{38}$ & $\alpha_{39}$ & $\alpha_{40}$ & $\alpha_{41}$ & $\alpha_{42}$ & $\alpha_{43}$ & $\alpha_{44}$ & $\alpha_{45}$ \\
&    Physical Symbol & $K_{m9glu}$ & $K_{m9gln}$ & $K_{m11asp}$ & $K_{m11asn}$ & $K_{m12nh3}$ & $K_{m16c}$ & $v_{max35}$ & $K_{m35a}$ & $K_{m35b}$ \\
&    Values (True) & 0.7519 & 4.1330 & 6.4439 & 6.5433 & 0.2607 & 0.0108 & 2.371 & 6.8779 & 9.1087 \\
&    Units & mmol/L & mmol/L & mmol/L & mmol/L & mmol/L & mmol/L & mmol/d & mmol/L & mmol/L \\
    \bottomrule
\end{tabular}
\caption{Names and ground-truth values for all parameters $\param = (\alpha_1, \alpha_2, \cdots, \alpha_{45})^T$, as well as the discovered values of several parameters by white-box and gray-box models. Here, meanings of "2PK" and others can be checked in \tabref{tab:experiments}.
}
\label{tab:parameters}
\end{table}
\endgroup

\newpage
\section{The ODE Expressions $\ddt{\state} = \fODE(\state; \flux(\state; \param))$}
\label{sec:fODE}
\begin{equation*}
\begin{aligned}
    \ddt{C_1} &= v_{18} \frac{C_{14}}{1000} \\
    \ddt{C_2} &= v_{19} \frac{C_{14}}{1000} \\
    \ddt{C_3} &= -v_{20} \frac{C_{14}}{1000} \\
    \ddt{C_4} &= v_{21} \frac{C_{14}}{1000} \\
    \ddt{C_5} &= v_{22} \frac{C_{14}}{1000} \\
    \ddt{C_6} &= -v_{23} \frac{C_{14}}{1000} \\
    \ddt{C_7} &= v_{24} \frac{C_{14}}{1000} \\
    \ddt{C_8} &= -v_{25} \frac{C_{14}}{1000} \\
    \ddt{C_9} &= -v_{26} \frac{C_{14}}{1000} \\
    \ddt{C_{10}} &= v_{27} \frac{C_{14}}{1000} \\
    \ddt{C_{11}} &= -v_{28} \frac{C_{14}}{1000} \\
    \ddt{C_{12}} &= v_{29} \frac{C_{14}}{1000} \\
    \ddt{C_{13}} &= (v_{25} + v_{33}) \frac{C_{14}}{1000} \\
    \ddt{C_{14}} &= \frac{\ddt{C_1}}{2.31} - 0.003 C_{14} \\ 
\end{aligned}
\end{equation*}

\newpage
\section{List of all Reactions}
\label{sec:reactions}

\newcommand{\IntFx}[1]{\textcolor{blue}{#1}}
\newcommand{\ExFxI}[1]{\textcolor{red}{\underline{#1}}}
\newcommand{\ExFxR}[1]{\textcolor{olive}{\textoverline{#1}}}

\begingroup
\renewcommand{\arraystretch}{0.75}
\begin{table}[!ht]
\centering
\begin{tabular}{c|cccl}
    \toprule
ID & Kin. & Stoi. & \makecell{Kin. -\\Stoi.} & Reaction \\
    \hline
\IntFx{1} & $1.18$ & $2.22$ & $-1.04$ & $\mathrm{G6P \rightarrow 2PYR + 3ATP + 2NADH (Cytosolic)}$  \\
    \hline
\IntFx{2} & $0.0801$ & $-0.901$ & $0.981$ & $\mathrm{PYR + NADH (Cytosolic) \leftrightarrow LAC}$  \\
    \hline
\IntFx{3} & $30.9$ & $28.6$ & $2.32$ & $\mathrm{PYR + GLU \leftrightarrow ALA + AKG}$  \\
    \hline
\ExFxI{4} & \makecell{$1.024$\\$\times10^{-12}$} & \makecell{$1.153$\\$\times10^{-15}$} & \makecell{$1.023$\\$\times10^{-12}$} & $\mathrm{PYR + OXA \rightarrow AKG + 2CO_2 + 2NADH (Mitochondrial)}$  \\
    \hline
\ExFxI{5} & $22.5$ & $20.4$ & $2.06$ & $\mathrm{AKG \rightarrow MAL + CO_2 + NADH (Mitochondrial) + FADH_2 + ATP}$  \\
    \hline
\ExFxI{6} & \makecell{$1.151$\\$\times10^{-13}$} & \makecell{$7.718$\\$\times10^{-16}$} & \makecell{$1.144$\\$\times10^{-13}$} & $\mathrm{MAL \rightarrow OXA + NADH (Mitochondrial)}$  \\
    \hline
\ExFxI{7} & $25.7$ & $22.3$ & $3.35$ & $\mathrm{MAL \rightarrow PYR + CO_2}$  \\
    \hline
\IntFx{8} & $1.36$ & $1.36$ & \makecell{$-2.852$\\$\times10^{-5}$} & $\mathrm{GLN \leftrightarrow GLU + NH_3}$  \\
    \hline
\IntFx{9} & $0.354$ & $3.78$ & $-3.42$ & $\mathrm{AKG + NH_3 + NADH (Mitochondrial) \leftrightarrow GLU}$  \\
    \hline
\IntFx{10} & $5.23$ & $5.23$ & \makecell{$-3.530$\\$\times10^{-5}$} & $\mathrm{ASN \leftrightarrow ASP + NH_3}$  \\
    \hline
\IntFx{11} & $1.21$ & $1.82$ & $-0.61$ & $\mathrm{ASP + AKG \leftrightarrow OXA + GLU}$  \\
    \hline
\IntFx{12} & $0.332$ & $-0.597$ & $0.929$ & $\mathrm{SER + CO_2 + NH_3 + NADH (Cytosolic) \leftrightarrow 2GLY}$  \\
    \hline
\IntFx{13} & $4.34$ & $3.41$ & $0.929$ & $\mathrm{Cystine + NADH (Cytosolic) \rightarrow 2Cysteine}$  \\
    \hline
\ExFxI{14} & $14.8$ & $19.4$ & $-4.55$ & $\mathrm{NADH (Mitochondrial) + 0.5 O_2 \rightarrow 2.5 ATP}$  \\
    \hline
\ExFxI{15} & $16.6$ & $20.5$ & $-3.89$ & $\mathrm{FADH_2 + 0.5 O_2 \rightarrow 1.5 ATP}$  \\
    \hline
\IntFx{16} & $1.14$ & $4.27$ & $-3.13$ & \makecell{$\mathrm{0.0838 ALA + 0.041 ASN + 0.0804 ASP + 8.6825 ATP + 0.0261 Cysteine + 0.452 G6P}$ \\ $\mathrm{+ 0.0873 GLN + 0.056 GLY + 0.427 OXA + 0.096 SER \rightarrow BIOM + 0.004 FADH_2}$ \\ $\mathrm{+ 0.0082 GLU + 0.4445 MAL + 0.6391 NADH (Mitochondrial) + 0.2085 PYR}$}  \\
    \hline
\IntFx{17} & $7.74$ & $7.34$ & $0.399$ & \makecell{$\mathrm{0.0614 ALA + 0.0344 ASN + 0.0389 ASP + 9.2 ATP + 0.024 Cysteine }$ \\ $\mathrm{+ 0.0479 GLU + 0.0449 GLN + 0.0719 GLY + 0.1 SER \rightarrow ANTI}$}  \\
    \hline
\IntFx{18} & $3.17$ & $4.27$ & $-1.1$ & $\mathrm{BIOM \rightarrow BIOM (Extracellular)}$  \\
    \hline
\IntFx{19} & $2.68$ & $7.34$ & $-4.66$ & $\mathrm{ANTI \rightarrow ANTI (Extracellular)}$  \\
    \hline
\ExFxI{20} & $4.28$ & $2.89$ & $1.39$ & $\mathrm{GLC(Extracellular) + ATP \rightarrow G6P}$  \\
    \hline
\ExFxR{21} & $0.0801$ & $-0.901$ & $0.981$ & $\mathrm{LAC \leftrightarrow LAC(Extracellular)}$  \\
    \hline
\ExFxR{22} & $30.3$ & $27.8$ & $2.56$ & $\mathrm{ALA \leftrightarrow ALA(Extracellular)}$  \\
    \hline
\ExFxI{23} & $5.54$ & $5.65$ & $-0.115$ & $\mathrm{ASN(Extracellular) \rightarrow ASN}$  \\
    \hline
\ExFxR{24} & $3.62$ & $2.78$ & $0.846$ & $\mathrm{ASP \leftrightarrow ASP(Extracellular)}$  \\
    \hline
\ExFxI{25} & \makecell{$5.756$\\$\times10^{-15}$} & $3.41$ & $-3.41$ & $\mathrm{Cystine (Extracellular) + GLU \rightarrow Cystine + GLU(Extracellular)}$  \\
    \hline
\ExFxR{26} & $1.8$ & $2.06$ & $-0.255$ & $\mathrm{GLN(Extracellular) \leftrightarrow GLN}$  \\
    \hline
\ExFxR{27} & $0.0431$ & $-1.96$ & $2.01$ & $\mathrm{GLY \leftrightarrow GLY(Extracellular)}$  \\
    \hline
\ExFxI{28} & $0.249$ & $3.07$ & $-2.82$ & $\mathrm{SER(Extracellular) \rightarrow SER}$  \\
    \hline
\ExFxR{29} & $5.9$ & $3.4$ & $2.5$ & $\mathrm{NH_3 \leftrightarrow NH_3(Extracellular)}$  \\
    \hline
\ExFxI{30} & $19.9$ & $23.2$ & $-3.26$ & $\mathrm{O_2(Extracellular) \leftrightarrow O_2}$  \\
    \hline
\ExFxI{31} & $47.8$ & $43.4$ & $4.47$ & $\mathrm{CO_2 \rightarrow CO_2(Extracellular)}$  \\
    \hline
\ExFxI{32} & $4.23$ & $3.27$ & $0.965$ & $\mathrm{2Cysteine + O_2 \rightarrow 2Cystine}$  \\
    \hline
\ExFxI{33} & \makecell{$8.539$\\$\times10^{-15}$} & $-22.8$ & $22.8$ & $\mathrm{GLU \rightarrow GLU(Extracellular)}$  \\
    \hline
\ExFxR{34} & $-5.6$ & \makecell{$4.368$\\$\times10^{-15}$} & $-5.6$ & $\mathrm{NADH (Cytosolic) \rightarrow 0.5NADH (Mitochondrial) + 0.5 FADH_2}$  \\
    \hline
\IntFx{35} & $0.483$ & $-1.26$ & $1.75$ & $\mathrm{G6P + ATP + 2GLU \rightarrow 2NADH (Cytosolic) + 2SER + 2AKG}$  \\
    \bottomrule
\end{tabular}
\caption{Expressions of all reactions in the metabolic network. Note that we distinguish between cytosolic and mitochondrial NADH,  while all metabolites are intracelluar  unless indicated otherwise.
Reactions with ``$\leftrightarrow$" are reversible, while ``$\rightarrow$" is used if and only if the reaction is irreversible.
ID numbers are colored as described in \cref{sec:stoiMat}.
Fluxes (numeric columns) are evaluated at the same state as used in \cref{tab:concentrations}.
}
\label{tab:reactions}
\end{table}
\endgroup

\newpage
\section{The Kinetic Expressions $\vIhat = \fkin(\state; \param)$}
\label{sec:fkin}
\begin{equation*}
\begin{aligned}
    \hat{v}_{I, 1} &= \frac{v_{max1} \frac{\text{GLC}}{K_{m1}}}{1 + \frac{\text{LAC}}{K_{i1}} + \frac{\text{GLC}}{K_{m1}} + \frac{\text{LAC}}{K_{i1}} \frac{\text{GLC}}{K_{m1}}} \\
    \hat{v}_{I, 2} &= \frac{v_{max2f} \frac{\text{GLC}}{K_{m2glc}} - v_{max2r} \frac{\text{LAC}}{K_{m2lac}}}{1 + \frac{\text{GLC}}{K_{m2glc}} + \frac{\text{LAC}}{K_{m2lac}}} \\
    \hat{v}_{I, 3} &= \frac{v_{max3f} \frac{\text{GLC}}{K_{m3glc}} - v_{max3r} \frac{\text{ALA}}{K_{m3ala}}}{1 + \frac{\text{GLC}}{K_{m3glc}} + \frac{\text{ALA}}{K_{m3ala}}} \\
    \hat{v}_{I, 4} &= \frac{v_{max8f} \frac{\text{GLN}}{K_{m8gln}} - v_{max8r} \frac{\text{GLU}}{K_{m8glu}} \frac{\text{NH3}}{K_{m8nh3}}}{1 + \frac{\text{GLN}}{K_{m8gln}} + \frac{\text{GLU}}{K_{m8glu}} + \frac{\text{NH3}}{K_{m8nh3}} + \frac{\text{GLU}}{K_{m8glu}} \frac{\text{NH3}}{K_{m8nh3}}} \\
    \hat{v}_{I, 5} &= \frac{v_{max9f} \frac{\text{NH3}}{K_{m9nh3}} - v_{max9r} (\frac{\text{GLU}}{K_{m9glu}} + \frac{\text{GLN}}{K_{m9gln}})}{1 + \frac{\text{NH3}}{K_{m9nh3}} + \frac{\text{GLU}}{K_{m9glu}} + \frac{\text{GLN}}{K_{m9gln}} +  \frac{\text{GLU}}{K_{m9glu}} \frac{\text{GLN}}{K_{m9gln}}} \\
    \hat{v}_{I, 6} &= \frac{v_{max10f} \frac{\text{ASN}}{K_{m10asn}} - v_{max10r} \frac{\text{ASP}}{K_{m10asp}} \frac{\text{NH3}}{K_{m10nh3}}}{1 + \frac{\text{ASN}}{K_{m10asn}} + \frac{\text{ASP}}{K_{m10asp}} + \frac{\text{NH3}}{K_{m10nh3}} +  \frac{\text{ASP}}{K_{m10asp}} \frac{\text{NH3}}{K_{m10nh3}}} \\
    \hat{v}_{I, 7} &= \frac{v_{max11} (\frac{\text{ASP}}{K_{m11asp}} + \frac{\text{ASN}}{K_{m11asn}})}{1 + \frac{\text{ASN}}{K_{m11asn}} + \frac{\text{ASP}}{K_{m11asp}} + \frac{\text{ASP}}{K_{m11asp}} \frac{\text{ASN}}{K_{m11asn}}} \\
    \hat{v}_{I, 8} &= \frac{v_{max12f} \cdot \hat{v}_{I, 10} \frac{\text{SER}}{K_{m12ser}} \frac{\text{NH3}}{K_{m12nh3}} - v_{max12r} (\frac{\text{GLY}}{K_{m12gly}})^2}{1 + \frac{\text{SER}}{K_{m12ser}} + \frac{\text{GLY}}{K_{m12gly}} + \frac{\text{NH3}}{K_{m12nh3}} + (\frac{\text{GLY}}{K_{m12gly}})^2} \\
    \hat{v}_{I, 9} &= \frac{v_{max13} \frac{\text{CYS}}{K_{m13}}}{1 + \frac{\text{CYS}}{K_{m13}}} \\
    \hat{v}_{I, 10} &= \frac{v_{max16} \frac{\text{GLN}}{K_{m16a}} \frac{\text{ASN}}{K_{m16b}} \frac{\text{ALA}}{K_{m16c}}}{1 + \frac{\text{GLN}}{K_{m16a}} + \frac{\text{ASN}}{K_{m16b}} + \frac{\text{ALA}}{K_{m16c}} + \frac{\text{GLN}}{K_{m16a}} \frac{\text{ASN}}{K_{m16b}} + \frac{\text{GLN}}{K_{m16a}} \frac{\text{ALA}}{K_{m16c}} + \frac{\text{ASN}}{K_{m16b}} \frac{\text{ALA}}{K_{m16c}} + \frac{\text{GLN}}{K_{m16a}} \frac{\text{ASN}}{K_{m16b}} \frac{\text{ALA}}{K_{m16c}}} \\
    \hat{v}_{I, 11} &= \frac{v_{max17}}{1 + \frac{\text{LAC}}{K_{i17}}} \\
    \hat{v}_{I, 12} &= \hat{v}_{I, 10} \\
    \hat{v}_{I, 13} &= \hat{v}_{I, 11} \\
    \hat{v}_{I, 14} &= \frac{v_{max35} \frac{\text{GLC}}{K_{m35a}} (\frac{\text{GLU}}{K_{m35b}})^2}{1 + \frac{\text{GLC}}{K_{m35a}} + \frac{\text{GLU}}{K_{m35b}} + (\frac{\text{GLU}}{K_{m35b}})^2} 
\end{aligned}
\end{equation*}

\newpage
\section{Deviations in Kinetic Expressions from \cite{NOLAN2011108}}
\label{sec:Nolan}

All kinetically defined intracellular reactions have been modeled using the Michaelis Menten format, based on kinetic expressions from Nolan and Lee \cite{NOLAN2011108}. Some of the modifications are summarized below and will be described in detail in a separate publication (Khare et al., to be published).

The temperature dependent constants $TC$ which were used to scale the maximum forward reaction rate - $v_{max}$, have been removed from the kinetic expressions as have the inhibition exponential constants $exp_i$. Secondly, the redox variable $R$ which was used to account for the redox state of the cell in the reaction kinetics was removed. Instead, reactions were modeled solely based on the extracellular metabolites with consumption and production rates being dependent on the concentrations of the associated extracellular metabolites. Reaction stoichiometries were revised and additional reactions were included based on literature and $^13$C-labeled (intracellular and extracellular) metabolite tracking data. Reasons for the modifications are listed below:

\begin{itemize}
    \item Reaction 1 (corresponding to $\hat{v}_{I, 1}$): $\mathrm{G6P \rightarrow 2PYR + 3ATP + 2NADH}$

    Temperature dependent constant $TC_1$ and inhibition exponential constant $exp_1$ (which is also temperature dependent) were removed from the kinetic expression.
    
    \item Reaction 2 (corresponding to $\hat{v}_{I, 2}$): $\mathrm{PYR + NADH \leftrightarrow LAC}$

    Redox variable $R$ removed and reaction made continuous.
    
    \item Reaction 3 (corresponding to $\hat{v}_{I, 3}$): $\mathrm{PYR + GLU \leftrightarrow ALA + AKG}$
    
    Temperature constant $TC_{3b}$ removed.
    
    \item Reaction 8 (corresponding to $\hat{v}_{I, 4}$): $\mathrm{GLN \leftrightarrow GLU + NH_3}$
    
    Temperature constant $TC_8$ removed. 
    
    \item Reaction 9 (corresponding to $\hat{v}_{I, 5}$): $\mathrm{AKG + NH_3 + NADH \leftrightarrow GLU}$ 
    
    Dependence on intracellular metabolite AKG (not dynamically tracked) removed, to make reaction expression adhere to Michaelis Menten kinetics. 
    
    \item Reaction 10 (corresponding to $\hat{v}_{I, 6}$): $\mathrm{ASN \leftrightarrow ASP + NH_3}$
    
    Temperature dependent constant $TC_{10}$ removed from kinetic expression.
    
    \item Reaction 11 (corresponding to $\hat{v}_{I, 7}$): $\mathrm{ASP + AKG \leftrightarrow OXA + GLU}$

    The reaction expression has been changed to be irreversible.

    \item Reaction 12 (corresponding to $\hat{v}_{I, 8}$): $\mathrm{SER + CO_2 + NH_3 + NADH \leftrightarrow 2GLY}$

    Forward reaction made dependent on ammonia in addition to serine.

    \item Reaction 13 (corresponding to $\hat{v}_{I, 9}$): $\mathrm{Cystine + NADH \rightarrow 2Cysteine}$ 

    Variable $r$, which is temperature dependent, has been removed. 

    \item Reaction 16 (corresponding to $\hat{v}_{I, 10}$): Biomass production 
    
    Temperature constant $TC_{16}$ and temperature dependency removed.

    \item Reaction 17 (corresponding to $\hat{v}_{I, 11}$): Antibody synthesis 

    Exponential inhibition constant $exp_{17}$ removed. 

    \item Reaction 33: Removed from our expressions since we have defined expressions for only intracellular reactions as Michaelis-Menten and not transport reactions. The transport reactions are all calculated stoichiometrically.

    \item Reaction 35 (corresponding to $\hat{v}_{I, 14}$): $\mathrm{G6P + ATP + 2GLU \rightarrow 2NADH (Cytosolic) + 2SER + 2AKG}$

    This reaction was added based on the presence of the serine synthesis pathway (SSP) originating from glycolysis.

\end{itemize}

\newpage
\section{The Stoichiometric Matrix $\Smat$} \label{sec:stoiMat}

\small{
    \IntFx{Blue:} indices of intracellular fluxes \\
    \ExFxR{Olive Overlined:} indices of reversible extracellular fluxes\\
    \ExFxI{Red Underlined:} indices of irreversible extracellular fluxes 
}

\rotatebox{90}{\tiny{\begingroup
\setlength\arraycolsep{1.85pt} \renewcommand{\arraystretch}{2.65}
{$\begin{bNiceMatrix}[first-row, first-col]
& \IntFx{1} & \IntFx{2} & \IntFx{3} & \ExFxI{4} & \ExFxI{5} & \ExFxI{6} & \ExFxI{7} & \IntFx{8} & \IntFx{9} & \IntFx{10} & \IntFx{11} & \IntFx{12} & \IntFx{13} & \ExFxI{14} & \ExFxI{15} & \IntFx{16} & \IntFx{17} & \IntFx{18} & \IntFx{19} & \ExFxI{20} & \ExFxR{21} & \ExFxR{22} & \ExFxI{23} & \ExFxR{24} & \ExFxI{25} & \ExFxR{26} & \ExFxR{27} & \ExFxI{28} & \ExFxR{29} & \ExFxI{30} & \ExFxI{31} & \ExFxI{32} & \ExFxI{33} & \ExFxR{34} & \IntFx{35} \\
\mathrm{AKG} & 0 & 0 & 1 & 1 & -1 & 0 & 0 & 0 & -1 & 0 & -1 & 0 & 0 & 0 & 0 & 0 & 0 & 0 & 0 & 0 & 0 & 0 & 0 & 0 & 0 & 0 & 0 & 0 & 0 & 0 & 0 & 0 & 0 & 0 & 2 \\
\mathrm{ALA} & 0 & 0 & 1 & 0 & 0 & 0 & 0 & 0 & 0 & 0 & 0 & 0 & 0 & 0 & 0 & -0.0838 & -0.0614 & 0 & 0 & 0 & 0 & -1 & 0 & 0 & 0 & 0 & 0 & 0 & 0 & 0 & 0 & 0 & 0 & 0 & 0 \\
\mathrm{ANTI} & 0 & 0 & 0 & 0 & 0 & 0 & 0 & 0 & 0 & 0 & 0 & 0 & 0 & 0 & 0 & 0 & 1 & 0 & -1 & 0 & 0 & 0 & 0 & 0 & 0 & 0 & 0 & 0 & 0 & 0 & 0 & 0 & 0 & 0 & 0 \\
\mathrm{ASN} & 0 & 0 & 0 & 0 & 0 & 0 & 0 & 0 & 0 & -1 & 0 & 0 & 0 & 0 & 0 & -0.041 & -0.0344 & 0 & 0 & 0 & 0 & 0 & 1 & 0 & 0 & 0 & 0 & 0 & 0 & 0 & 0 & 0 & 0 & 0 & 0 \\
\mathrm{ASP} & 0 & 0 & 0 & 0 & 0 & 0 & 0 & 0 & 0 & 1 & -1 & 0 & 0 & 0 & 0 & -0.0804 & -0.0389 & 0 & 0 & 0 & 0 & 0 & 0 & -1 & 0 & 0 & 0 & 0 & 0 & 0 & 0 & 0 & 0 & 0 & 0 \\
\mathrm{ATP} & 3 & 0 & 0 & 0 & 1 & 0 & 0 & 0 & 0 & 0 & 0 & 0 & 0 & 2.5 & 1.5 & -8.6825 & -9.2 & 0 & 0 & -1 & 0 & 0 & 0 & 0 & 0 & 0 & 0 & 0 & 0 & 0 & 0 & 0 & 0 & 0 & -1 \\
\mathrm{BIOM} & 0 & 0 & 0 & 0 & 0 & 0 & 0 & 0 & 0 & 0 & 0 & 0 & 0 & 0 & 0 & 1 & 0 & -1 & 0 & 0 & 0 & 0 & 0 & 0 & 0 & 0 & 0 & 0 & 0 & 0 & 0 & 0 & 0 & 0 & 0 \\
\mathrm{CO_2} & 0 & 0 & 0 & 2 & 1 & 0 & 1 & 0 & 0 & 0 & 0 & -1 & 0 & 0 & 0 & 0 & 0 & 0 & 0 & 0 & 0 & 0 & 0 & 0 & 0 & 0 & 0 & 0 & 0 & 0 & -1 & 0 & 0 & 0 & 0 \\
\mathrm{Cysteine} & 0 & 0 & 0 & 0 & 0 & 0 & 0 & 0 & 0 & 0 & 0 & 0 & 2 & 0 & 0 & -0.0261 & -0.024 & 0 & 0 & 0 & 0 & 0 & 0 & 0 & 0 & 0 & 0 & 0 & 0 & 0 & 0 & -2 & 0 & 0 & 0 \\
\mathrm{Cystine} & 0 & 0 & 0 & 0 & 0 & 0 & 0 & 0 & 0 & 0 & 0 & 0 & -1 & 0 & 0 & 0 & 0 & 0 & 0 & 0 & 0 & 0 & 0 & 0 & 1 & 0 & 0 & 0 & 0 & 0 & 0 & 0 & 0 & 0 & 0 \\
\mathrm{FADH_2} & 0 & 0 & 0 & 0 & 1 & 0 & 0 & 0 & 0 & 0 & 0 & 0 & 0 & 0 & -1 & 0.004 & 0 & 0 & 0 & 0 & 0 & 0 & 0 & 0 & 0 & 0 & 0 & 0 & 0 & 0 & 0 & 0 & 0 & 0.5 & 0 \\
\mathrm{G6P} & -1 & 0 & 0 & 0 & 0 & 0 & 0 & 0 & 0 & 0 & 0 & 0 & 0 & 0 & 0 & -0.452 & 0 & 0 & 0 & 1 & 0 & 0 & 0 & 0 & 0 & 0 & 0 & 0 & 0 & 0 & 0 & 0 & 0 & 0 & -1 \\
\mathrm{GLU} & 0 & 0 & -1 & 0 & 0 & 0 & 0 & 1 & 1 & 0 & 1 & 0 & 0 & 0 & 0 & 0.0082 & -0.0479 & 0 & 0 & 0 & 0 & 0 & 0 & 0 & -1 & 0 & 0 & 0 & 0 & 0 & 0 & 0 & -1 & 0 & -2 \\
\mathrm{GLN} & 0 & 0 & 0 & 0 & 0 & 0 & 0 & -1 & 0 & 0 & 0 & 0 & 0 & 0 & 0 & -0.0873 & -0.0449 & 0 & 0 & 0 & 0 & 0 & 0 & 0 & 0 & 1 & 0 & 0 & 0 & 0 & 0 & 0 & 0 & 0 & 0 \\
\mathrm{GLY} & 0 & 0 & 0 & 0 & 0 & 0 & 0 & 0 & 0 & 0 & 0 & 2 & 0 & 0 & 0 & -0.056 & -0.0719 & 0 & 0 & 0 & 0 & 0 & 0 & 0 & 0 & 0 & -1 & 0 & 0 & 0 & 0 & 0 & 0 & 0 & 0 \\
\mathrm{LAC} & 0 & 1 & 0 & 0 & 0 & 0 & 0 & 0 & 0 & 0 & 0 & 0 & 0 & 0 & 0 & 0 & 0 & 0 & 0 & 0 & -1 & 0 & 0 & 0 & 0 & 0 & 0 & 0 & 0 & 0 & 0 & 0 & 0 & 0 & 0 \\
\mathrm{MAL} & 0 & 0 & 0 & 0 & 1 & -1 & -1 & 0 & 0 & 0 & 0 & 0 & 0 & 0 & 0 & 0.4445 & 0 & 0 & 0 & 0 & 0 & 0 & 0 & 0 & 0 & 0 & 0 & 0 & 0 & 0 & 0 & 0 & 0 & 0 & 0 \\
\mathrm{NADH\,(Cytosolic)} & 2 & -1 & 0 & 0 & 0 & 0 & 0 & 0 & 0 & 0 & 0 & -1 & -1 & 0 & 0 & 0 & 0 & 0 & 0 & 0 & 0 & 0 & 0 & 0 & 0 & 0 & 0 & 0 & 0 & 0 & 0 & 0 & 0 & -1 & 2 \\
\mathrm{NADH\,(Mitochondrial)} & 0 & 0 & 0 & 2 & 1 & 1 & 0 & 0 & -1 & 0 & 0 & 0 & 0 & -1 & 0 & 0.6391 & 0 & 0 & 0 & 0 & 0 & 0 & 0 & 0 & 0 & 0 & 0 & 0 & 0 & 0 & 0 & 0 & 0 & 0.5 & 0 \\
\mathrm{NH_3} & 0 & 0 & 0 & 0 & 0 & 0 & 0 & 1 & -1 & 1 & 0 & -1 & 0 & 0 & 0 & 0 & 0 & 0 & 0 & 0 & 0 & 0 & 0 & 0 & 0 & 0 & 0 & 0 & -1 & 0 & 0 & 0 & 0 & 0 & 0 \\
\mathrm{O_2} & 0 & 0 & 0 & 0 & 0 & 0 & 0 & 0 & 0 & 0 & 0 & 0 & 0 & -0.5 & -0.5 & 0 & 0 & 0 & 0 & 0 & 0 & 0 & 0 & 0 & 0 & 0 & 0 & 0 & 0 & 1 & 0 & -1 & 0 & 0 & 0 \\
\mathrm{OXA} & 0 & 0 & 0 & -1 & 0 & 1 & 0 & 0 & 0 & 0 & 1 & 0 & 0 & 0 & 0 & -0.427 & 0 & 0 & 0 & 0 & 0 & 0 & 0 & 0 & 0 & 0 & 0 & 0 & 0 & 0 & 0 & 0 & 0 & 0 & 0 \\
\mathrm{PYR} & 2 & -1 & -1 & -1 & 0 & 0 & 1 & 0 & 0 & 0 & 0 & 0 & 0 & 0 & 0 & 0.2085 & 0 & 0 & 0 & 0 & 0 & 0 & 0 & 0 & 0 & 0 & 0 & 0 & 0 & 0 & 0 & 0 & 0 & 0 & 0 \\
\mathrm{SER} & 0 & 0 & 0 & 0 & 0 & 0 & 0 & 0 & 0 & 0 & 0 & -1 & 0 & 0 & 0 & -0.096 & -0.1 & 0 & 0 & 0 & 0 & 0 & 0 & 0 & 0 & 0 & 0 & 1 & 0 & 0 & 0 & 0 & 0 & 0 & 2 \\
\end{bNiceMatrix}.$} \endgroup}}

\section{Closed-form Solution for Stoichiometric-based Approach (w/o Inequality Bounds)}
\label{sec:stoi_sol}
The objective function for Equation \cref{eqn:stoi_based} could be rewritten as:
\begin{equation}
    ||\vI - \vIhat||_2^2 = \vI^T \cdot \vI - 2 \vIhat \cdot \vI + \vIhat^T \cdot \vIhat. \label{eqn:obj}
\end{equation}
Notice that the last term of \cref{eqn:obj} is not the function of $\vI$ or $\vE$, which will be further dropped, and lead to a revised form of the optimization problem:
\begin{equation}
    \label{eqn:opt}
    \min_{\vI, \vE} \vI^T \cdot \vI - 2 \vIhat \cdot \vI, \text{ s.t. } \SmatI \cdot \vI + \SmatE \cdot \vE = 0.
\end{equation}
In order to solve this optimization problem, we define the Lagrangian function
\begin{equation}
    \mathcal{L} = \vI^T \cdot \vI - 2 \vIhat \cdot \vI + \LM^T \cdot (\SmatI \cdot \vI + \SmatE \cdot \vE), \label{eqn:lag}
\end{equation}
where $\LM \in \mathbb{R}^{\numMeta}$ is the vector of Lagrange multiplier. A necessary condition for $\vI$ to be an extremum is
\begin{equation}
    \frac{\partial \mathcal{L}}{\partial \vI} = 2 \vI - 2 \vIhat + \SmatI^T \LM = 0. \label{eqn:grad_vI}
\end{equation}
A necessary condition for $\vE$ to be an extremum is
\begin{equation}
    \frac{\partial \mathcal{L}}{\partial \vE} = \SmatE^T \LM = 0. \label{eqn:grad_vE}
\end{equation}
A necessary condition for $\LM$ to be an extremum is just the constraint
\begin{equation}
    \frac{\partial \mathcal{L}}{\partial \LM} = \SmatI \cdot \vI + \SmatE \cdot \vE = 0. \label{eqn:grad_gamma}
\end{equation}
Notice that \cref{eqn:grad_vI}, \cref{eqn:grad_vE} and \cref{eqn:grad_gamma} are linear equations of $\vE$, $\vI$ and $\LM$, which leads to a linear system of $\numExt + \numInt + \numMeta = 21 + 14 + 24 = 59$ equations and variables:
\begin{equation}
    \begin{bmatrix}
        2\mat{I}_{\numInt} & 0 & \SmatI^T \\
        0 & 0 & \SmatE^T \\
        \SmatI & \SmatE & 0 \\
    \end{bmatrix} \begin{bmatrix}
        \vI \\ \vE \\ \LM
    \end{bmatrix} = \begin{bmatrix}
        2 \vIhat \\ 0 \\ 0
    \end{bmatrix}, \label{eqn:lag_sys}
\end{equation}
where $\mat{I}_{\numInt} \in \mathbb{R}^{\numInt \times \numInt}$ is the identity matrix. As we computed that the determinant of
\begin{equation}
    \mat{A} = \begin{bmatrix}
    2 \mat{I}_{\numInt} & 0 & \SmatI^T \\
    0 & 0 & \SmatE^T \\
    \SmatI & \SmatE & 0 \\
    \end{bmatrix} \label{eqn:lag_lhs}
\end{equation}
is nonzero (aka $\mat{A}$ is invertible), we know \cref{eqn:lag_sys} has a unique solution:
\begin{equation}
    \begin{bmatrix}
        \vI \\ \vE \\ \LM
    \end{bmatrix} = \begin{bmatrix}
    2 \mat{I}_{\numInt} & 0 & \SmatI^T \\
    0 & 0 & \SmatE^T \\
    \SmatI & \SmatE & 0 \\
    \end{bmatrix}^{-1} \begin{bmatrix}
    2 \vIhat \\ 0 \\ 0
    \end{bmatrix}. \label{eqn:lag_sol}
\end{equation}
If we denote $\mat{I}_{\numExt} \in \mathbb{R}^{\numExt \times \numExt}$ as an identity matrix, then
\begin{equation}
    \begin{bmatrix}
    \vI \\ \vE 
    \end{bmatrix} = \begin{bmatrix}
    \mat{I}_{\numInt} & 0 & 0 \\
    0 & \mat{I}_{\numExt} & 0 \\
    \end{bmatrix} \begin{bmatrix}
    \vI \\ \vE \\ \LM
    \end{bmatrix} = \begin{bmatrix}
    \mat{I}_{\numInt} & 0 & 0 \\
    0 & \mat{I}_{\numExt} & 0 \\
    \end{bmatrix} \begin{bmatrix}
    2 \mat{I}_{\numInt} & 0 & \SmatI^T \\
    0 & 0 & \SmatE^T \\
    \SmatI & \SmatE & 0 \\
    \end{bmatrix}^{-1} \begin{bmatrix}
    2 \vIhat \\ 0 \\ 0
    \end{bmatrix} \label{eqn:fpssbform}
\end{equation}
is the closed-form expression for stoichiometric-based approach (w/o bounds).

\newpage
\section{Additional figures}
\label{sec:appendixDataFig}

\begin{figure}[!ht]
    \centering
        \includegraphics[width=0.96\textwidth]{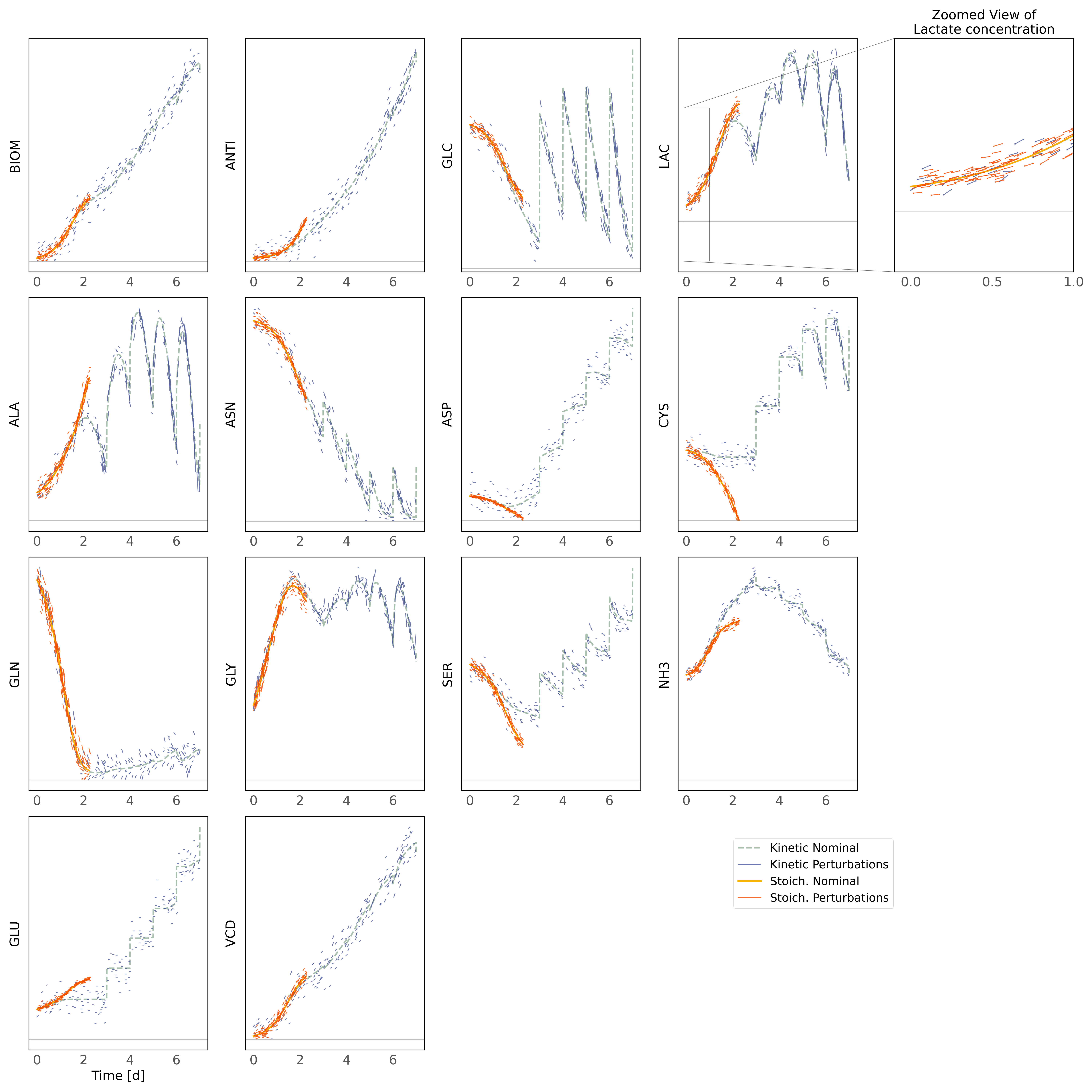}
        \caption{
            Nominal trajectories overlaid on sampled short-time flows from perturbed initial conditions.
            Color is used to distinguish between different curves described in the legend.
        }
    \label{fig:nominalAll}
\end{figure}

\begin{figure}[!ht]
    \centering
        \includegraphics[width=0.96\textwidth]{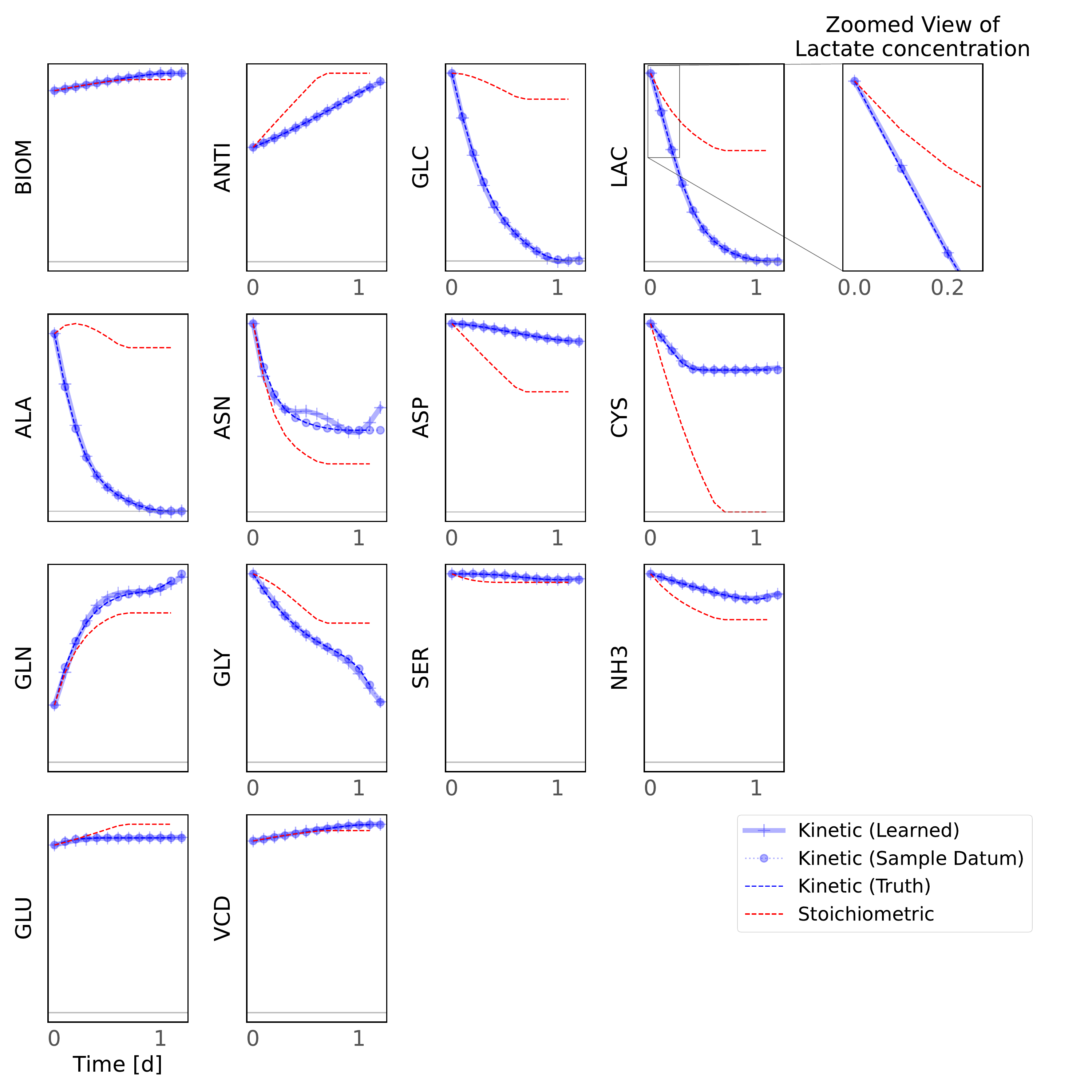}
        \caption{
            Black-box predictions all variables (\typeOne). See also \figref{fig:bb_results_kin}.
            Color is used to distinguish between different curves described in the legend.
        }
    \label{fig:bb_traj_comparison-KINETIC-14}
\end{figure}

\begin{figure}[!ht]
    \centering
        \includegraphics[width=0.96\textwidth]{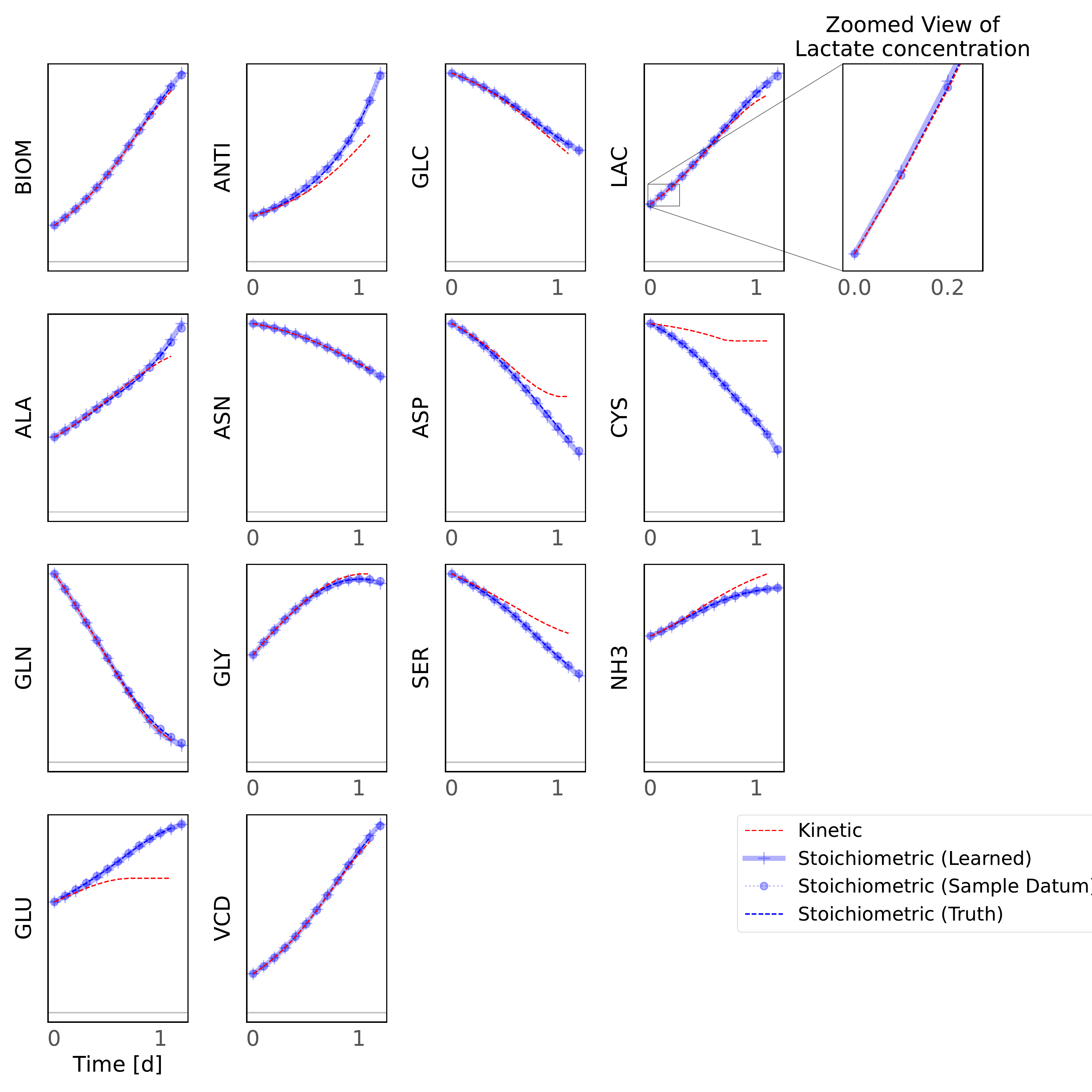}
        \caption{
            Black-box predictions; all variables (\typeTwo).
            Color is used to distinguish between different curves described in the legend.
        }
    \label{fig:bb_traj_comparison-STOICHIOMETRIC-14}
\end{figure}

\begin{figure}[!ht]
    \centering
        \begin{subfigure}[b]{0.45\textwidth}
            \includegraphics[width=\textwidth]{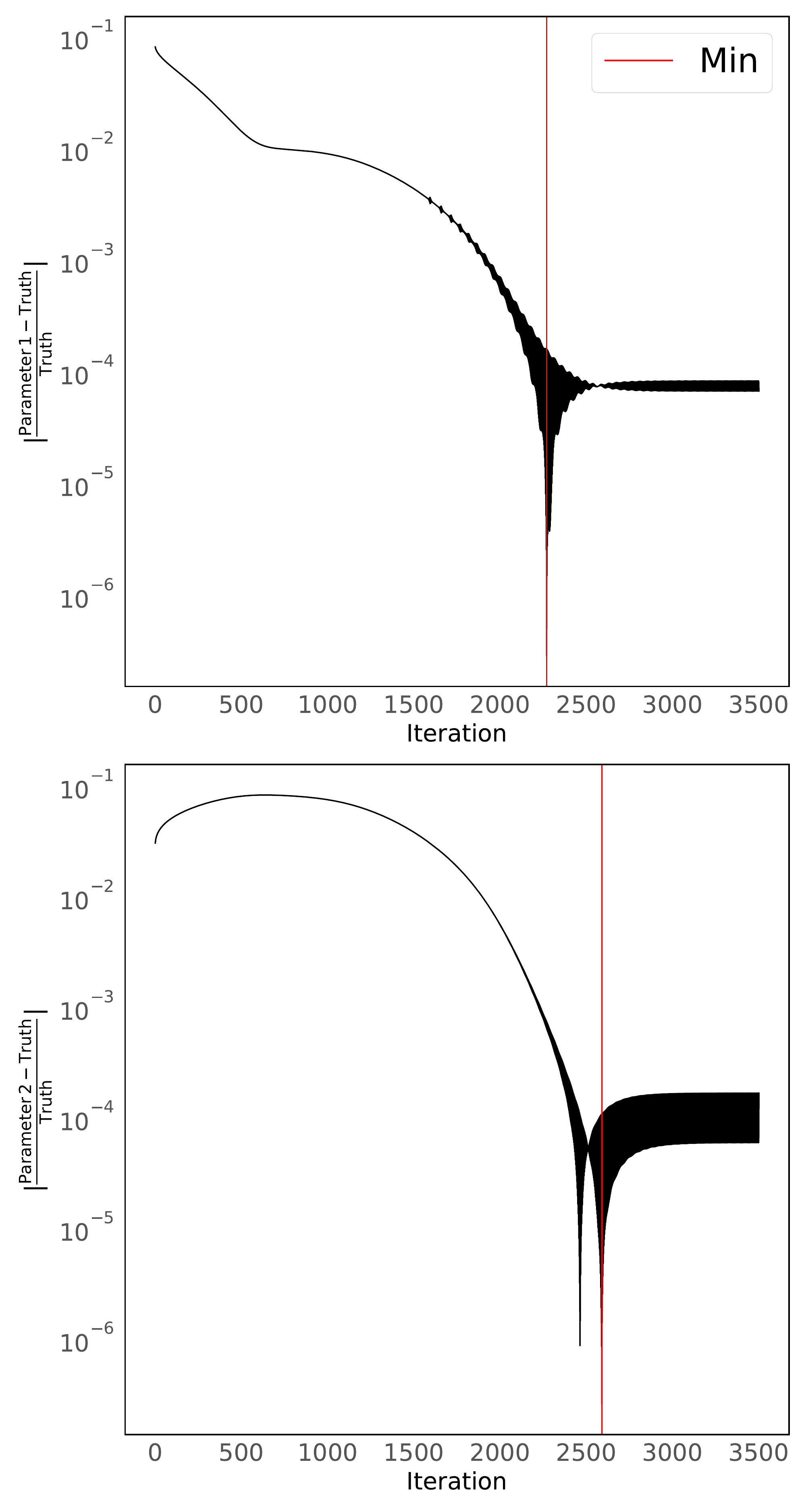}
            \caption{Kinetic}
        \end{subfigure}
        \begin{subfigure}[b]{0.45\textwidth}
            \includegraphics[width=\textwidth]{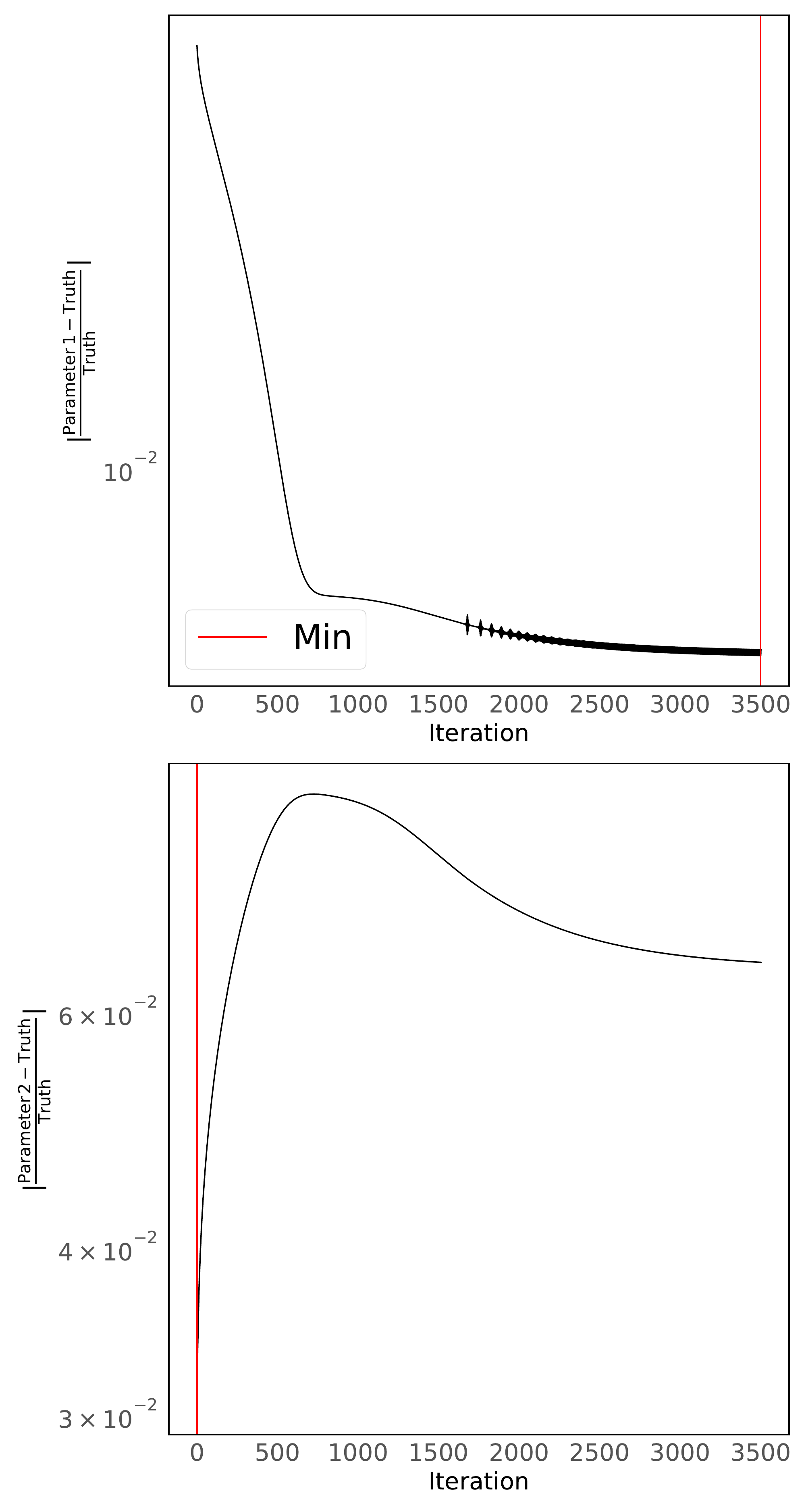}
            \caption{Stoichiometric}
        \end{subfigure}
    \caption{
        Absolute parameter error history.
        (Two-parameter white-box case.)
    }
    \label{fig:absrelTwoParam}
\end{figure}

\begin{figure}[!ht]
    \centering
    \begin{subfigure}[b]{0.45\textwidth}
            \includegraphics[width=\textwidth]{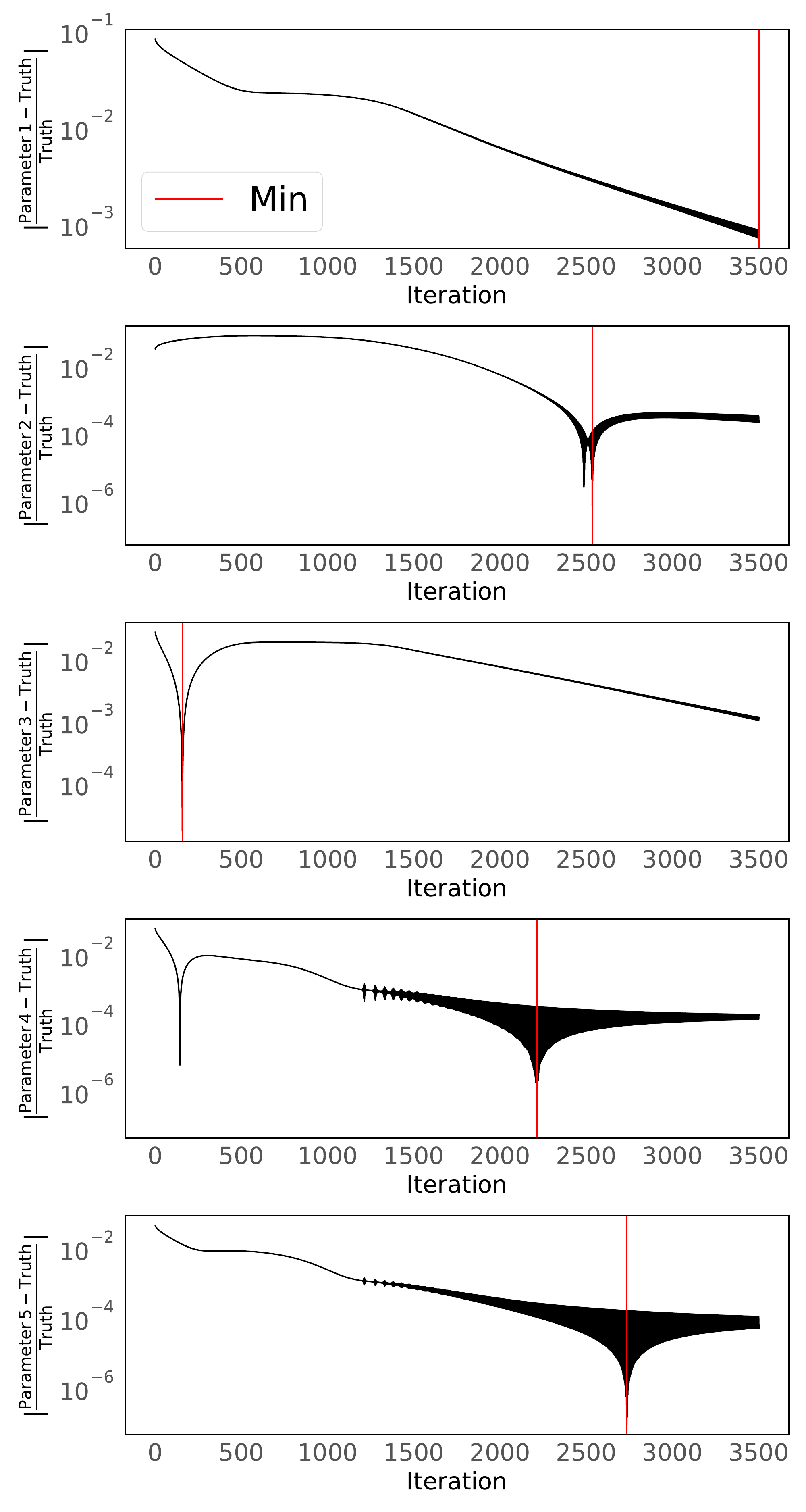}
            \caption{Kinetic}
        \end{subfigure}
        \begin{subfigure}[b]{0.45\textwidth}
            \includegraphics[width=\textwidth]{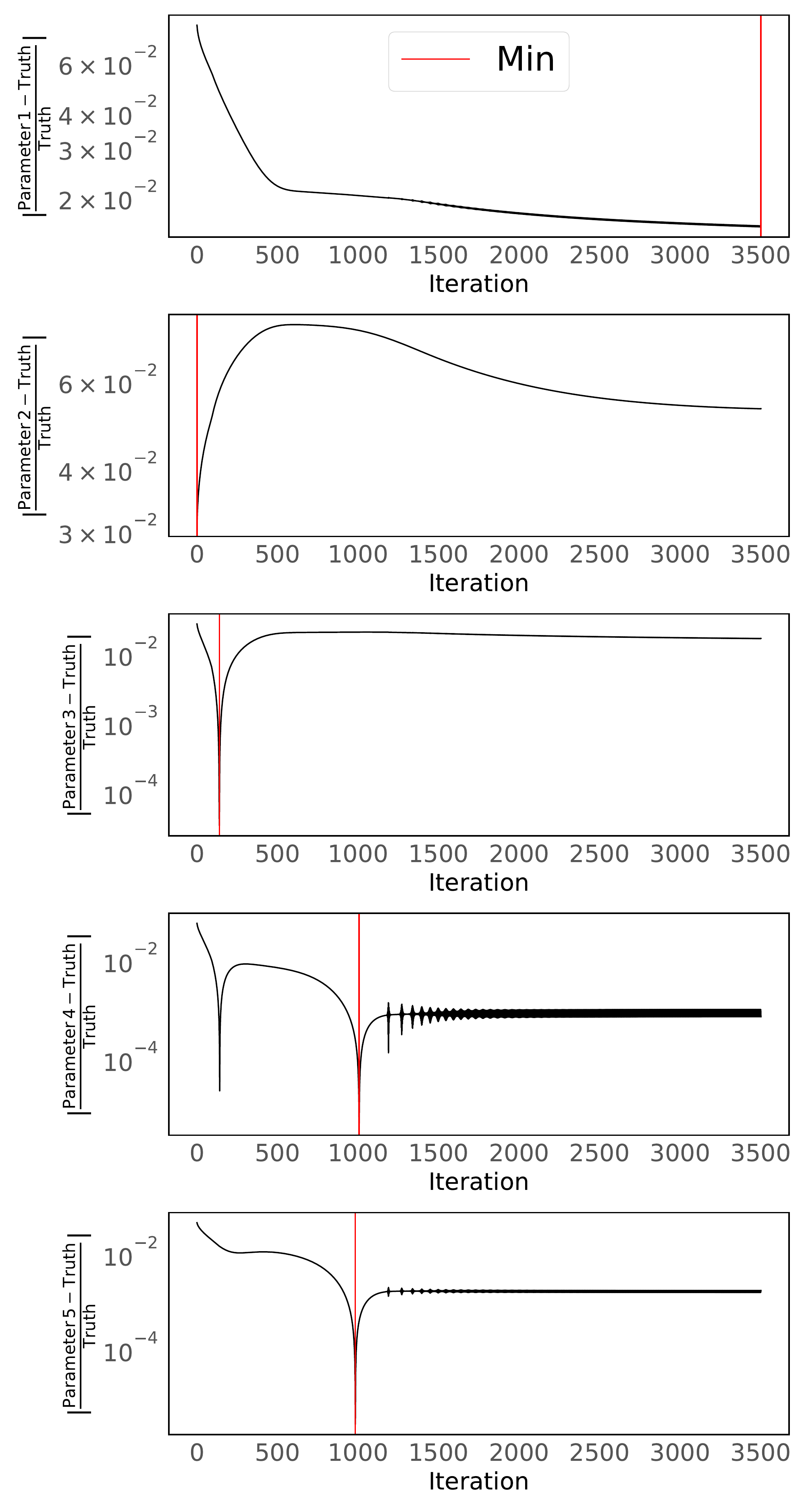}
            \caption{Stoichiometric}
        \end{subfigure}
        \caption{Absolute parameter error history, five-parameter white-box case.}
    \label{fig:absrelFiveParam}
\end{figure}

\end{document}